\documentclass[preprint,aps,floats,nofootinbib,amssymb]{revtex4}
\usepackage{epsf,epsfig}

\newcommand{\be}{\begin{equation}}
\newcommand{\ee}{\end{equation}}
\newcommand{\bea}{\begin{eqnarray}}
\newcommand{\eea}{\end{eqnarray}}
\newcommand{\nn}{\nonumber}

\newcommand{\lsp}{\chi^0_1}
\renewcommand{\check}{({\bf This statement needs to be checked!!})}
\newcommand{\lsim}{
\mathrel{\hbox{\rlap{\hbox{\lower4pt\hbox{$\sim$}}}\hbox{$<$}}}}
\newcommand{\gsim}{
\mathrel{\hbox{\rlap{\hbox{\lower4pt\hbox{$\sim$}}}\hbox{$>$}}}}

\newcommand{\half}{\frac{1}{2}}

\newcommand{\ba}{\begin{array}}
\newcommand{\ea}{\end{array}}

\preprint{
\hbox to \hsize{
\hfill$\vcenter{\hbox{\bf MADPH-06-1254}
        \hbox{\bf UFIFT-HEP-06-08}
            \hbox{\bf UPR-1143T}
                \hbox{\bf hep-ph/0603247}
                \hbox{March 2006}}$}
}

\begin{document}

\title{\vspace*{.75in}
Higgs Sector in Extensions of the MSSM}

\author{
Vernon Barger$^1$, Paul Langacker$^2$, Hye-Sung Lee$^3$, and Gabe Shaughnessy$^1$}

\affiliation{
$^1$Department of Physics, University of Wisconsin,
Madison, WI 53706 \\
$^2$Department of Physics and Astronomy, University of Pennsylvania,
Philadelphia, PA 19104 \\
$^3$Department of Physics, University of Florida,
Gainesville, FL 32608
\vspace*{.5in}}

\thispagestyle{empty}

\begin{abstract}
\noindent

Extensions of the Minimal Supersymmetric Standard Model (MSSM) with additional singlet scalar fields solve the important $\mu$-parameter fine tuning problem of the MSSM.  We compute and compare the neutral Higgs boson mass spectra, including one-loop corrections, of the following MSSM extensions: Next-to-Minimal Supersymmetric Standard Model (NMSSM), the nearly-Minimal Supersymmetric Standard Model (nMSSM), and the $U(1)'$-extended Minimal Supersymmetric Standard Model (UMSSM) by performing scans over model parameters.  We find that the Secluded $U(1)'$-extended Minimal Supersymmetric Standard Model (sMSSM) is identical to the nMSSM if three of the additional scalars decouple.  The dominant part of the one-loop corrections are model-independent since the singlet field does not couple to MSSM particles other than the Higgs doublets.  Thus, model-dependent parameters enter the masses only at tree-level.  We apply constraints from LEP bounds on the Standard Model and MSSM Higgs boson masses and the MSSM chargino mass, the invisible $Z$ decay width, and the $Z-Z'$ mixing angle.  Some extended models permit a Higgs boson with mass substantially below the SM LEP limit or above theoretical limits in the MSSM.  Ways to differentiate the models via masses, couplings, decays and production of the Higgs bosons are discussed.

\end{abstract}
\maketitle

\newpage
\section{Introduction}

Supersymmetry (SUSY) is a leading candidate for physics beyond the Standard Model (SM).  It is the only extension of the bosonic spacetime Poincar$\acute{e}$ symmetry to include a fermionic spacetime.  Superstring theory, the currently prevailing paradigm of quantum gravity, generally includes SUSY, though not necessarily at the weak scale.  The cancellation of the quadratic divergence in the Higgs mass-squared radiative correction, requiring fine-tuning in the SM, strongly motivates SUSY at the TeV scale. TeV-scale SUSY also unites the gauge coupling constants at the GUT scale and provides an attractive cold dark matter candidate, the lightest neutralino, when R-parity is conserved.  The simplest supersymmetric extension of the SM is the Minimal Supersymmetric Standard Model (MSSM).  

The MSSM suffers from the $\mu$-problem \cite{muproblem}.  The $\mu$-parameter is the only dimensionful parameter in the SUSY conserving sector.  Naively, in a top down approach, one would expect the $\mu$-parameter to be either zero or at the Planck scale, ${\cal O}(10^{19})$ GeV.  At tree-level, the MSSM gives the relation \cite{ref:sugrawgc}
\be 
\half M_Z^2={m_{d}^2 - m_{u}^2 \tan^2\beta \over \tan^2\beta -1} - \mu^2,
\ee
where $m_d$ and $m_u$ are the soft mass parameters for the down-type and up-type Higgs, respectively.  With the soft parameters at the EW/TeV scale, $\mu$ must be at the same scale, while LEP constraints on the chargino mass require $\mu$ to be non-zero \cite{ref:lepsusy}.  A simple solution is to promote the $\mu$-parameter to a dynamical field in extensions of the MSSM that contain an additional singlet scalar field that does not interact with MSSM fields other than the two Higgs doublets.  Extended models thereby circumvent the need for a fine tuning of the $\mu$-parameter to the electroweak scale.  

The discovery of Higgs bosons is a primary goal of the Tevatron and the  Large Hadron Collider (LHC) experiments.  Although Higgs boson signals at colliders have been extensively studied, most of these studies were based on the assumption that the Higgs bosons occur only in doublet fields \cite{review:run2:atlas:cms}.  The few case studies of the Higgs sector in the extensions of the MSSM have not been as comprehensive as the SM and MSSM Higgs studies \cite{ref:studies,ref:htoaa}.  With the addition of singlet scalar fields, the properties of the Higgs bosons can be substantially different from those in the SM or the MSSM.  Moreover, with SUSY, there are also one or more extra neutralinos and there may be an extra neutral gauge boson in some models. 

In this paper we consider models with an extra Higgs singlet field that yield a dynamical solution to the $\mu$-problem.  The dynamical field that gets a vacuum expectation value (VEV) generates an effective $\mu$-parameter that is associated with a new symmetry.  These models have a third CP-even Higgs boson and, in some cases, an extra CP-odd Higgs boson.  The mixing with the extra scalar state alters the masses and couplings of the physical Higgs bosons.  We evaluate the phenomenological consequences of an extra scalar for the Higgs masses, couplings, decays and production.  We include one-loop radiative corrections to the Higgs masses, which to a good approximation turn out to be common among the models at this order for the neutral and charged Higgs boson sector.  While performing our systematic study on the Higgs sector alone, we consider indirect consequences from the neutralino sector in anticipation of a later full treatment including both sectors.  Detailed studies of the neutralino sector in these models have been done by examining the lightest neutralino \cite{xMSSM_neutralino}.  We translate the constraints from LEP experiments on the SM (lightest MSSM) Higgs into limits on the CP-even (CP-even and CP-odd) Higgs boson masses in the extended models and include constraints from the LEP chargino mass limit,  the invisible $Z$ width and the $Z-Z'$ mixing angle. 


The extended models of present interest\footnote{Many of the ideas of some of the models appeared already in Ref. \cite{Fayet}.  For a recent review of supersymmetric singlet models, see Ref. \cite{cpnsh}.} are the Next-to-Minimal Supersymmetric Standard Model (NMSSM) \cite{NMSSM}, the Minimal Nonminimal Supersymmetric Standard Model (MNSSM) or the nearly Minimal Supersymmetic Standard Model (nMSSM) \cite{nMSSM}, the $U(1)'$-extended Minimal Supersymmetric Standard Model (UMSSM) \cite{UMSSM}, and the Secluded $U(1)'$-extended Minimal Supersymmetric Standard Model (sMSSM) \cite{SUMSSM}.  A common $\mu$-generating term, $h_s \hat H_u \cdot \hat H_d \hat S$, is contained in the superpotentials of these models, which are listed in Table \ref{table:model}.  After the $S$ field gets a VEV, the effective $\mu$-parameter is identified as 
\be
\mu_{\rm eff} = h_s \langle S \rangle.
\ee
where $\langle S \rangle$ denotes the VEV of the singlet field.  

The defining feature of each model is the symmetry that is allowed by the superpotential.  The NMSSM has a discrete $\mathbb Z_3$ symmetry, allowing the $S^3$ term \cite{NMSSM,NMSSM_Higgs}.  With any discrete symmetry, the possibility of domain walls exists.  It has been shown that domain walls can be viewed as a source of dark energy \cite{darkenergydomainwall}.  In the NMSSM, the equation of the state, $p=w\rho$, of dark energy is predicted to have $w = -2/3$ which is disfavored by a recent analysis of WMAP data that place $w=-1.062^{+0.128}_{-0.079}$ \cite{Spergel:2006hy}.

The domain walls may be eliminated if the $\mathbb Z_3$ symmetry is broken by higher dimensional operators, but these may lead to very large destabilizing tadpole operators \cite{domainwall}; one possibility for avoiding this problem is described in Ref. \cite{NMSSMwodomain}.  The nMSSM with a $\mathbb Z_5^R$ or $\mathbb Z_7^R$ symmetry has a  tadpole term of $\hat S$ that breaks the discrete symmetries and is thus free from domain walls \cite{nMSSM, nMSSM_Higgs}.  The harmful tadpole divergences can destabilize the gauge hierarchy, but the discrete symmetries $\mathbb Z^R_5$ or $\mathbb Z^R_7$ allow the divergences to exist only at six and seven-loop order, respectively \cite{nMSSM_Higgs}.  At these orders, the divergences are suppressed at scales below $M_{\rm Planck}$.

An extra $U(1)$ gauge symmetry, $U(1)'$, is motivated by many models beyond the SM, including grand unified theories (GUT) \cite{gutu1mot,ref:hewettrizzo}, extra dimensions \cite{U1_xd}, superstrings \cite{U1_string}, little Higgs \cite{U1_littleHiggs}, dynamical symmetry breaking \cite{U1_strongdynamics} and the Stueckelberg mechanism \cite{U1_stueckelberg}.  The UMSSM and sMSSM each contains a $U(1)'$ gauge symmetry and its gauge boson, $Z'$, that can mix with the SM after symmetries are broken $Z$ \cite{UMSSM, UMSSM_Higgs}.  While the continuous $U(1)'$ symmetry is free from domain wall constraints, the UMSSM may require exotic fields \cite{Erler:2000wu,Batra,Morrissey:2005uz} to cancel chiral anomalies related to the $U(1)'$ symmetry\footnote{Exotic fermions can be avoided in a family non-universal $U(1)'$ model \cite{Demir:2005ti}. }.  There are constraints on the UMSSM from the strict experimental limits on $Z-Z'$ mixing that are at the mil-level \cite{LEPmixingangle}.  The $Z'$ mass must be above 600-900 GeV to satisfy the Tevatron dilepton search results, with the precise experimental limit dependent on the $U(1)'$ model \cite{ref:zpmasslim}.  With a leptophobic $Z'$, these mass limits are evaded.  

\begin{table}[t]
\caption{Higgs bosons of the MSSM and several of its extensions.  We denote the single CP-odd state in the MSSM and UMSSM by $A_2^0$ for easier comparison with the other models.
\label{table:model}}
\begin{tabular}{|r|c|l|l|l|c|}
\hline
Model~~ & Symmetry & ~~~~~~~Superpotential & ~~~~~~~~~CP-even & ~~~~CP-odd & Charged\\
\hline
MSSM   & --                             & $\mu \hat H_u \cdot \hat H_d$                                               & $H_1^0, H_2^0$        & $A_2^0$        & $H^\pm$ \\
NMSSM  & $\mathbb Z_3$                  & $h_s \hat S \hat H_u \cdot \hat H_d + \frac{\kappa}{3} \hat S^3$            & $H_1^0, H_2^0, H_3^0$ & $A_1^0, A_2^0$ & $H^\pm$\\
nMSSM  & $\mathbb Z^R_5, \mathbb Z^R_7$ & $h_s \hat S \hat H_u \cdot \hat H_d + \xi_F M_{\rm n}^2 \hat S$          & $H_1^0, H_2^0, H_3^0$ & $A_1^0, A_2^0$ & $H^\pm$\\
UMSSM  & $U(1)'$                        & $h_s \hat S \hat H_u \cdot \hat H_d$                                        & $H_1^0, H_2^0, H_3^0$ & $A_2^0$        & $H^\pm$ \\
 sMSSM & $U(1)'$                        & $h_s \hat S \hat H_u \cdot \hat H_d + \lambda_s \hat S_1 \hat S_2 \hat S_3$ & $H_1^0, H_2^0, H_3^0, H_4^0, H_5^0, H_6^0$ & $A_1^0, A_2^0, A_3^0, A_4^0$ & $H^\pm$\\
\hline
\end{tabular}
\end{table}

The Higgs field content of the above listed models is given in Table \ref{table:model}.  In the MSSM, the usual 2 Higgs doublets give two CP-even ($H^0_1$, $H^0_2$), a CP-odd ($A_2$), and a pair of charged ($H^\pm$) Higgs bosons\footnote{We ignore the possibility of CP-violating mixing effects.}.  The extended models include additional CP-even Higgs bosons and CP-odd Higgs bosons or a $Z'$ gauge boson, depending on the model.  The sMSSM contains three additional singlets that allow six CP-even and four CP-odd Higgs states.  However, the additional Higgs fields decouple if $\lambda$ is small and the vacuum expectation values $\langle S_1\rangle,\langle S_2\rangle,\langle S_3\rangle$ are large.  The decoupling limit eliminates the $D$-terms in the mass-squared matrix for the $S,H^0_d$, and $H^0_u$ fields and yields a model similar to the nMSSM with three CP-even and two CP-odd Higgs bosons.  This is shown in Appendix \ref{apx:sumssmdecoup}.  We shall therefore refer to the nMSSM as n/sMSSM since the results of the nMSSM correspond to the sMSSM in the decoupling regime.  The charged Higgs sector for all of these models remains the same as in the MSSM due to the assumption that the number of Higgs doublets is unchanged.

We present an overview of the Higgs mass-squared matrices including radiative corrections due to top and stop loops in Section \ref{sect:massmtx}. We discuss the experimental and theoretical constraints applied in Section \ref{sect:constraints} and the details of the parameter scans in Section \ref{sect:scan}. In Section \ref{sect:results}, we discuss the Higgs spectra and couplings for various models, while implications for collider phenomenology are presented in Section \ref{sect:collpheno}.  Finally, we summarize our results in Section \ref{sect:concl}.  We provide details of decoupling of the sMSSM in Appendix \ref{apx:sumssmdecoup}.  The derivation of the mass-squared matrices of each model are presented in Appendix \ref{apx:Higgs} and the neutralino mass matrices are given in Appendix \ref{apx:neut}.  In Appendix \ref{apx:masses}, important limits in the Higgs sector are addressed, while additional information on the heavier states is given in Appendix \ref{apx:addparm}.

\section{Higgs Mass Matrices}\label{sect:massmtx}
\subsection{Tree-level}\label{sect:tree-level}

The tree-level Higgs mass-squared matrices are found from the potential, $V$, which is a sum of the $F$-term, $D$-term and soft-terms in the lagrangian, as follows.
\bea
V_F &=& |h_s H_u\cdot H_d+\xi_F M_{\rm n}^2+ \kappa S^2|^2 + |h_s S|^2 \left(|H_d|^2+|H_u|^2 \right), \\
V_D &=& \frac{G^2}{8}\left( |H_d|^2-|H_u|^2 \right)^2+ \frac{g_{2}^2}{2} \left( |H_d|^2|H_u|^2-|H_u \cdot H_d|^2 \right),\\
 &+& {{g_{1'}}^2\over2}\left(Q_{H_d} |H_d|^2+Q_{H_u} |H_u|^2+Q_{S} |S|^2\right)^2\\ \nn
V_{\rm soft}&=&m_{d}^{2}|H_d|^2 + m_{u}^{2}|H_u|^2+ m_s^{2}|S|^2 + \left( A_s h_s S H_u\cdot H_d + {\kappa \over 3} A_{\kappa} S^3+\xi_S M_{\rm n}^3 S + h.c. \right).
\label{eq:potential}
\eea
Here, the two Higgs doublets with hypercharge $Y=-1/2$ and $Y=+1/2$, respectively, are
\be
H_d = \left( \begin{array}{c} H_d^0 \\ H^- \end{array} \right), \qquad
H_u = \left( \begin{array}{c} H^+ \\ H_u^0 \end{array} \right).
\ee
and $H_u \cdot H_d = \epsilon_{ij} H_u^i H_d^j$.  For a particular model, the parameters in $V$ are understood to be turned-off appropriately according to Table \ref{table:model}
\bea
{\rm NMSSM}&:& g_{1'}=0, M_{\rm n} = 0,\nn\\
{\rm nMSSM}&:& g_{1'}=0, \kappa=0, A_\kappa = 0, \\
{\rm UMSSM}&:& M_{\rm n} = 0, \kappa = 0, A_\kappa = 0.\nn
\eea
The couplings $g_1,g_2$, and ${g_{1'}}$ are for the $U(1)_Y,SU(2)_L$, and $U(1)'$ gauge symmetries, respectively, and the parameter $G$ is defined as $G^2=g_1^2+g_2^2$.  
The NMSSM model-dependent parameters are $\kappa$ and $A_\kappa$ while the free nMSSM parameters are $\xi_F$ and $\xi_S$ with $M_{\rm n}$ being fixed near the SUSY scale.  The model dependence of the UMSSM is expressed by the $D$-term that has the $U(1)'$ charges of the Higgs fields, $Q_{H_d}, Q_{H_u}$ and $Q_S$.  In general, these charges are free parameters with the restriction\footnote{Additional restrictions on the charges of the ordinary and exotic particles come from the cancellation of anomalies.} that $Q_{H_d}+Q_{H_u}+Q_{S}=0$ to preserve gauge invariance.  In any particular $U(1)'$ construction, the charges have specified values.  We assume the charges of a $E_6$ model that breaks via the chain $E_6 \to SO(10)\times U(1)_\psi \to SU(5)\times U(1)_\chi \times U(1)_\psi$ \cite{ref:hewettrizzo}.  At some high energy scale, the $U(1)_{\chi}\times U(1)_{\psi}$ symmetry is assumed to  break into one $U(1)'$\footnote{This is the same breaking scheme as in the Exceptional Supersymmetric Standard Model (ESSM) \cite{ESSM}. In the ESSM, among three pairs of SU(2) doublet scalars with MSSM Higgs quantum numbers and three singlet scalars, only one pair of doublets and one singlet develop VEVs due to an extra $Z_2^H$ symmetry and imposed hierarchical structure of the Yukawa interactions, yielding a model similar to the UMSSM.}.  The above breaking scenario results in the charges
\be
Q_{H_d} = {-1\over \sqrt{10}} \cos \theta_{E_6}-{1\over\sqrt6} \sin\theta_{E_6},\qquad 
Q_{H_u} = {1\over \sqrt{10}} \cos \theta_{E_6}-{1\over\sqrt6} \sin\theta_{E_6},
\ee
where $\theta_{E_6}$ is the mixing angle between the two $U(1)$s and is the only model-dependent parameter. 

The $F$-term and the soft terms contain the model dependence of the NMSSM and n/sMSSM.  The soft terms $A_{\kappa}$ of the NMSSM and $\xi_S M_{\rm n}^3$ of the n/sMSSM are new to $V_{\rm soft}$.  The $B$-term of the MSSM is expressed in $V_{\rm soft}$ as $A_s h_sSH_u\cdot H_d$ after we identify 
\be
B \mu = A_s \mu_{\rm eff}.
\ee
The other terms in $V_{\rm soft}$ are the usual MSSM soft mass terms. 

The minimum of the potential is found explicitly using the minimization conditions found in Appendix \ref{apx:Higgs}.  The conditions found allow us to express the soft mass parameters in terms of the VEVs of the Higgs fields.  At the minimum of the potential, the Higgs fields are expanded as  
\be
H_d^0 = \frac{1}{\sqrt{2}} \left( v_d + \phi_d + i \varphi_d \right), \quad
H_u^0 = \frac{1}{\sqrt{2}} \left( v_u + \phi_u + i \varphi_u \right), \quad
S     = \frac{1}{\sqrt{2}} \left( s + \sigma + i \xi \right).
\label{eq:fieldexp}
\ee
with $v^2 \equiv v_d^2+v_u^2 = (246{\rm ~GeV})^2$ and $\tan\beta \equiv v_u / v_d$.    
We write the Higgs mass-squared matrix in a compact form that includes all the extended models under consideration.  The CP-even tree-level matrix elements in the $H_d^0, H_u^0, S$ basis are:
\bea
\label{eq:cpetree1}
\left({\mathcal{M}_{+}^0}\right)_{11} &=&  \left[\frac{G^2}{4} + Q_{H_d}^{2} {g_{1'}}^{2}\right] v_d^2 + \left(\frac{h_s A_s}{\sqrt{2}} + \frac{h_s \kappa s}{2} + \frac{h_s \xi_F M_{\rm n}^2}{s}\right) \frac{v_u s}{v_d}, \\
\left({\mathcal{M}_{+}^0}\right)_{12} &=& -\left[\frac{G^2}{4} - h_s^{2} - Q_{H_d} Q_{H_u} {g_{1'}}^{2}\right] v_d v_u - \left(\frac{h_s A_s}{\sqrt{2}} + \frac{h_s \kappa s}{2} + \frac{ h_s \xi_F M_{\rm n}^2}{s}\right) s,\\
\left({\mathcal{M}_{+}^0}\right)_{13} &=&  \left[h_s^{2} + Q_{H_d} Q_{S} {g_{1'}}^{2}\right] v_d s - \left(\frac{h_s A_s}{\sqrt{2}} + h_s \kappa s\right) v_u,\\
\left({\mathcal{M}_{+}^0}\right)_{22} &=&  \left[\frac{G^2}{4} + Q_{H_u}^{2} {g_{1'}}^{2}\right] v_u^2 + \left(\frac{h_s A_s}{\sqrt{2}} + \frac{h_s \kappa s}{2} + \frac{h_s \xi_F M_{\rm n}^2}{s} \right) \frac{v_d s}{v_u}, \\
\left({\mathcal{M}_{+}^0}\right)_{23} &=&  \left[h_s^{2} + Q_{H_u} Q_{S} {g_{1'}}^{2}\right] v_u s - \left(\frac{h_s A_s}{\sqrt{2}} + h_s \kappa s\right) v_d, \\
\left({\mathcal{M}_{+}^0}\right)_{33} &=&  \left[Q_S^{2} {g_{1'}}^{2} + 2 \kappa^2 \right] s^2 + \left(\frac{h_s A_s}{\sqrt{2}} - \frac{\sqrt{2} \xi_S M_{\rm n}^3}{v_d v_u}\right) \frac{v_d v_u}{s} + \frac{\kappa A_\kappa}{\sqrt{2}} s.
\label{eq:cpetree2}
\eea
The tree-level CP-odd matrix elements are:
\bea
\left({\mathcal{M}_{-}^0}\right)_{11} &=& \left(\frac{h_s A_s}{\sqrt{2}} + \frac{  h_s \kappa s}{2} + \frac{h_s \xi_F M_{\rm n}^2}{s}\right) \frac{v_u s}{v_d}, \\
\left({\mathcal{M}_{-}^0}\right)_{12} &=& \left(\frac{h_s A_s}{\sqrt{2}} + \frac{  h_s \kappa s}{2} + \frac{h_s \xi_F M_{\rm n}^2}{s}\right) s, \\
\left({\mathcal{M}_{-}^0}\right)_{13} &=& \left(\frac{h_s A_s}{\sqrt{2}} - h_s \kappa s\right) v_u, \\
\left({\mathcal{M}_{-}^0}\right)_{22} &=& \left(\frac{h_s A_s}{\sqrt{2}} + \frac{  h_s \kappa s}{2} + \frac{h_s \xi_F M_{\rm n}^2}{s}\right) \frac{v_d s}{v_u}, \\
\left({\mathcal{M}_{-}^0}\right)_{23} &=& \left(\frac{h_s A_s}{\sqrt{2}} - h_s \kappa s\right) v_d, \\
\left({\mathcal{M}_{-}^0}\right)_{33} &=& \left(\frac{h_s A_s}{\sqrt{2}} + 2 h_s \kappa s - \frac{\sqrt{2} \xi_S M_{\rm n}^3}{v_d v_u}\right) \frac{v_d v_u}{s} - \frac{3 \kappa A_\kappa}{\sqrt{2}} s.
\eea
The tree-level charged Higgs mass-squared matrix elements are:
\bea
\left({\mathcal{M}^\pm}\right)_{11} &=& {v_u^2 \left(g_2^2-2 h_s^2\right) \over 4}+\left({1 \over \sqrt 2} A_s h_s s+{1\over 2} h_s \kappa s^2+h_s \xi_F M_{\rm n}^2\right){v_u \over v_d},\\
\left({\mathcal{M}^\pm}\right)_{12} &=& -{v_d v_u \left(g_2^2-2 h_s^2\right) \over 4}- \left({1 \over \sqrt 2} A_s h_s s+{1\over 2} h_s \kappa s^2+h_s \xi_F M_{\rm n}^2\right),\\
\left({\mathcal{M}^\pm}\right)_{22} &=& {v_d^2 \left(g_2^2-2 h_s^2\right) \over 4}+ \left({1 \over \sqrt 2} A_s h_s s+{1\over 2} h_s \kappa s^2+h_s \xi_F M_{\rm n}^2\right){v_d\over v_u}.
\eea

The physical Higgs boson masses are found by diagonalizing the mass-squared matrices, ${\cal M}_D=R {\cal M} R^{-1}$, where ${\cal M}$ also includes the radiative corrections discussed below.  The rotation matrices for the diagonalization of the CP-even and CP-odd mass-squared matrices, $R_{\pm}^{ij}$, and for the charged Higgs matrix, ${\cal R}^{ij}$, may then be used to construct the physical Higgs fields.
 \bea
 H_{i} &=& R_{+}^{i1} \phi_d+R_{+}^{i2} \phi_u+R_{+}^{i3} \sigma, \\
 A_{i} &=& R_{-}^{i1} \varphi_d+R_{-}^{i2} \varphi_u+R_{-}^{i3} \xi, \\
 H^\pm_{i} &=& {\cal R}^{i1} H^-+{\cal R}^{i2} H^+ .
 \eea
where the physical states are ordered by their mass as $M_{H_1}\le M_{H_2}\le M_{H_3}$ and $M_{A_1}\le M_{A_2}$.  Many features of the models are apparent by inspection of the mass-squared matrix elements.  We discuss these aspects in Section \ref{sect:results}.
 
\subsection{Radiative Corrections}
An accurate analysis of the Higgs masses requires loop corrections.  The dominant contributions at one-loop are from the top and scalar top loops due to their large Yukawa coupling.  In the UMSSM, the gauge couplings are small compared to the top quark Yukawa coupling so the one-loop gauge contributions can be dropped.  Corrections unique to the NMSSM and n/sMSSM begin only at the two-loop level.  Thus all contributions that are model-dependent do not contribute significantly at one-loop order and the usual one-loop SUSY top and stop loops are universal in these models.  A similar approach has been done in studying extended Higgs sectors with many additional singlet fields \cite{Ham:2004pd}.  These one-loop corrections to the potential can be found from the Coleman-Weinberg potential \cite{Coleman:1973jx} and are reviewed in Appendix \ref{apx:loop}.

The mass squared matrix elements become 
\be
{\cal M}_{\pm} = {\cal M}_{\pm}^0+{\cal M}_{\pm}^1,
\ee
where the radiative corrections to the CP-even mass-squared matrix elements are given by 
\bea
\label{eq:cpemassmtx1}
({\mathcal{M}_{+}^1})_{11} &=& k \left[ \left( \frac{({\widetilde m}^2_1)^2}{(m_{\widetilde t_1}^2 - m_{\widetilde t_2}^2)^2} {\mathcal G} \right) v_d^2 + \left( \frac{h_s h_t^2 A_t}{2 \sqrt{2}} \mathcal{F} \right) \frac{v_u s}{v_d} \right], \\
({\mathcal{M}_{+}^1})_{12} &=&k \left[ \left( \frac{{\widetilde m}^2_1 {\widetilde m}^2_2}{(m_{\widetilde t_1}^2 - m_{\widetilde t_2}^2)^2} {\mathcal G} + \frac{h_t^2 {\widetilde m}^2_1}{m^2_{\widetilde t_1} + m^2_{\widetilde t_2}} (2-{\mathcal G}) \right) v_d v_u -  \left( \frac{h_s h_t^2 A_t}{2 \sqrt{2}} \mathcal{F} \right)s \right], \\
({\mathcal{M}_{+}^1})_{13} &=&k \left[ \left(\frac{{\widetilde m}^2_1 {\widetilde m}^2_s}{(m_{\widetilde t_1}^2 - m_{\widetilde t_2}^2)^2} {\mathcal G} + \frac{h_s^2 h_t^2}{2} {\mathcal F} \right) v_d s - \left( \frac{h_s h_t^2 A_t}{2 \sqrt{2}} \mathcal{F} \right) v_u\right], \\
({\mathcal{M}_{+}^1})_{22} &=& k \left( \frac{({\widetilde m}^2_2)^2}{(m_{\widetilde t_1}^2 - m_{\widetilde t_2}^2)^2} {\mathcal G} + \frac{2 h_t^2 {\widetilde m}^2_2}{m^2_{\widetilde t_1} + m^2_{\widetilde t_2}} (2-{\mathcal G}) + h_t^4 \ln \frac{m^2_{\widetilde t_1} m^2_{\widetilde t_2}}{m_t^4} \right) v_u^2 \\
&+& k \left( \frac{h_s h_t^2 A_t}{2 \sqrt{2}} \mathcal{F} \right) \frac{v_d s}{v_u}, \nn \\ 
({\mathcal{M}_{+}^1})_{23} &=& k \left[ \left(\frac{{\widetilde m}^2_2 {\widetilde m}^2_s}{(m_{\widetilde t_1}^2 - m_{\widetilde t_2}^2)^2} {\mathcal G} + \frac{h_t^2 {\widetilde m}^2_s}{m^2_{\widetilde t_1} + m^2_{\widetilde t_2}} (2-{\mathcal G}) \right) v_u s - \left( \frac{h_s h_t^2 A_t}{2 \sqrt{2}} \mathcal{F} \right)v_d\right], \\
({\mathcal{M}_{+}^1})_{33} &=& k \left[ \left(\frac{({\widetilde m}^2_s)^2}{(m_{\widetilde t_1}^2 - m_{\widetilde t_2}^2)^2} {\mathcal G} \right) s^2 +  \left( \frac{h_s h_t^2 A_t}{2 \sqrt{2}} \mathcal{F} \right) \frac{v_d v_u}{s}\right].
\label{eq:cpemassmtx2}
\eea
where $k={3\over(4\pi)^2}$ and the loop factors are
\be
{\cal G}(m^2_{\tilde t_1}, m^2_{\tilde t_2}) =  2\left[1- \frac{m^2_{\tilde t_1}+ m^2_{\tilde t_2}}{m^2_{\tilde t_1}- m^2_{\tilde t_2}} \log \left( {m_{\tilde t_1} \over m_{\tilde t_2}} \right)\right], \quad\quad\quad {\cal F} = \log \left( \frac{m^2_{\tilde t_1} m^2_{\tilde t_2}}{Q^4}\right) - {\cal G}(m^2_{\tilde t_1}, m^2_{\tilde t_2}).
\ee
Here we have defined
\bea
\widetilde{m}_1^{2} &=& h_t^2 \mu_{\rm eff} \left(\mu_{\rm eff} - A_t \tan\beta \right),\\ 
\widetilde{m}_2^{2} &=& h_t^2 A_t \left(A_t - \mu_{\rm eff} \cot\beta \right),\\ 
\widetilde{m}_s^{2} &=& \frac{v_d^2}{s^2} h_t^2 \mu_{\rm eff} \left(\mu_{\rm eff} - A_t \tan\beta \right),
\eea
with $Q$ being the $\overline{\text{DR}}$ renormalization scale and $A_t$ is the stop trilinear coupling.

The corrections to the CP-odd mass-squared matrices are given by
\be
(\mathcal{M}^1_{-})_{ij} = \frac{ h_s v_d v_u s}{\sqrt 2 v_i v_j}\frac{k h_t^2 A_t}{2} {\cal F}(m^2_{\tilde t_1}, m^2_{\tilde t_2}),
\label{eq:cpomassmtx}
\ee
where we identify $v_1\equiv v_d, v_2\equiv v_u$, and $v_3 \equiv s$.  These one-loop corrections agree with those of \cite{NMSSM_Higgs,nMSSM_Higgs,UMSSM_Higgs}.

The one-loop corrections to the charged Higgs mass are equivalent to those in the MSSM and can be significant for large $\tan \beta$.  The charged Higgs boson in the MSSM has a tree-level mass
\be
(M^{(0)}_{H^\pm})^2=M_W^2+M_Y^2,
\ee
and the extended-MSSM charged Higgs boson mass is 
\be
(M^{(0)}_{H^\pm})^2 = M_W^2+M_Y^2-{h_s^2 v^2\over 2}+h_s{ \sqrt2  ( 2\xi_F M_{\rm n}^2+\kappa s^2) \over\sin 2 \beta},
\ee
where $M_Y^2={ \sqrt 2 h_s s A_s \over \sin 2 \beta}$ is the tree-level mass of the MSSM CP-odd Higgs boson.  The case of large $M_Y$ (or $M_A$ in the MSSM) yields a large charged Higgs mass and is consistent with the MSSM decoupling limit yielding a SM Higgs sector.  Radiative corrections in the MSSM shift the mass by
\be
(M^{(1)}_{H^\pm})^2={h_s A_t s k h_t^2 {\cal F}\over \sqrt 2 \sin 2 \beta} + \delta M_{H^\pm}^2,
\ee
where after including $\tan \beta$ dependent terms, $\delta M_{H^\pm}^2$ is given by the leading logarithm result of the full one-loop calculation \cite{ref:chghiggscorr}
\be
\delta M_{H^\pm}^2 ={N_{\rm c} g^2 \over 32 \pi^2 M_W^2}\left({2 m_t^2 m_b^2\over \sin^2\beta \cos^2\beta}-M_W^2\left({m_t^2\over \sin^2\beta}+{m_b^2\over \cos^2\beta}\right)+{2\over3}M_W^4\right) \log{ M_{SUSY}^2\over m_t^2},
\ee
where $N_{\rm c}=3$ is the number of colors and $M_{SUSY}$ is the supersymmetric mass scale, taken to be 1 TeV.  Model-dependent terms come in at tree-level, giving a charged Higgs mass after radiative corrections of
\be
M_{H^\pm}^2 = M_W^2+M_Y^2-{h_s^2 v^2\over 2}+h_s{ \sqrt2  ( 2\xi_F M_{\rm n}^2+\kappa s^2) \over\sin 2 \beta}  +\left({h_s A_t s k h_t^2 {\cal F}\over \sqrt 2 \sin 2 \beta}+\delta M_{H^\pm}^2 \right).
\label{eq:chghiggs}
\ee

\section{Constraints}\label{sect:constraints}

Both theoretical and experimental constraints are important in ensuring that the models are realistic.  In the following, we list the constraints that we apply in obtaining the allowed Higgs mass spectra.

To generate the Higgs boson masses, we scan over the relevant parameters of each model.  Theoretical constraints eliminate large regions of the parameter space.  To avoid solutions that contain unstable saddle-points of the potential, we require that the mass-squared eigenvalues are positive-definite, i.e. $M_{A_i}^2, M_{H_i}^2,M_{H^\pm}^2 \ge 0$.  We also exclude solutions which give $m^2_{\widetilde t_i} <0$.

\subsection{Direct constraints}
\label{sect:dirlimits}

The direct constraints are provided by collider data.  Currently, LEP  gives the best experimental bound on the mass of the SM Higgs boson, $h$, of 114.4 GeV at 95\% C.L \cite{Sopczak:2005mc}.  We translate this to limit the mass of the lightest Higgs boson of the extended models by using the $ZZh$ coupling limits from LEP, as reproduced in Fig. \ref{fig:lep}a of Section \ref{sect:distresults}, that consider all SM particle decay modes down to $M_{h} = 12$ GeV \cite{Sopczak:2005mc}\footnote{These limits actually assume standard model branching ratios for the $H_i$, which are dominantly into $b \bar b$ and $\tau^+ \tau^-$ in the relevant mass range.  As discussed in Section \ref{sect:decaybf}, for some of the parameter values in the extended models the dominant decays are into (invisible) neutralinos, or into two light CP-odd states, and for those points the  constraint in Figure \ref{fig:lep}a does not strictly apply. However, there are also quite stringent limits on the invisible $H_i$ decay modes \cite{Abdallah:2003ry}, and (weaker) limits on the decays into two CP-odd states which subsequently decay into  $b \bar b$ or $\tau^+ \tau^-$ \cite{Abdallah:2004wy,Abbiendi:2002in}.  These have not been given for the entire kinematic ranges of interest here, so we will simply take the conservative approach of allowing only those points satisfying the $ZZH_i$ coupling limit in Figure \ref{fig:lep}a.}.  The $ZZH_i$ coupling relative to the SM coupling is given by the factor
\be
\xi_{ZZH_i}= \left({g_{ZZH_i} / g^{SM}_{ZZh}}\right)^2 = (R_{+}^{i1} \cos \beta + R_{+}^{i2} \sin \beta)^2.
\label{eq:zzh}
\ee
Since the $ZZH_i$ coupling in extended models is reduced by doublet-singlet mixing effects, it is possible to have Higgs bosons lighter than the SM bound of 114.4 GeV.  The reach of the $ZZH_i$ coupling limit extends only to 12 GeV, below which we do not enforce this constraint.  However, this low mass region is well constrained by bounds on $M_A$ and $M_h$ in the MSSM discussed below.  The LEP bound is also applied to $H_2$ and $H_3$ since a heavier Higgs boson may violate the bound even if the lightest does not.

Another channel of relevance from LEP is $Z\to A_i H_j$ with $A_i\to b\bar b$ and $H_j\to b\bar b$.  Current limits place the lightest possible CP-even and odd MSSM masses at $M_{H_1}=92.9$ GeV and $M_{A_2}=93.4$ GeV, respectively and are calculated assuming maximal stop mixing, yielding the most conservative limit on the lightest Higgs masses in the MSSM \cite{ref:Amasslep}.  An estimation of the corresponding limits in extended models may be obtained by comparing the expected production cross section of the extended-MSSM models at the maximum LEP energy, $\sqrt s=209$ GeV, to that of the MSSM \cite{SUMSSM_Higgs}.  At this energy the mass limits of the CP-even and CP-odd Higgs bosons provide an upper bound of the cross section at 40 fb.  In practice, we find that the LEP $Z\to A_i H_j$ constraint eliminates a significant fraction of the points generated with a low CP-odd Higgs mass.  In Fig. \ref{fig:lep}b, we show $\cos^2(\alpha-\beta)$, the prefactor of the $ZAh$ coupling where $\alpha$ is the rotation angle required to diagonalize the MSSM CP-even Higgs mass-squared matrix, versus CP-odd Higgs mass for the MSSM.  A strong $ZAh$ coupling results in an enhanced $Ah$ production cross section.  In the extended-MSSM models, we calculate the cross section for $e^+ e^- \to A_i H_j$ where $A_i$ is the lightest nonzero CP-even Higgs for that model.  If it is above the calculated LEP limit of 40 fb, the generated point fails this constraint.  Mixing effects which maximize the $ZA_i H_j$ coupling in the MSSM also result in a lower value of $M_{A_2}$, so that the LEP limit implies a lower bound on  $M_{A_2}$.  With the two complementary limits on the neutral Higgs bosons and the charged Higgs mass ($M_{H^\pm}=78.6\text{ GeV}$ from LEP\cite{ref:chghiggslim}), the Higgs sector in the MSSM and extended-MSSM models are rather well constrained.

\subsection{Indirect constraints}

While we focus on the Higgs sector in our analysis, indirect constraints from the neutralino and chargino sectors also need to be considered.  The lightest chargino mass is currently limited by LEP to be $M_{\chi^\pm} > 104$ GeV at 95\% C.L.\cite{ref:lepsusy}.  The chargino masses are determined by the diagonalization of
\bea
{\cal M}_{\chi^\pm} = \left( \begin{array} {c c}
	M_2 	&	\sqrt 2 M_W \sin \beta \\
	\sqrt 2 M_W \cos \beta 	& 	\mu_{\rm eff}
	\end{array} \right), 
\eea
The $SU(2)_L$ gaugino mass, $M_2$, that enters the chargino sectors does not have a direct effect on the Higgs sector, but the lower bound on  $M_{\chi^\pm}$ does constrain possible parameter values.

Precision electroweak data also provide an upper bound on the new contributions to the invisible $Z$ decay width of 1.9 MeV at 95\% C.L.\footnote{This is based  on the constraint on new physics contributions to the invisible $Z$ width, $\Gamma^{new}_{inv}= -2.65 \pm 1.5$ MeV \cite{ref:pdg}, renormalizing the probability distribution to require that the true value is positive.  Strictly speaking, such decays may not be invisible, and slightly weaker constraints would be obtained using the total or hadronic widths.  We use the invisible width to be conservative and for simplicity, since it is also applied to decays of the $Z$ into neutralino pairs.}   Contributions to this decay width include $Z\to A_i H_j$ for $M_{A_i}+M_{H_j} \le M_Z$ and $Z\to Z^* H_i \to f \bar f H_i$ for $M_{H_i}\le M_Z$.  The decay widths are given by
\bea
\Gamma_{Z\to A_i H_j} &=& {\alpha\over 48 x_W (1-x_W)} M_Z \lambda^{3/2}\left({M_{A_i}^2 / M_Z^2},{M_{H_j}^2 / M_Z^2}\right)\left( R^{i1}_{+} R^{j1}_{-}-R^{i2}_{+} R^{j2}_{-}\right)^2,\\
{d\Gamma_{Z\to f\bar f H_i} \over dx_{H_i}}&=& \Gamma_{Z \to SM} {\alpha\over 4\pi x_W(1-x_W)}{1+{2\over3} y_{H_i}^2-x_{H_i}+{1\over 12}x_{H_i}^2 \over (x_{H_i}-y_{H_i}^2)^2} \sqrt{x_{H_i}^2-4y_{H_i}^2}.
\eea
where  $\lambda(x,y)=1-x^2-y^2-2(xy-x-y)$, $x_{H_i}=2 E_{H_i}/M_Z$ and $y_{H_i}=M_{H_i}/M_Z$ and where the SM $Z$ decay width is $\Gamma_{Z\to SM} = 2.50$ GeV \cite{ref:pdg}.  Here we assume massless fermions in the $Z\to f\bar f H_i$ decay, which is a good approximation at low $M_{H_i}$.  This decay mode complements the $ZZH_i$ coupling constraint quite well as it is valid below the reach of the LEP limit on $\xi_{ZZH_i}$.  Since the masses of $H_2,H_3$ and $A_2$ are typically larger than $M_Z$, we only consider $Z \to H_1 A_1$ and $Z\to f \bar f H_1$ decay modes\footnote{Singlet mixing may allow $H_2$ or $A_2$ to be slightly less than $M_Z$ but the decay is still kinematically inaccessible.}.

The neutralino sector also provides constraints on the allowed parameter space via $Z$ boson decay.  If $M_{\lsp} \leq M_Z/2$, then $Z$ decays into neutralino pairs and this decay contributes to the invisible $Z$-decay width.  Since the $Z$ does not couple to the singlino, the superpartner of the Higgs singlet, the decay width formula in the extended models is similar to that of the MSSM, except for mixing effects \cite{Hesselbach:2001ri}.  The $Z$ decay width to neutralino pairs, when kinematically accessible, is 
\be
\Gamma_{Z \to \lsp \lsp} = \frac{g_2^2+g_1^2}{96 \pi M_Z^2} (|N_{13}|^2-|N_{14}|^2)^2\left(M_Z^2-(2 M_{\lsp})^2\right)^{3/2}.
\ee
The neutralino rotation matrix elements, $N_{ij}$, are found by diagonalizing the model-dependent neutralino mass-squared matrices in Appendix \ref{apx:neut}.  

The $Z-Z'$ mixing angle,
\be
\alpha_{ZZ'} = \half \tan^{-1} \left( {2 M^2_{ZZ'} \over M_{Z'}^2 - M_Z^2}\right),
\label{eq:mixing}
\ee 
is also constrained by electroweak precision data to be less than ${\cal O}(10^{-3})$, where the exact value is dependent on the $U(1)'$ model.  The $Z'$ mass parameters are 
\be
M^2_{Z'} = {g_{1'}}^2(Q_{H_d}^2 v_d^2+Q_{H_u}^2 v_u^2+Q_{S}^2 s^2),\quad M^2_{ZZ'} = {g_{1'}} \sqrt{g_1^2+g_2^2} \left( v_d^2 Q_{H_d} - v_u^2 Q_{H_u}\right).
\label{eq:mzp}
\ee
Eq. (\ref{eq:mixing}) bounds what types of $Z'$ models and associated Higgs sector parameters are allowed; it translates into a high value of $s$, typically at the TeV scale.  There do, however, exist isolated points that allow a suppression of $\alpha_{ZZ'}$ at low $s$ such as the following
\begin{enumerate}
\item If $Q_{H_d},Q_{H_u}$ have the same sign, a cancellation occurs at $\tan \beta = \sqrt{Q_{H_d} \over Q_{H_u}}$.
\item If $Q_{H_u}$ is small and $\tan \beta$ is large, the mixing term is suppressed.
\end{enumerate}
The $Z'$ mass is also constrained \cite{LEPmixingangle,ref:zpmasslim}, but the limits are very model-dependent on the quark and lepton couplings and can be eliminated entirely in the leptophobic $Z'$ case \cite{ref:zpleptphob}.  In any case, the large $s$ limit yields a $Z'$ with mass typically large enough to avoid existing experimental constraints.  Therefore, we only apply the $Z-Z'$ mixing constraint in our study.

Constraints due to the possibility of electroweak baryogenesis have been previously explored in the NMSSM \cite{NMSSM}, the sMSSM \cite{ref:Kang} and the nMSSM \cite{Menon:2004wv}. The cubic ($A_s$) term in the tree-level potential makes it much easier to achieve the needed strongly first-order phase transition in these models than in the MSSM \cite{Carena:2002ss}. However, we do not consider CP-violating phases in the Higgs sector, which is also a necessary condition for baryogenesis.  Furthermore, there are other possibilities for baryogenesis.  Therefore, electroweak baryogenesis constraints are not included here.

\section{Parameter scans}\label{sect:scan}

To generate the Higgs boson masses, we perform both grid and random scans over the allowed available parameter space of each model.  In the random scan, we evaluate 500000 points in the available parameter space for each model.  Our grid scan gives a reproducible catalogue of the Higgs masses of each model.  However, due to the large number of parameters, a finely spaced grid on individual parameters is not feasible.  The results from the grid scan serve as a useful guide of the allowed Higgs boson masses but do not provide definitive upper or lower mass limits.  

\begin{table}[t]
\caption{Parameter ranges in scans.  (a) Model-independent parameters. (b) Model-dependent parameters. Parameters not scanned assume the values $M_{\widetilde Q} = M_{\widetilde U} = 1$ TeV and $Q = 300$ GeV.}
\begin{center}
\begin{tabular}{|c|c|c|c|c|c|c|}
\hline
	&$\tan \beta$ & $s$ & $\mu_{\rm eff}$  & $A_s$ & $A_t$ & $M_2$\\ \hline
	Range	&1, 50 &   50 , 2000 GeV & 50 , 1000 GeV & 0 , 1 TeV & -1 , 1 TeV& -500, 500 GeV \\ \hline
	Step size	&-- &   100 GeV & 100 GeV & 100 GeV & 250 GeV & 100 GeV \\ \hline
\end{tabular}
\end{center}
(a)
\begin{center}
\begin{tabular}{|c|c|c|c|c|c|}
\hline
	&$\kappa$ & $A_\kappa$& $\xi_S$  & $\xi_F$ & $\theta_{E_6}$ \\ \hline
	Range	&-0.75 , 0.75&   -1 , 1 TeV & -1 , 1 & -1 , 1 & $0 , \pi$ \\ \hline
	Step size	&0.25 &   250 GeV & 0.2 & 0.2 & ${\pi\over10}$ \\ \hline
\end{tabular}
\end{center}
(b)
\label{table:scan}
\end{table}%

The model-independent parameters scanned over are $\tan \beta$, $s$, $\mu_{\rm eff}$, $A_s$, $A_t$,  and $M_2$, where we always assume gaugino mass unification $M_1=M_{1'}={5 g_1^2 \over 3 g_2^2} M_2$.  The masses $M_{\widetilde U}$ and $M_{\widetilde Q}$ are the soft masses of the up-type squarks and doublet-type squarks, respectively, and are fixed at 1 TeV; $M_{1'}$ is the mass of the $Z'$-ino in the UMSSM.  The model-dependent parameters are $\kappa$ and $A_\kappa$ for the NMSSM, $\xi_F$ and $\xi_S$ for the n/sMSSM, and $\theta_{E_6}$ for the UMSSM.  

In the parameter scans, we veto points that fail the direct and indirect constraints of Section \ref{sect:constraints}.  We choose the phase convention $A_s>0$, $\mu_{\rm eff}>0$, with all the VEVs real and positive.  We limit $h_s$ to be real and positive and allow the gaugino mass $M_2$ and coupling $\kappa$ to be real with either sign, although more generally these parameters could be complex.  With complex parameters, CP violation could occur.  If phases were included, the Higgs sector would be further complicated with up to five states for the NMSSM and n/sMSSM (four states for the UMSSM) that can intermix.  The Higgs sector with an arbitrary number of additional singlets and CP violation was studied in Ref. \cite{Ham:2004pd}.

The couplings run as the energy scale is varied.  Naturalness and the requirement that the couplings remain perturbative at the GUT scale limits $0.1\le h_s \le 0.75$ or $0.1\le \sqrt{\kappa^2+h_s^2}< 0.75$ for the NMSSM.  The couplings in the n/sMSSM are real and fixed in the interval $-1 \le \xi_{S},\xi_{F}\le 1$ with $M_{\rm n} = 500$ GeV.  We also constrain $\mu_{\rm eff}$ to the range $50 \le \mu_{\rm eff} \le 1000$ GeV to avoid fine tuning.  A summary of the scan ranges over model parameters are given in Table \ref{table:scan}.  For the grid scan, the step size for each parameter is given and we specifically scan $\tan \beta=1,1.5,2,10,50$.

\section{Discussion of the Higgs Mass spectra}\label{sect:results}

Throughout most of the parameter space, model distinguishing features are apparent in the Higgs masses.  However, different models can produce similar masses and mixings in certain limits.  Characteristics that are a direct consequence of how the singlet states mix affect the limits placed on the lightest Higgs boson mass.

\subsection{Common Characteristics}
\label{sect:comchar}

If the model-dependent parameters in the Higgs mass-squared matrices are set to zero, we obtain common mass-squared matrices and an additional symmetry that applies for each model.  For the n/sMSSM, this is a Peccei-Quinn (PQ) symmetry which protects the mass of one CP-odd Higgs.  Depending on what parameters vanish, the NMSSM may either have a PQ or a $U(1)_R$ symmetry \cite{Dobrescu:2000yn}, the global invariance of supersymmetry.  Near these limits, the $A_1$ mass in these extended models is small, allowing decay modes involving light CP-odd Higgs bosons; this is addressed in more detail in Section \ref{sect:decaybf}.  

In the UMSSM in the $g_{1'}\to 0$ limit, the gauged $U(1)'$ turns into a global $U(1)_{PQ}$ symmetry for the matter fields.  A massless CP-odd state, $A_1$, emerges, which is just the Goldstone boson of the broken $U(1)$ while the other CP-odd state, present for $g_{1'}\neq 0$, remains massive.  The $Z'$ decouples and remains massless in this limit.  In Table \ref{tbl:limits}, we summarize the common limits of the extended models.  

\begin{table}[t]
\begin{center}
\caption{Common Higgs mass-squared matrix limits of various models and their effects.  Note that in the UMSSM, the $U(1)$ is a global symmetry and not a remaining $U(1)'$ symmetry.  In these limiting cases, two of the CP-even Higgs bosons of each model are equivalent to the MSSM Higgs bosons if $s \gg \mu_{\rm eff}$, while the third decouples and is heavy for the NMSSM, or light for the n/sMSSM or UMSSM.}
\begin{tabular}{|c|c|c|c|}
\hline
Model &	Limits	& Symmetry	& 	Effects	\\
\hline
MSSM &	$B\to0$&	$U(1)_{PQ}$ & $M_{A_2} \to 0$\\
NMSSM & $\kappa, A_\kappa \to 0$ & $U(1)_{PQ}$ & $M_{A_1} \to 0$ \\
NMSSM & $A_s, A_\kappa \to 0$ & $U(1)_R$  & $M_{A_1} \to 0$ \\
n/sMSSM & $\xi_F$, $\xi_S  \to 0$ & $U(1)_{PQ}$ & $M_{A_1} \to 0$ \\
UMSSM & ${g_{1'}} \to 0$ & $U(1)_{PQ}$ & $M_{Z'},M_{A_1}\to 0$\\
\hline
\end{tabular}
\label{tbl:limits}
\end{center}
\end{table}%

In the PQ limits (and for the UMSSM for all $g_{1'}$), the CP-odd Higgs mass-squared matrix factors into a tree-level matrix times the one-loop correction. Such a form is required by the $U(1)$ symmetries to require the existence of two massless CP-odd goldstones, one of which is eaten by the $Z$ and the second by the $Z'$ in the UMSSM after radiative corrections are included.  Thus, $M_A$ is elevated by a factor of $1+\frac{k h_t^2 A_t}{2 A_s}{\cal F}$, where the ${\cal F}$ term is the loop contribution, i.e., 
\be
M^2_A = \frac{h_s A_s}{\sqrt{2}} \left( 1+k \frac{h_t^2}{2} \frac{A_t}{A_s} {\mathcal F} \right) \left( \frac{v_d v_u}{s} + \frac{v_u s}{v_d} + \frac{v_d s}{v_u} \right).
\label{eq:maradcorr}
\ee
Effectively the soft mass is increased by
\be
A_s \to A_s+k {h_t^2 \over 2} A_t {\cal F}
\label{eq:Ahscale}
\ee 
to promote the tree-level mass of the CP-odd Higgs boson to the radiatively corrected one.  In the $U(1)_R$ limit of the NMSSM, the radiative correction to the CP-odd masses vanishes.

Another limit, the $s$-decoupling limit, $s\to\infty$ while keeping $\mu_{\rm eff} ={h_s s\over \sqrt 2} \sim {\cal O}(\text{EW})$, gives similar EW/TeV scale Higgs boson masses for all models.  In this limit there is little mixing among Higgs states.  For the NMSSM and UMSSM, two CP-even Higgs correspond to the MSSM Higgs states, while the remaining Higgs boson is dominantly singlet with the mass ordering depending on $A_s, \kappa, A_\kappa$ and $g_{1'}$.  In the n/sMSSM, the lightest Higgs boson has vanishing mass and is singlet dominated while $H_2$ and $H_3$ correspond to the MSSM Higgs bosons.  Mass expressions in this limit cay be found in Appendix \ref{apx:masses}.  The Higgs boson that is dominantly singlet couples weakly to MSSM particles.  

The strength of a particular Higgs boson, $H_i$, coupling to fields in the MSSM may be quantified as the MSSM fraction
\be
\xi^{H_i}_{\text{MSSM}} = \sum_{j=1}^{2} (R^{ij}_{+})^2.
\ee
This quantity is not to be confused with the scaled $ZZH_i$ coupling $\xi_{ZZH_i}$.  Since $R$ is unitary, a sum rule exists
\be
\sum_{i=1}^{3} \xi_{\text{MSSM}}^{H_i} = 2,
\label{eq:summssmfrac}
\ee
which implies that at most two CP-even Higgs bosons can be MSSM-like; equal mixing scenarios have $\xi^{H_{1,2,3}}=\frac{2}{3}$.  A similar quantity can be found for the CP-odd Higgs bosons.  In the UMSSM and in the limits in Table \ref{tbl:limits} for the NMSSM and n/sMSSM, the MSSM fraction of the massive CP-odd Higgs boson is 
\be
\xi^{A_2}_{\text{MSSM}} = \left(1+\frac{v^2}{s^2} \cos^2 \beta \sin^2 \beta\right)^{-1},
\ee
consistent with the $s$-decoupling limit.  

Since the trace is invariant under rotations, a mass-squared sum rule exists.  The limits in Table \ref{tbl:limits} lead to a common sum rule of the tree-level Higgs masses:
\be
Tr \left[{\cal M}_{+}^0-{\cal M}_{-}^0\right] = M^2_{H^0_1}+M^2_{H^0_2}+M^2_{H^0_3} - M^2_{A_2} = M^2_Z,
\ee
where the $Z$ mass is given by $M_Z^2 = \frac{G^2}{4} (v_d^2 + v_u^2)$.  The sum rule for the MSSM is realized by taking the $s$-decoupling limit in the n/sMSSM, and additionally requires $g_{1'}\to 0$ in the UMSSM and $\kappa \to 0$ in the NMSSM.  In the CP-even and CP-odd mass-squared matrices of Section \ref{sect:massmtx}, we see that the upper left $2 \times 2$ submatrix is that of the MSSM while the third column/row vanishes.  Then, the decoupled $M_{H_1}$ and $M_{A_1}$ ($M_{Z'}$ for the UMSSM) become massless at tree-level and the Higgs mass-squared sum rules become MSSM like:
\be
M^2_{h^0}+M^2_{H^0} - M^2_{A} = M^2_Z,
\ee
where $h^0 = H^0_2$ and $H^0 = H^0_3$ are the usual MSSM CP-even Higgses.

\subsection{Distinguishing Characteristics}
\label{sect:distresults}
 
The introduction of the singlet Higgs field in MSSM extensions produces Higgs boson properties that are distinct from those of the MSSM.  Each model has additional defining characteristics that may be used to distinguish one model from another.  In this section, we give bounds on the lightest CP-even Higgs mass and provide expressions for the masses utilizing the hierarchy of matrix elements given in Appendix \ref{apx:masses}.  We scan over relevant model parameters to determine their effects on the Higgs masses.  Finally, we summarize the results of the complete random and grid scans.  


\subsubsection{Lightest CP-even Higgs Mass Bounds}
\label{sect:massbounds}

In any supersymmetric theory that is perturbative at the GUT scale, the lightest Higgs boson mass has an upper limit \cite{ref:upperlimhiggs}.  Since the mass-squared CP-even matrix ${\cal M}_{+}$ is real and symmetric, an estimation of the upper bound on the smallest mass-squared eigenvalue may be obtained by the Rayleigh Quotient 
\be
M_{H_1}^2 \le \frac{u^T {\cal M}_{+} u}{u^T u},
\ee
where $u$ is an arbitrary nonzero vector.  With the choices
\bea
u^T &=& (\cos \beta, \quad \sin \beta) \qquad \mbox{~~~~~[MSSM]} \nn \\
&=&  (\cos \beta, \quad \sin \beta, \quad 0) \qquad \mbox{[extended models]}
\eea
the well-known upper bound of the lightest Higgs mass-squared from the mass-squared matrices of Eq. (\ref{eq:cpetree1}-\ref{eq:cpetree2}) and (\ref{eq:cpemassmtx1}-\ref{eq:cpemassmtx2}) are given as 

(i) MSSM \cite{MSSM_Higgs}:
\be\label{eq:masslimits1}
M^2_{H^0_1} \le M^2_Z \cos^2 2\beta +\tilde {\cal M}^{(1)},
\ee
where 
\be
\tilde {\cal M}^{(1)} = ({\cal M}^{(1)}_{+})_{11}\cos^2\beta +({\cal M}^{(1)}_{+})_{22}\sin^2\beta + ({\cal M}^{(1)}_{+})_{12}\sin 2\beta.
\ee
(ii) NMSSM, n/sMSSM, and Peccei-Quinn limits  \cite{Drees:1988fc}:
\be
M^2_{H^0_1} \le M^2_Z \cos^2 2\beta + \frac{1}{2} h_s^2 v^2 \sin^2 2\beta+\tilde {\cal M}^{(1)}.
\ee

(iii) UMSSM \cite{UMSSM_Higgsbound}:
\be\label{eq:masslimits2}
M^2_{H^0_1} \le M^2_Z \cos^2 2\beta + \frac{1}{2} h_s^2 v^2 \sin^2 2\beta + g_{1'}^2 v^2 (Q_{H_d} \cos^2 \beta + Q_{H_u} \sin^2 \beta)^2+\tilde {\cal M}^{(1)}.
\ee

Although the upper bounds change with the choice of the $u$ vector, these results indicate that extended models have larger upper bounds for the lightest Higgs due to the contribution of the singlet scalar.  The UMSSM can have the largest upper bound due to the quartic coupling contribution from the additional gauge coupling term, $g_{1'}$, in the $U(1)'$ extension.  In the MSSM, large $\tan \beta$ values are suggested by the conflict between the experimental lower bound and the theoretical upper bound on $M_{H_1}$.  Since the extended models contain additional terms which relax the theoretical bound, they allow smaller values for $\tan \beta$ than the MSSM.

\subsubsection{Numerical Evaluation of masses}
\label{sect:numeval}

\emph{a.  CP-even Higgs Masses}

\begin{figure}[h]
\begin{center}
\includegraphics[angle=-90,width=0.49\textwidth]{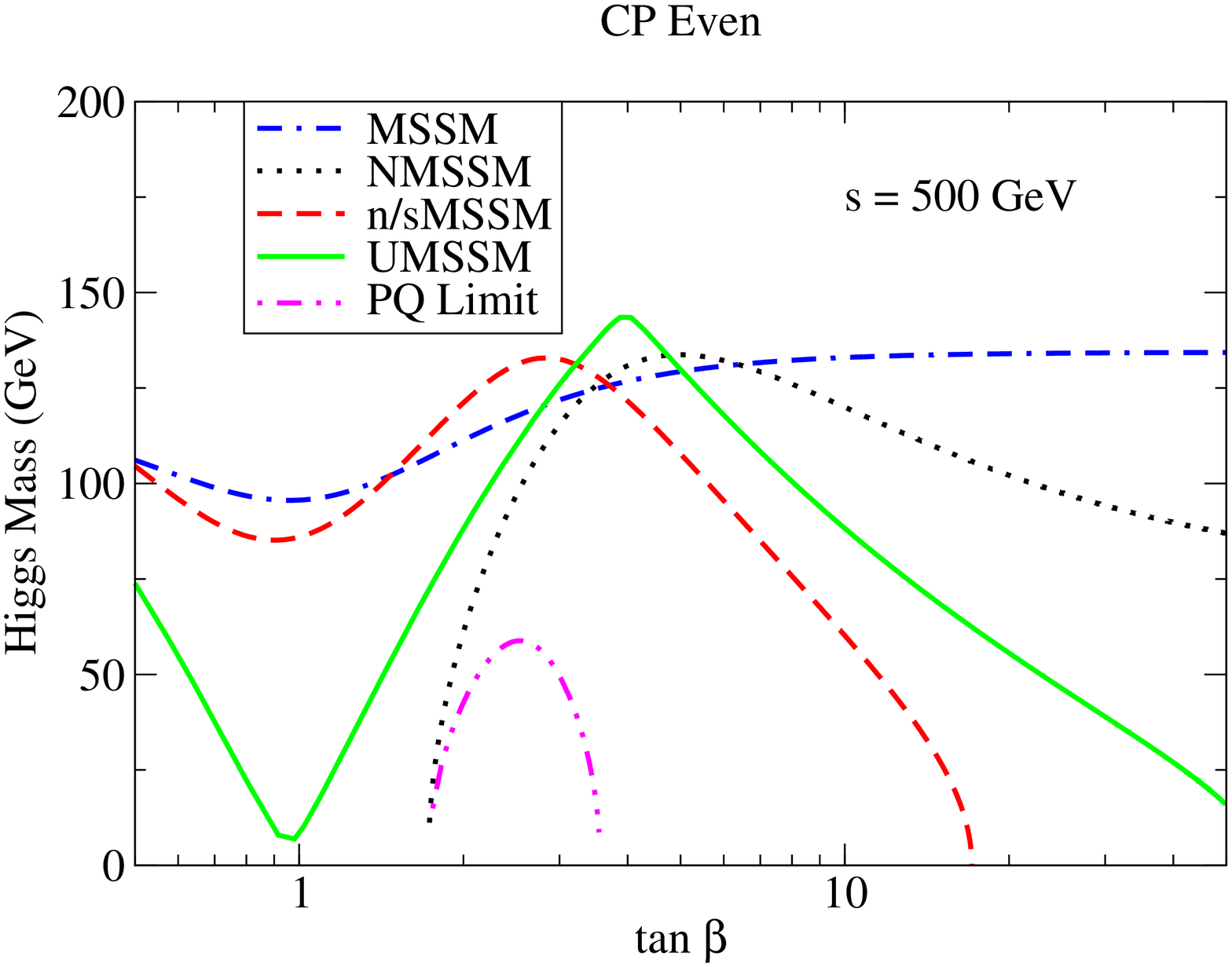}
\includegraphics[angle=-90,width=0.49\textwidth]{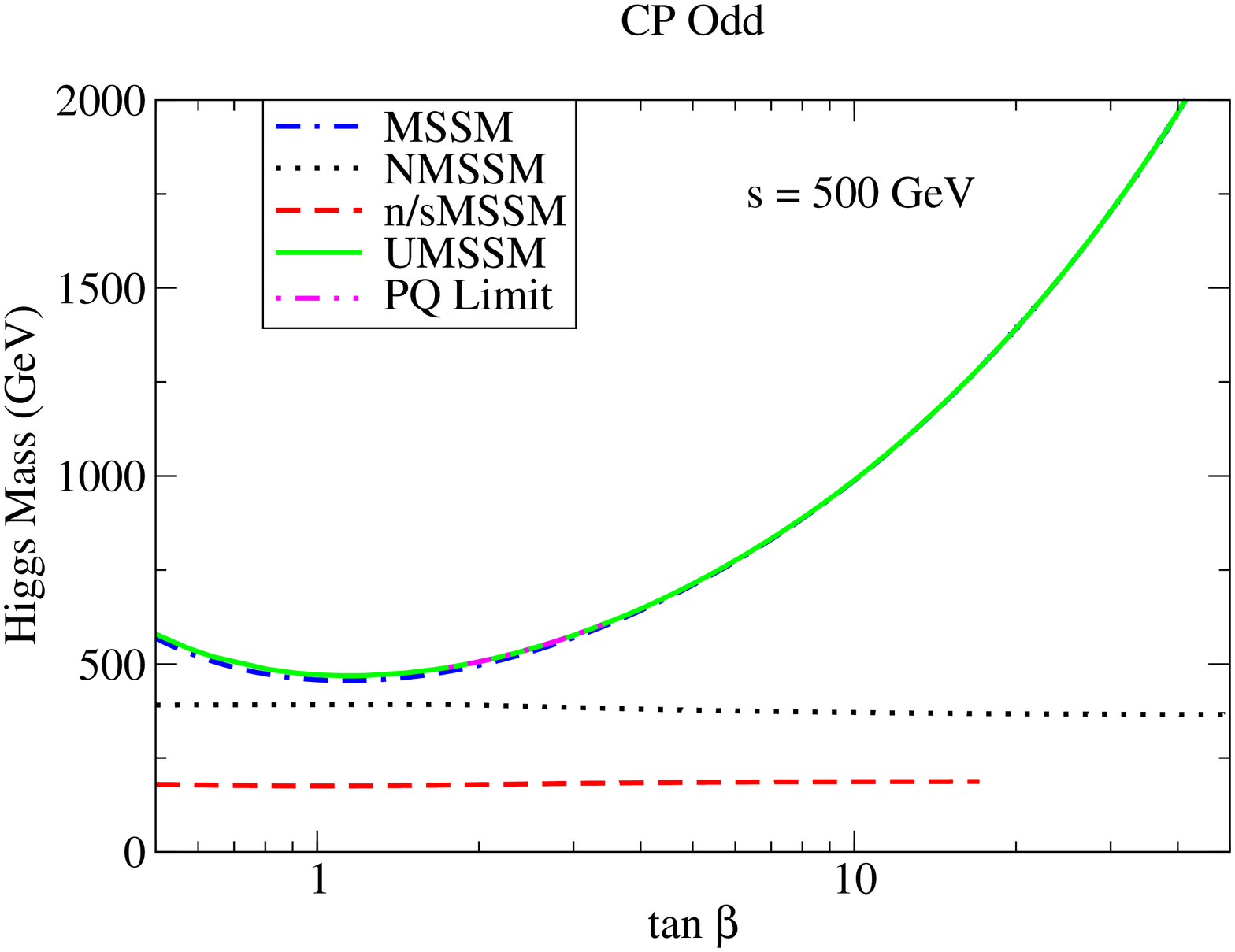}
(a)\hspace{0.48\textwidth}(b)
\includegraphics[angle=-90,width=0.49\textwidth]{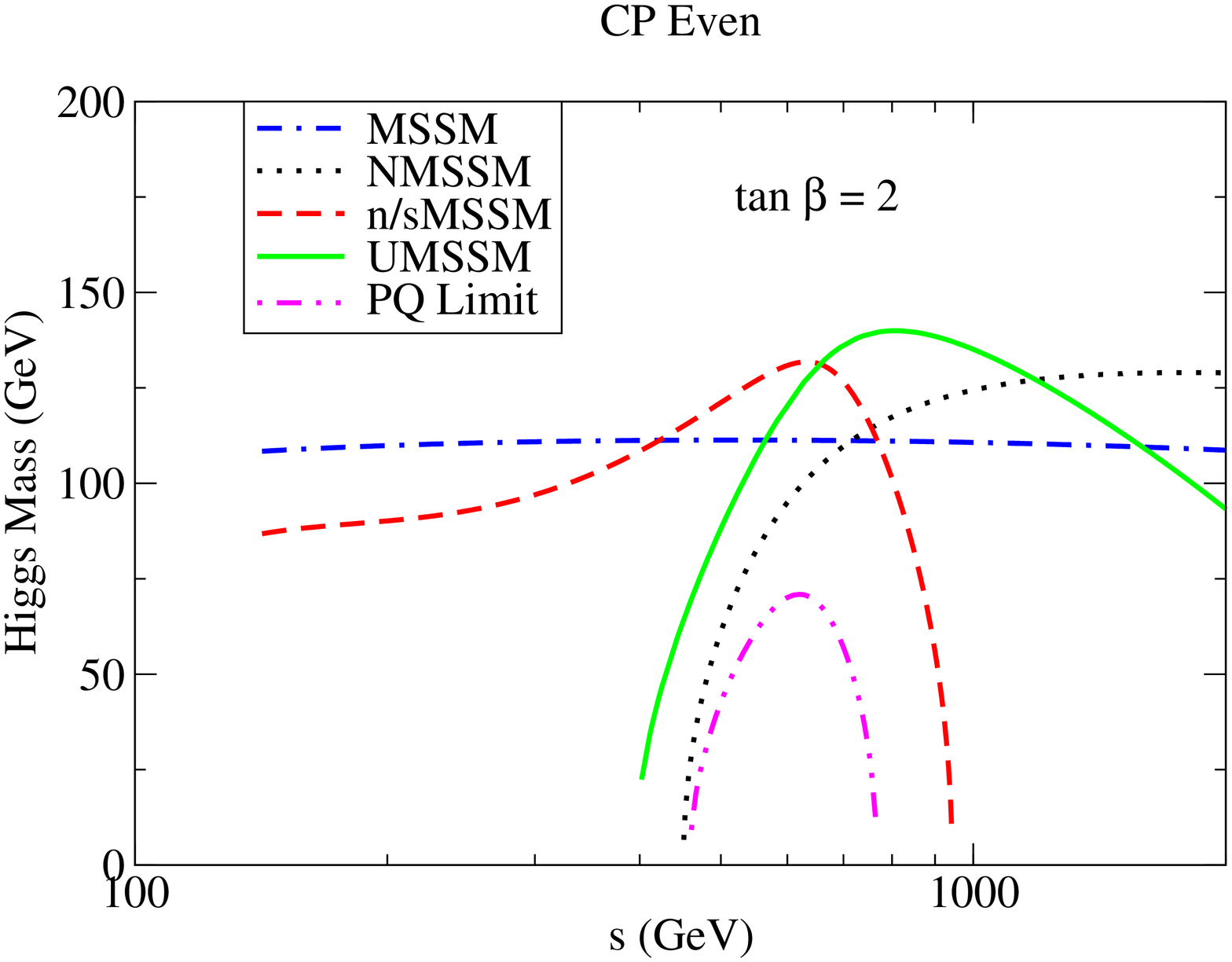}
\includegraphics[angle=-90,width=0.49\textwidth]{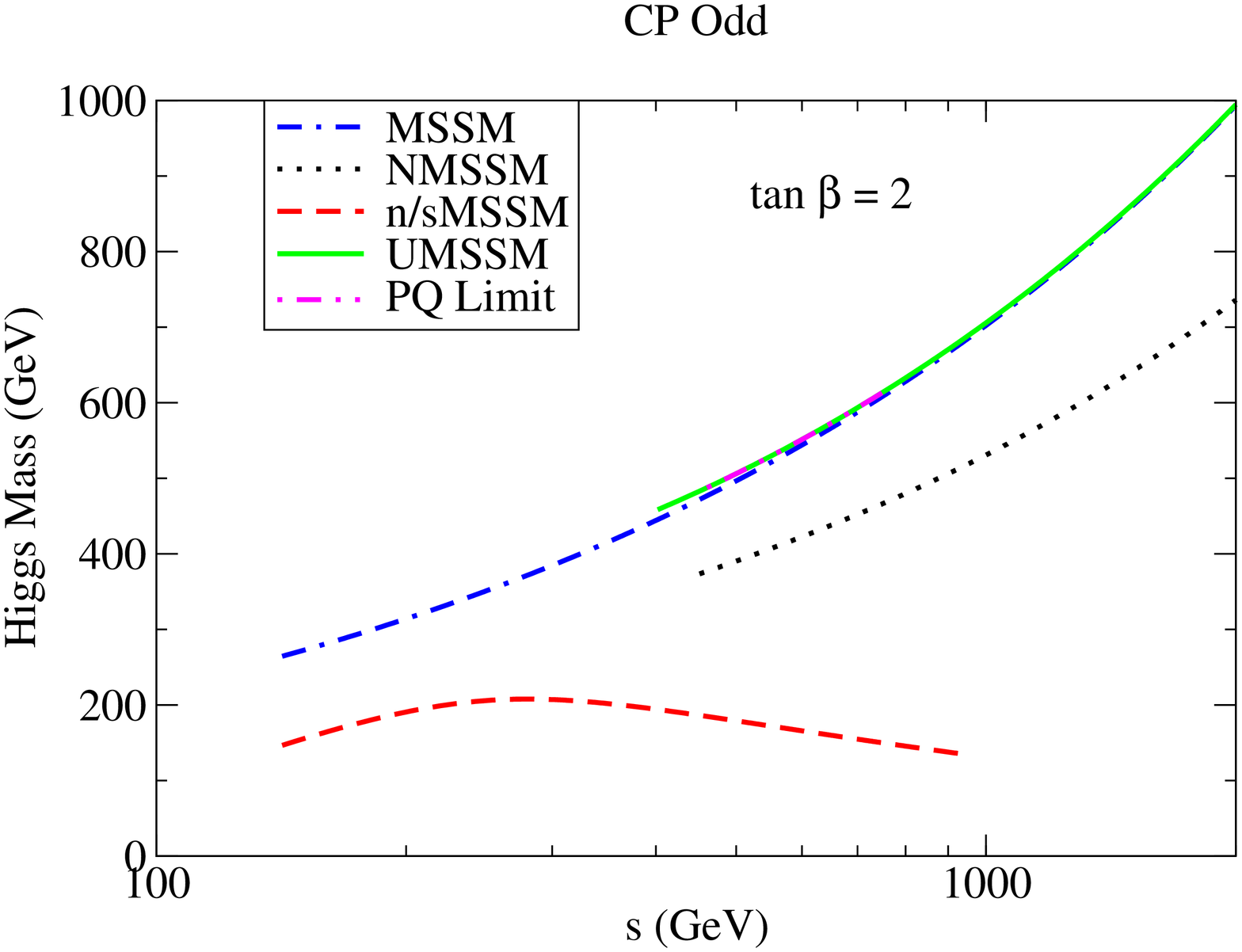}
(c)\hspace{0.48\textwidth}(d)
\caption{Lightest CP-even and lightest CP-odd Higgs masses vs. $\tan\beta$ and $s$ for the MSSM, NMSSM, n/sMSSM, UMSSM, and the PQ limits.  Only the theoretical constraints are applied with $s = 500$ GeV (for $\tan\beta$-varying curves), $\tan\beta = 2$ (for $s$-varying curves).  Input parameters of  $A_s = 500$ GeV, $A_t = 1$ TeV, $M_{\tilde Q} = M_{\tilde U} = 1$ TeV, $\kappa = 0.5$, $A_\kappa = -250$ GeV, $M_{\rm n}=500$ GeV, $\xi_F = -0.1$, $\xi_S = -0.1$, $h_s = 0.5$, $\theta_{E6} = -\tan^{-1}\sqrt{5\over3}$, and $Q = 300$ GeV, the renormalization scale, are taken. The $U(1)_{PQ}$ limit allows one massive CP-odd Higgs whose mass is equivalent to that of the UMSSM CP-odd Higgs.}
\label{fig:modelscans}
\end{center}
\end{figure}

In Fig. \ref{fig:modelscans} we show the variation of the lightest Higgs mass in the different models as functions of $s$ and $\tan \beta$ with the other parameters fixed.   (Similar plots for the heavier states are shown in
Appendix \ref{apx:addparm}.)  We only apply the theoretical constraints to these spectra to see the general trends of the models before experimental constraints are applied.  The UMSSM would fail to pass the $\alpha_{ZZ'}$ constraint in most of the plotted range of $s$.  

Note that the MSSM does not conform to the behavior of the extended models in the CP-even sector.  Since the MSSM contains only two CP-even Higgs bosons, the heavier of the two mass-squares increases with $\mu_{\rm eff}A_s$ at tree-level, similar to the CP-odd and charged Higgs masses.  Since we fix $h_s=0.5$, this Higgs mass-squared scales as the singlet VEV, $s$.  The radiative corrections do not contribute a significant $s$ dependence to the mass-squared matrix.  The tree-level dependence on $s$ prevents a level crossing between the $H_1$ and $H_2$ states.  However, in the extended models there are three CP-even Higgs bosons.  Level crossings are possible here as there is a Higgs boson of  intermediate mass: see Fig. \ref{fig:modelscans}(c).  We also see a significant difference between the MSSM and the extended-MSSM models in the $\tan \beta$ scan, which is expected since a moderate value of $s=500$ GeV is chosen.  The terms that differentiate the matrix elements in the extended models from that of the MSSM are not negligible at this value of $s$, giving different $s$-dependences of the Higgs mass.  

The MSSM $\tan \beta$ scan shows a dip in the Higgs mass at $\tan \beta = 1$ and a maximal mass is approached as $\tan \beta$ increases.  However, the extended-MSSM models have a decrease in mass after $\tan\beta$ of $~2-4$ due to the level crossing with the additional moderate mass CP-even Higgs present in these models.  The presence of the dip in the masses at $\tan \beta\sim 1$ for the UMSSM and n/sMSSM is not a consequence of a level crossing, but is due to the mass dependence on $\tan \beta$.  When $\tan \beta = 1$, the upper bound on the lightest CP-even Higgs mass decreases as seen in Eq. (\ref{eq:masslimits1}-\ref{eq:masslimits2}).  Overall, we see substantial differences in the spectra of the lightest higgs in the extended models compared to the MSSM.  

\emph{b.  CP-odd Higgs Masses}

Since only one massive CP-odd Higgs boson exists in the MSSM, UMSSM and the Peccei-Quinn limit of the extended models, the CP-odd masses generally behave the same over both scans and conform to the general scaling $M_{A_2}^2 \sim A_s s(\cot \beta+\tan\beta)$.  (The exact expression in these cases is given by Eq. (\ref{eq:maradcorr}), with the first term omitted for the MSSM.)  Further, we note that the CP-odd mass in the Peccei-Quinn limit is identical to that of the UMSSM, which may be understood by the absence of mixings and the resulting mass splittings that occur in the MSSM or other extended models.  However, the MSSM mass approaches the PQ/UMSSM mass as $s$ increases, a result consistent with the $s$-decoupling limit.  
The lightest CP-odd Higgs in the n/sMSSM and the NMSSM, however, does not share the similarities of the other models.  In these models, there are two CP-odd Higgs bosons, resulting in a different dependences on $s$ and $\tan \beta$.  Mixing effects tend to lower the lightest Higgs masses in these models, providing interesting phenomenological consequences.  These are further discussed in Section \ref{sect:collpheno}.

\emph{ c. Higgs Mass Ranges}

\begin{figure}[t]
\begin{center}
\includegraphics[angle=-90,width=.49\textwidth]{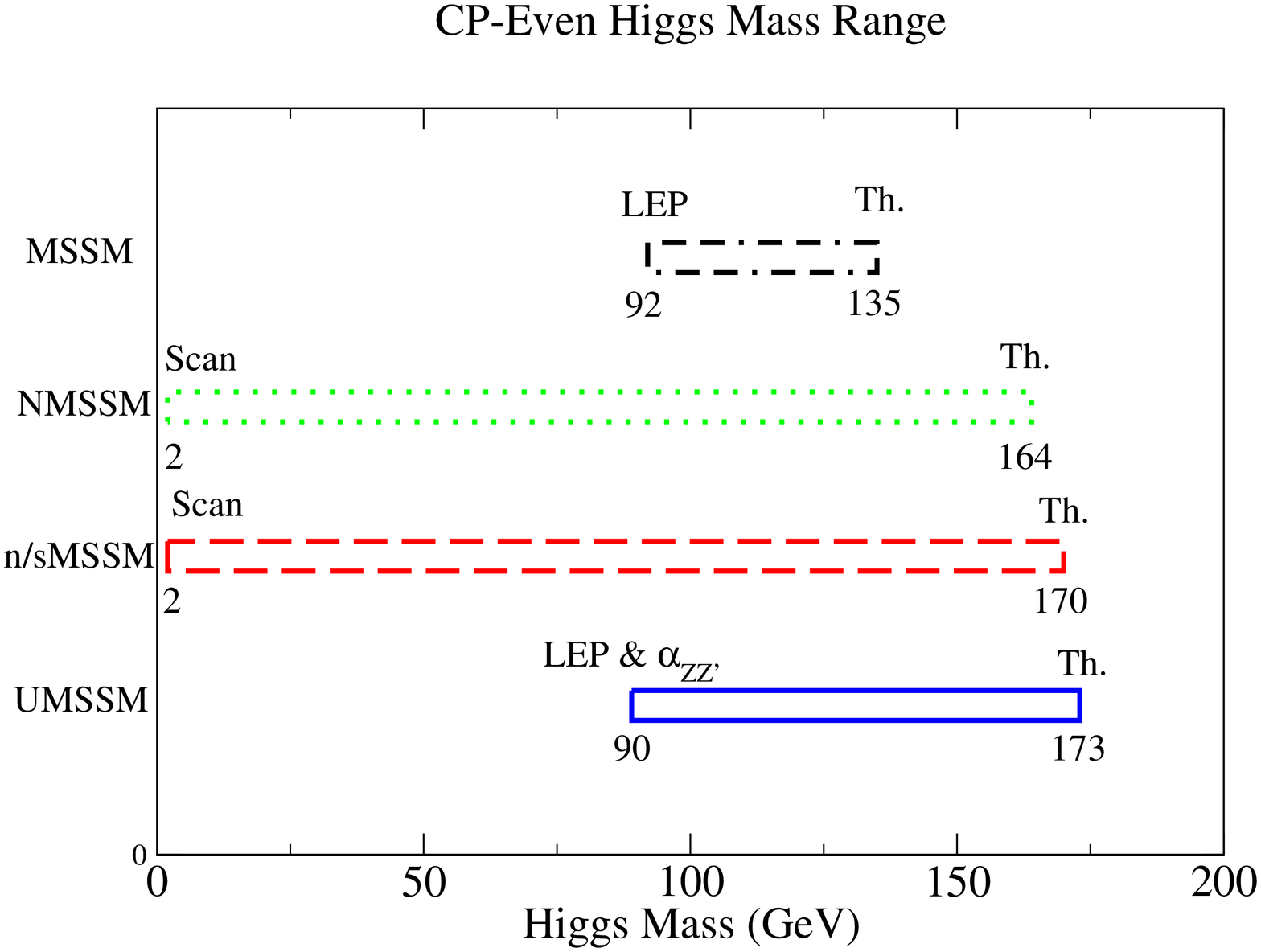}
\includegraphics[angle=-90,width=.49\textwidth]{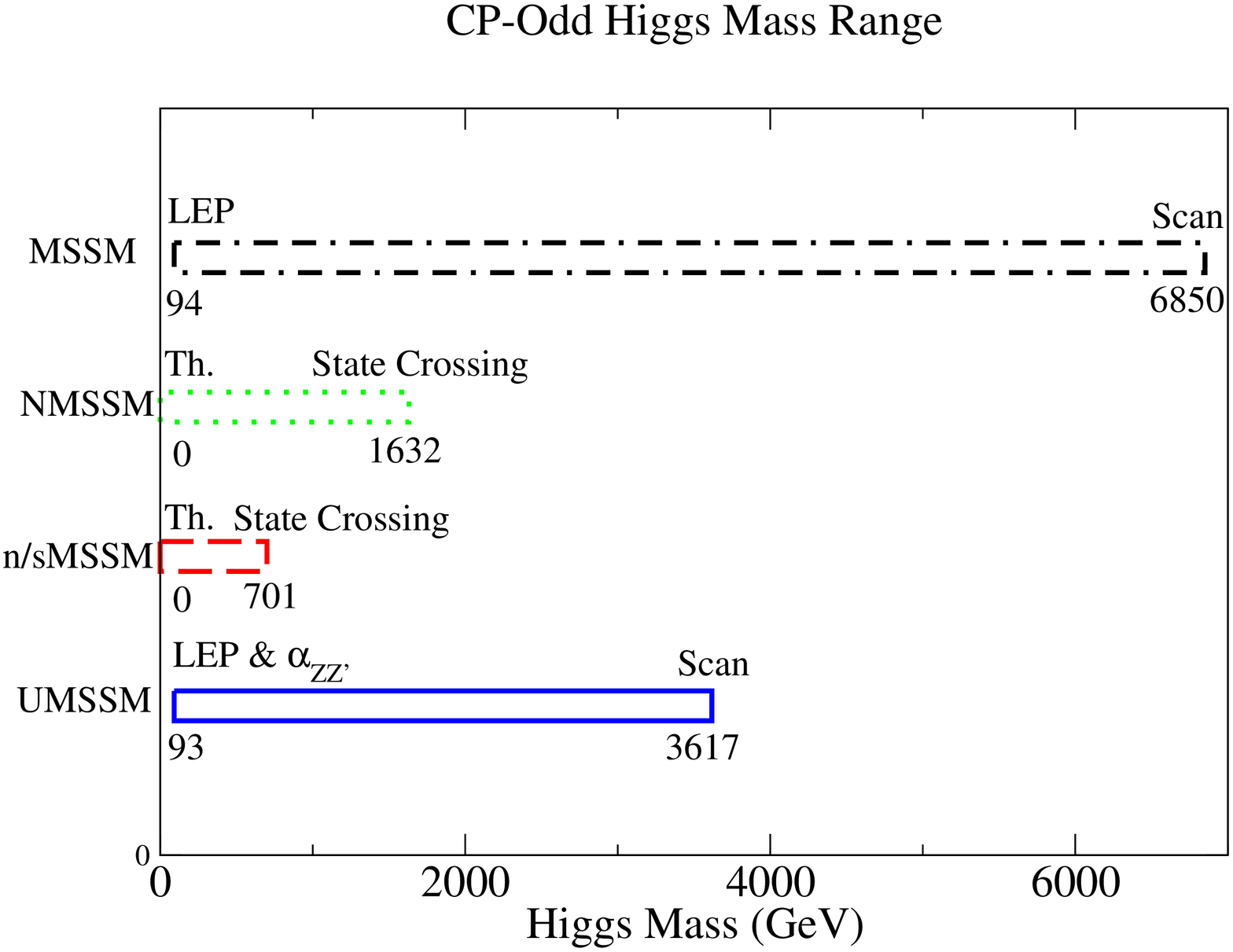}
\includegraphics[angle=-90,width=.49\textwidth]{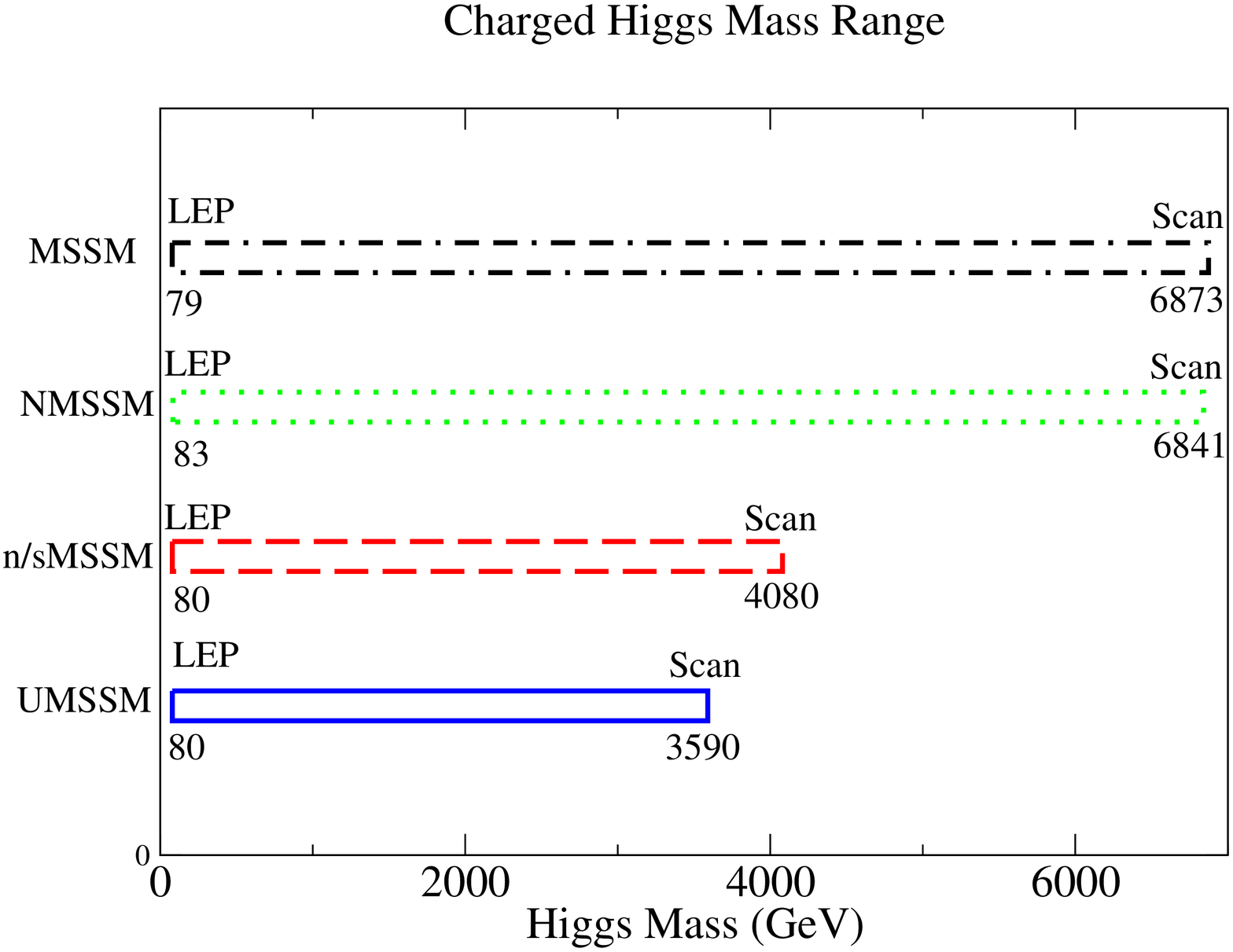}
\caption{Mass ranges of the lightest CP-even and CP-odd and the charged Higgs bosons in each extended-MSSM model from the grid and random scans. Explanation of extremal bounds and their values are provided for each model.  Explanations are Th. - theoretical bound met, value not sensitive to limits of the scan parameters; Scan - value sensitive to limits of the scan parameters; State Crossing - value has maximum when crossing of states occurs (specifically for $A_1$ and $A_2$ in the NMSSM and n/sMSSM); LEP - experimental constraints from LEP; $\alpha_{ZZ'}$ - experimental constraints in the UMSSM on the $Z-Z'$ mixing angle.}
\label{fig:mhrange}
\end{center}
\end{figure}

We summarize the available ranges found in the grid and random scans of the lightest CP-even, CP-odd  and charged Higgs boson masses that satisfy the applied constraints in Fig. \ref{fig:mhrange}.  For each model, the values of the maximum and minimum masses are given as well as the reason for the bounds.  

The lightest CP-even and CP-odd and the charged Higgs boson mass ranges differ significantly among the models.  The CP-even Higgs mass range is quite restricted in the MSSM and satisfies the upper theoretical mass bound and lower experimental bound from LEP discussed in Section \ref{sect:constraints}.   The upper limits for the CP-even Higgs masses in the extended models saturate the theoretical bounds and are extended by $30-40$ GeV compared to the MSSM while the upper limits in the lightest CP-odd Higgs masses are artificial in the MSSM and UMSSM as they change with the size of the scan parameters such as $A_s$ and $\tan \beta$.  The lower limits of the lightest CP-odd masses in the MSSM and UMSSM reflect the LEP limits on $M_{A_2}$; the UMSSM is similar to the MSSM since $s$ is required to be large by the strict $\alpha_{ZZ'}$ constraint, decoupling the singlet state and recovering a largely MSSM Higgs sector.  However, fine tuning the Higgs doublet charges under the $U(1)'$ gauge symmetry  and $\tan \beta$ allows the $Z-Z'$ mixing constraint on $s$ to be less severe, and can result in a lower Higgs mass with respect to the MSSM.  These instances along with the values $A_s=A_t=0$ GeV allow very low CP-even Higgs masses at ${\cal O}(1\text{ GeV})$ and a massless CP-odd state.  Since these points are distinct from the range of masses typically found in the UMSSM, we do not show these points in Fig. \ref{fig:mhrange} but simply note that they exist.  However, the NMSSM and n/sMSSM may have a massless CP-odd state due to global $U(1)$ symmetries discussed in Section \ref{sect:comchar} while the upper limit on the lightest CP-odd Higgs mass depends on the specifics of the state crossing with the heavier state, $A_2$, that has a scan-dependent mass.  In these models, the CP-odd masses extend to zero since the mixing of two CP-odd states allow one CP-odd Higgs to be completely singlet and avoid the constraints discussed above.   

The charged Higgs masses are found to be as low as 79 GeV in the scans, in agreement with the imposed experimental limit of 78.6 GeV.  In these cases where $M_{H^\pm}\sim 80$ GeV, the charged Higgs is often the lightest member of the Higgs spectrum.  However, these cases require fine tuning to obtain values of $\mu_{\rm eff}>100$ GeV \cite{cpnsh}.  The upper limit of the charged Higgs mass is dependent on the range of the scan parameters as seen in Eq. (\ref{eq:chghiggs}).  The discrepancy in the upper limit of the charged and CP-odd Higgs mass between the UMSSM and MSSM is a consequence of a lower $\mu_{\rm eff}$ in the UMSSM, resulting in a lower $M_Y$.  Large values of $\mu_{\rm eff}$ are more fine-tuned in the UMSSM than the MSSM since the additional gauge, $g_{1'}$, and Higgs, $h_s$, couplings often drive $M_{H_1}^2<0$.  Consequently, CP-odd and charged Higgs masses comparable to the higher MSSM limit are not present in the scan.  The upper bound on the charged Higgs mass in the NMSSM is relaxed due to the additional parameter of the model.

\emph{d.  Higgs Boson Searches}

\begin{figure}[t]
\begin{center}
\includegraphics[angle=-90,width=0.49\textwidth]{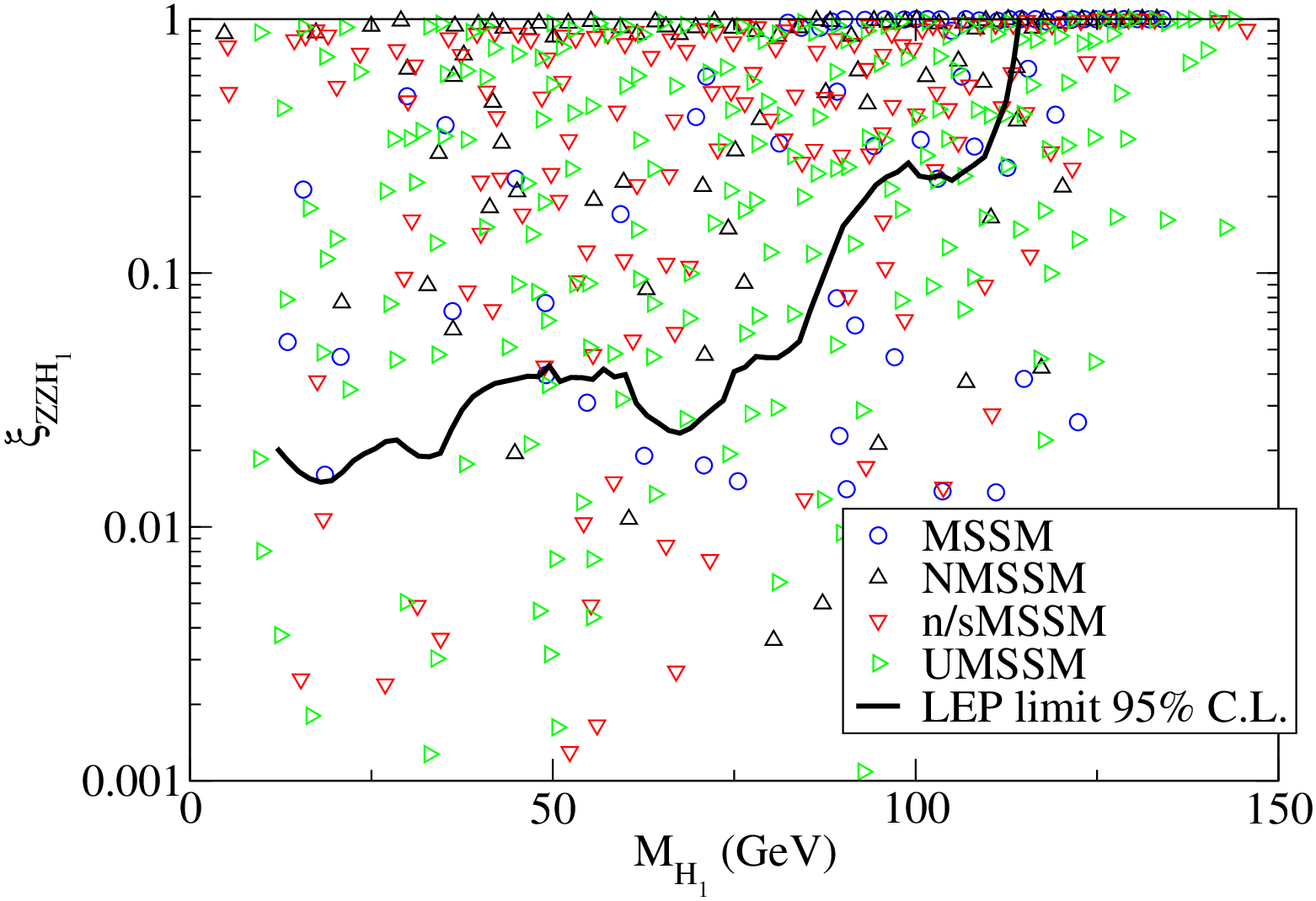}
\includegraphics[angle=-90,width=0.49\textwidth]{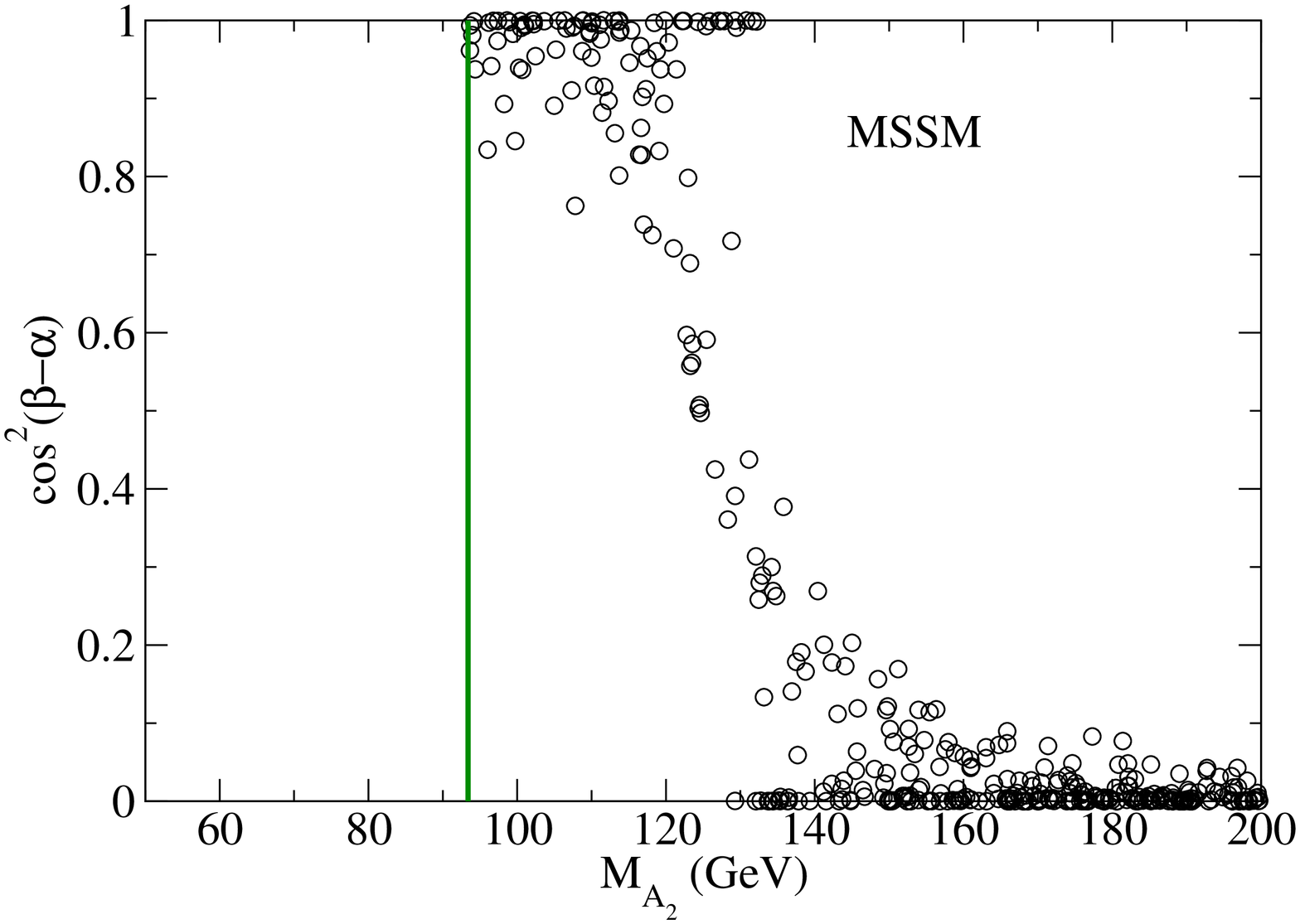}
(a)\hspace{0.48\textwidth}(b)
\caption{(a)  LEP limit \cite{Sopczak:2005mc} on $\xi_{ZZH_i} = \left({g_{ZZH_i} / g^{SM}_{ZZh}}\right)^2 = \Gamma_{Z \to Z H_i}/\Gamma^{SM}_{Z\to Zh}$, the scaled $ZZH_i$ coupling in new physics, vs. the light Higgs mass.  The solid black curve is the observed limit with a 95\% C.~L.  Points falling below this curve pass the $ZZH_i$ constraint.  (b) $\cos^2(\beta-\alpha)$ vs. $M_{A_2}$ in the MSSM.  The hard cutoff shown by the solid green line at $M_{A_2} = 93.4$ GeV is due to the constraint on $\sigma(e^+ e^- \to A_i H_1)$ discussed in Section \ref{sect:dirlimits}.}
\label{fig:lep}
\end{center}
\end{figure}

The focus of Higgs searches is most commonly the lightest CP-even Higgs boson.  In the models that we consider, the lightest CP-even Higgs boson can have different couplings than in the SM.  In Fig. \ref{fig:lep}a, we show the present limits from LEP on the scaled $ZZH_{i}$ coupling.\footnote{For clarity, in all the plots that follow we sample the passed points in the results from the random scans.}  Mixing effects can lower the $ZZH_i$ coupling and, in the MSSM, this occurs if $M_{A_2}$ is low, as seen in Fig. \ref{fig:lep}b where the $ZZH_i$ coupling is lowest for $\cos^2(\beta-\alpha)=1$.  However, an additional limit is placed on the mixing via the $e^+ e^- \to A_i H_1$ cross section discussed in Section \ref{sect:dirlimits}, eliminating low mass CP-even Higgs bosons in the MSSM, as seen in Fig. \ref{fig:lep}b.  In extended-MSSM models, additional mixing may occur with the singlet fields.  Due to this mixing and the subsequent evasion of the LEP limit on the $ZZH_i$ coupling, the lightest CP-even Higgs may then have a mass smaller than the SM Higgs mass limit.  Indeed, attempts to explain the $2.3 \sigma$ and $1.7 \sigma$ excess of Higgs events at LEP for masses of 98 GeV and 114.4 GeV, respectively, with light CP-even Higgs bosons in the UMSSM have been explored \cite{ref:umssmlep}.  This slight excess has also been studied in the NMSSM where a light Higgs with a SM coupling to $ZZ$ decays to CP-odd pairs \cite{NMSSMlepxs}.  

\begin{figure}[t]
\begin{center}
\includegraphics[angle=-90,width=0.49\textwidth]{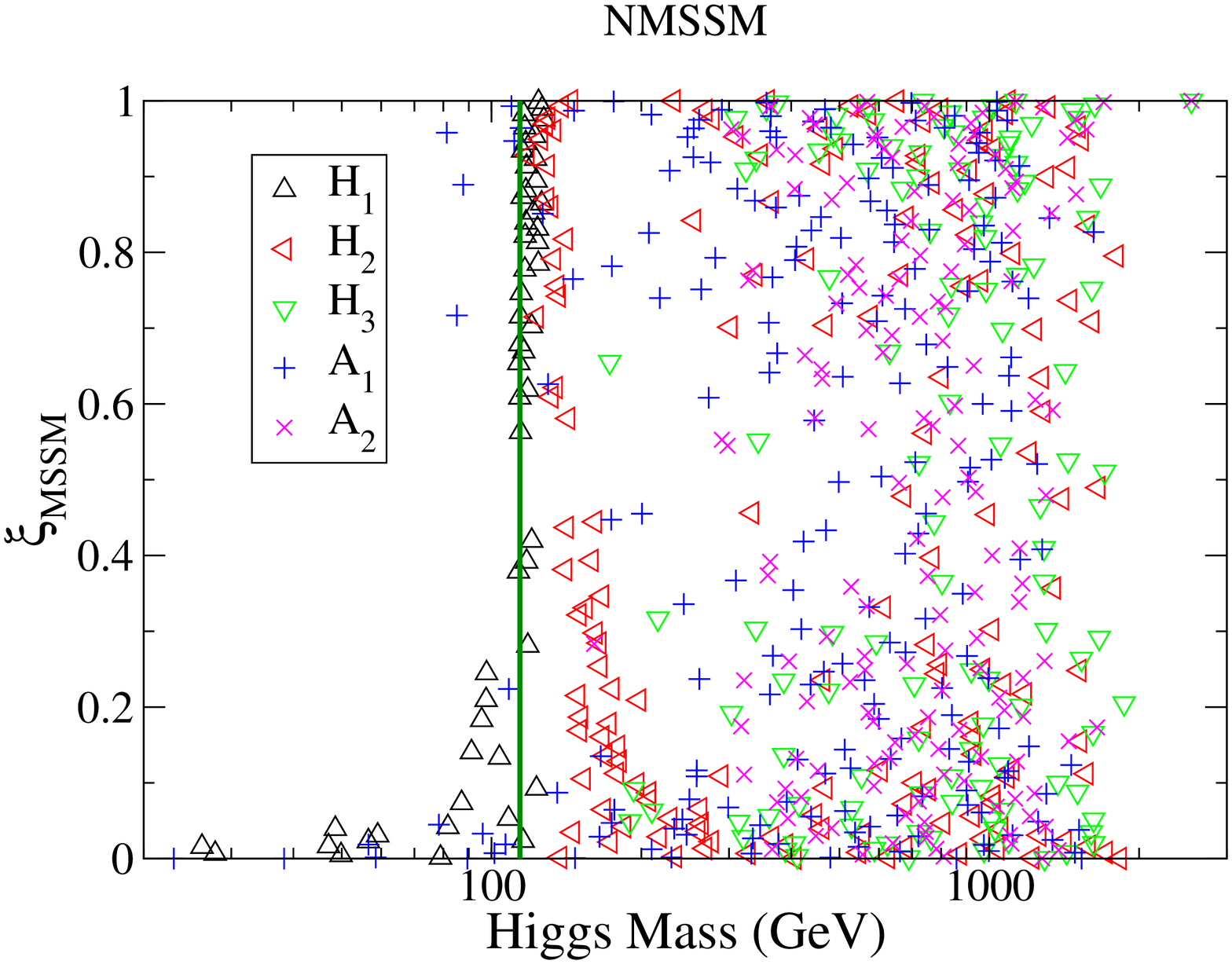}
\includegraphics[angle=-90,width=0.49\textwidth]{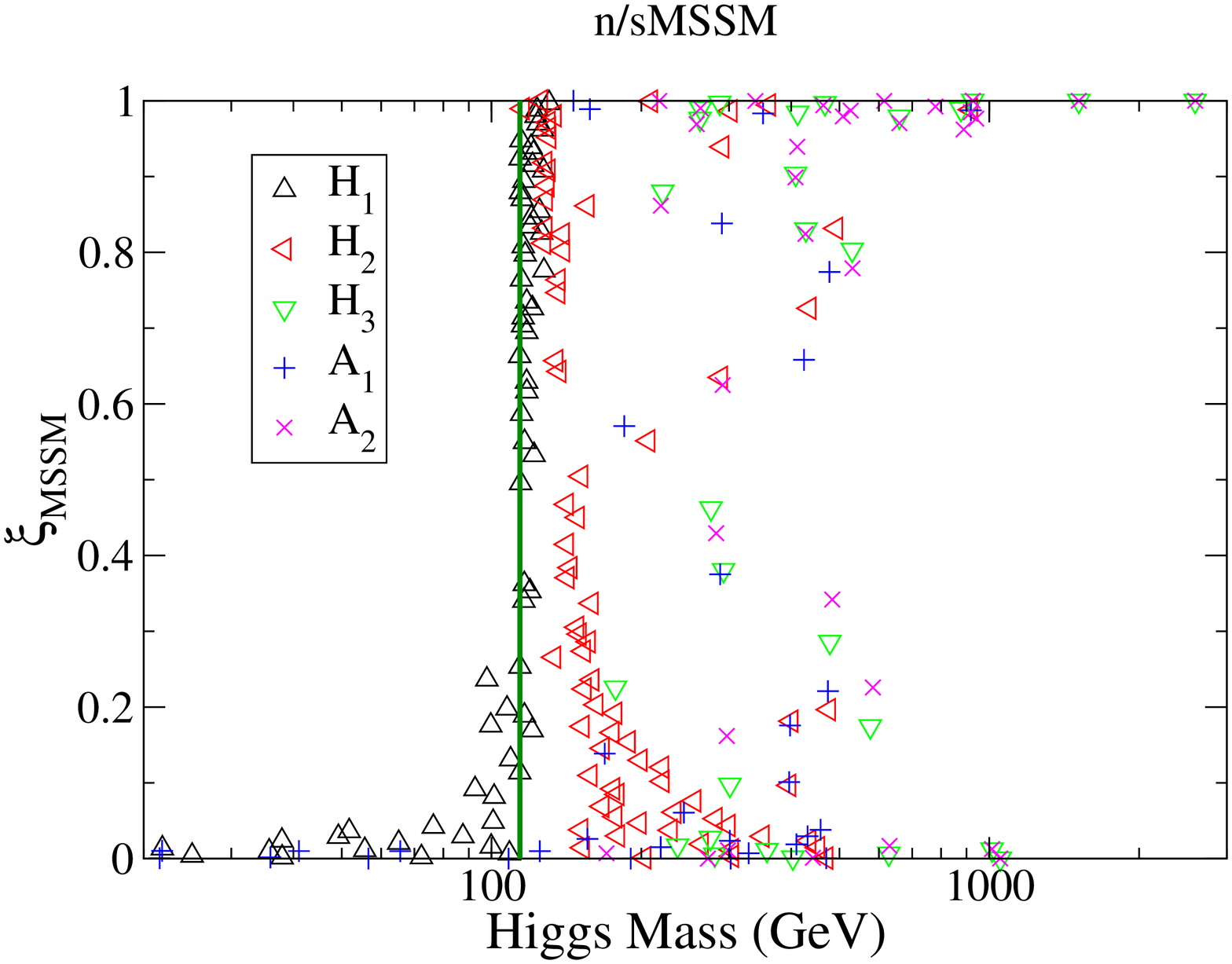}
(a)\hspace{0.48\textwidth}(b)
\includegraphics[angle=-90,width=0.49\textwidth]{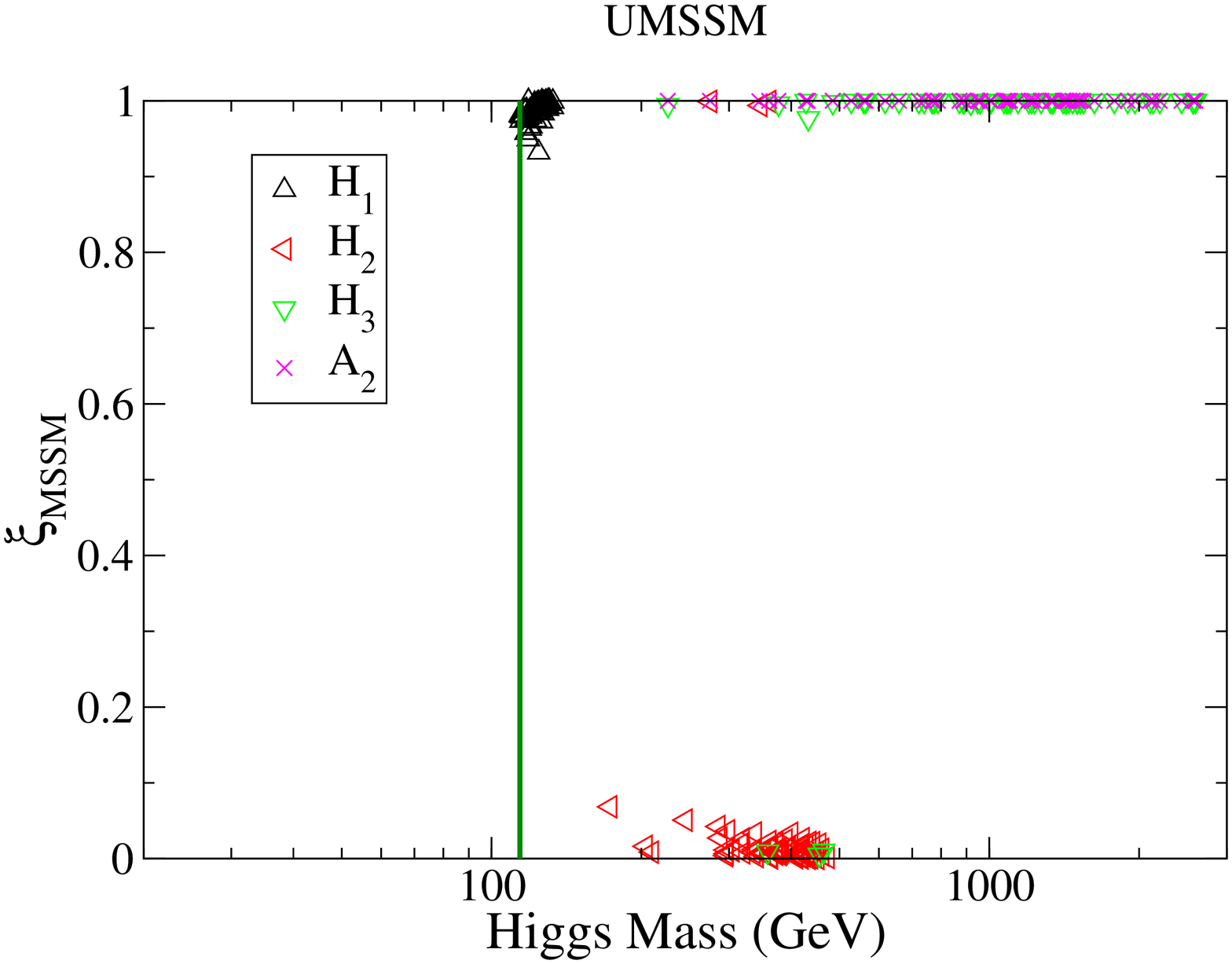}
\includegraphics[angle=-90,width=0.49\textwidth]{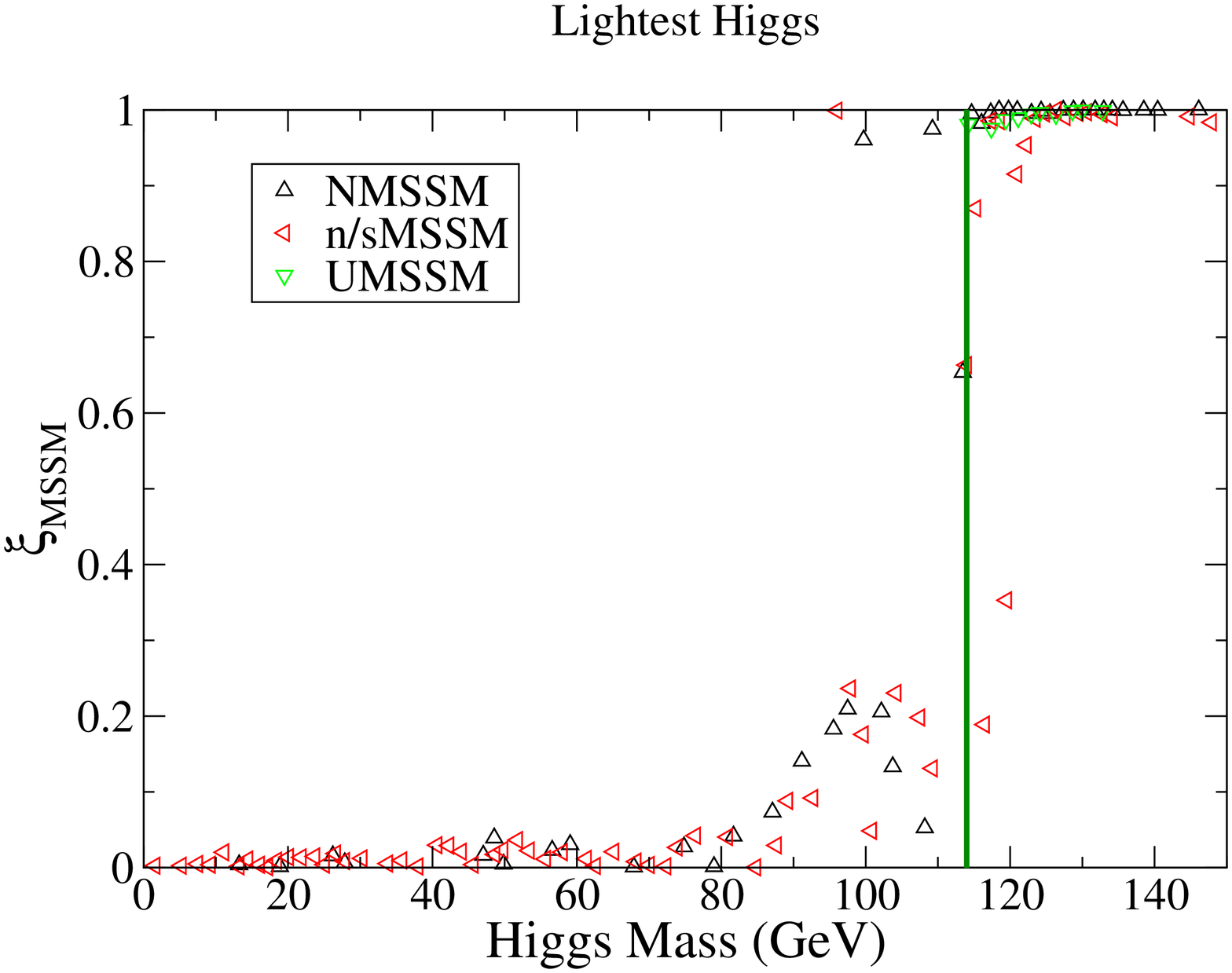}
(c)\hspace{0.48\textwidth}(d)
\caption{Higgs masses vs. $\xi_{\text{MSSM}}$ in the (a) NMSSM, (b) n/sMSSM, (c) UMSSM and (d) the lightest CP-even Higgs of all extended models. The vertical line is the LEP lower bound on the MSSM (SM-like) Higgs mass. }
\label{fig:mh-vs-xi}
\end{center}
\end{figure}

The reduction in the CP-even Higgs mass in extended models can be seen in Fig. \ref{fig:mh-vs-xi}, where we plot the MSSM fraction versus the Higgs boson mass.  When there is little mixing between the singlet and doublet Higgs fields, the MSSM limit is reached and the LEP bound applies, as seen by the MSSM cutoffs at $\xi_{\text{MSSM}} = 1$ and $M_{H_i} = 114.4$ GeV.  A common feature of each model is a CP-even Higgs boson with a mass range concentrated just above the LEP SM mass limit shown by the dark-green vertical line.  These Higgs bosons have a large MSSM fraction, for which the $ZZH_i$ coupling limit is effective in elimination of the generated points.  We note that there are cases where a Higgs boson mass below 114.4 GeV but with relatively high MSSM fraction is allowed due to cancellation between the rotation matrices in Eq. (\ref{eq:zzh}).  This cancellation permits the lightest MSSM Higgs boson to be below the SM limit, and has been taken as a possible explanation of the Higgs signal excess \cite{Drees:2005jg}.

\begin{figure}[t]
\begin{center}
\includegraphics[angle=-90,width=0.49\textwidth]{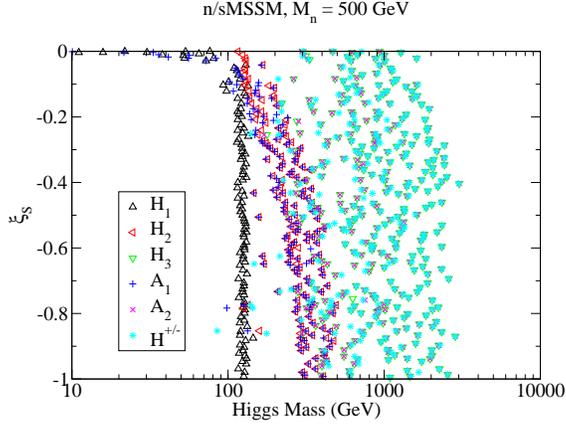}
\caption{Higgs mass dependence on $\xi_S$ in the n/sMSSM.  When $\xi_S\sim -0.1$, $H_2$ and $H_1$ switch content, allowing a light CP-even Higgs below the LEP limit.}
\label{fig:mhxis}
\end{center}
\end{figure}

By measuring the lightest Higgs boson couplings to MSSM fields, an estimation of the MSSM fraction may be obtained, providing important information on the singlet content.  In the NMSSM and especially the n/sMSSM the lightest CP-even Higgs boson may have both low MSSM fraction and low mass as seen in Fig. \ref{fig:mh-vs-xi}d.  Since $\mu_{\rm eff}$ is fixed at the EW scale, the matrix elements $({\cal M}_{+})_{i3}$ are suppressed in the n/sMSSM at large $s$.  This results in a low mass CP-even Higgs boson with high singlet composition; the other Higgs states have a high MSSM fraction due to the sum rule in Eq. (\ref{eq:summssmfrac}).  However in the n/sMSSM, the existence of a low mass CP-even Higgs boson depends on the value of $\xi_S$.  In appendix \ref{apx:lnmhexpr}, we show that the tree-level mass-squares of the singlet dominated CP-even and odd Higgs bosons in the n/sMSSM at large $s$ are
\be
M_{H_1}^2 \sim M_{A_1}^2 \sim -{\sqrt 2 \xi_S M_{\rm n}^3\over s},
\ee
which forces the parameter $\xi_S$ to be negative in this limit.  Therefore, a largely singlet CP-even Higgs boson can have a mass lower than the LEP limit if 
\be
-\xi_S <  {(114\text{ GeV})^2 s\over \sqrt 2 M_{\rm n}^3} \sim 0.1.
\ee
In Fig. \ref{fig:mhxis}, we show the Higgs mass dependence on this parameter, which exhibits the crossing of states at $\xi_S = -0.1$.

\begin{figure}[t]
\begin{center}
\includegraphics[angle=-90,width=0.49\textwidth]{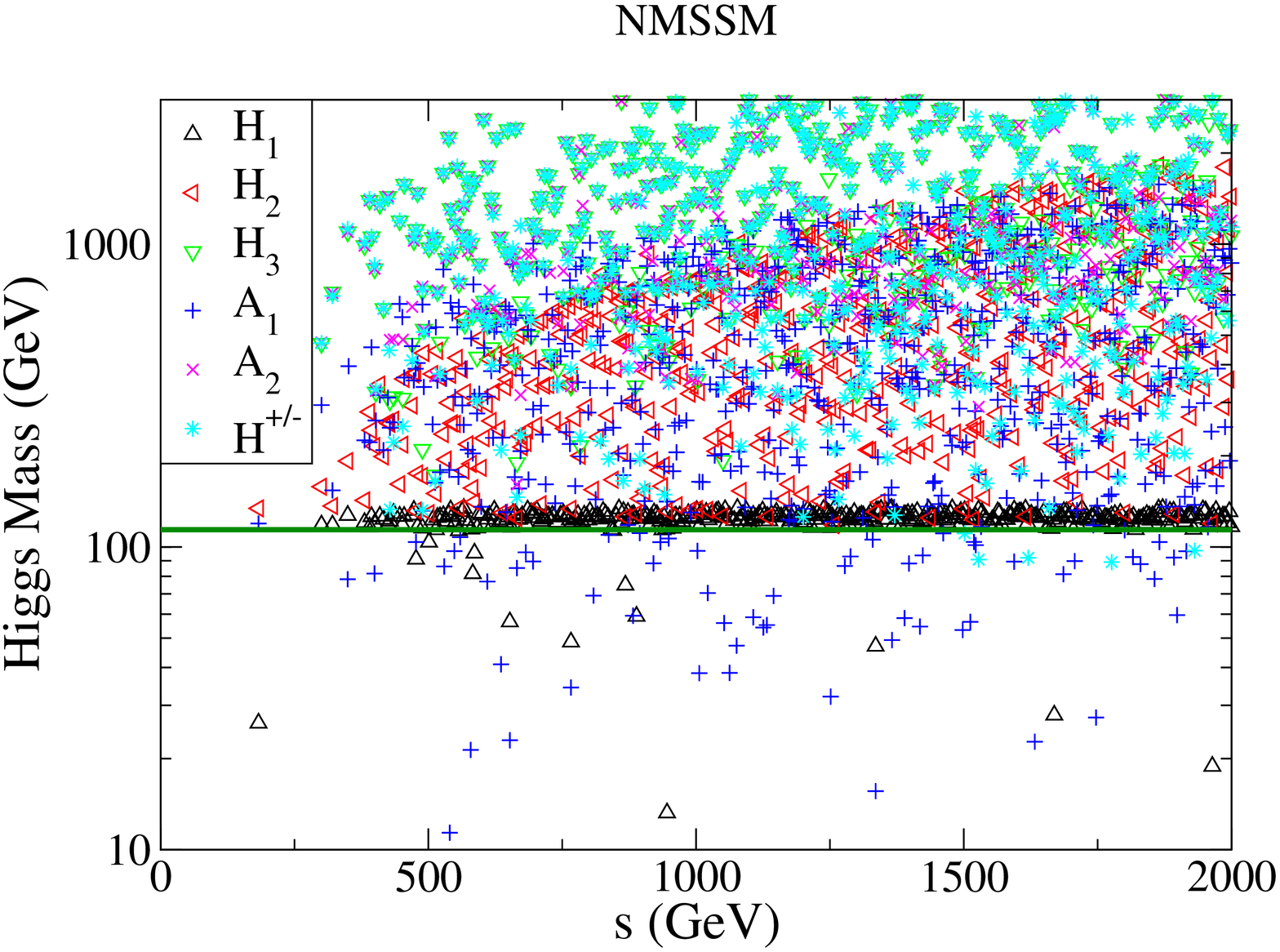}
\includegraphics[angle=-90,width=0.49\textwidth]{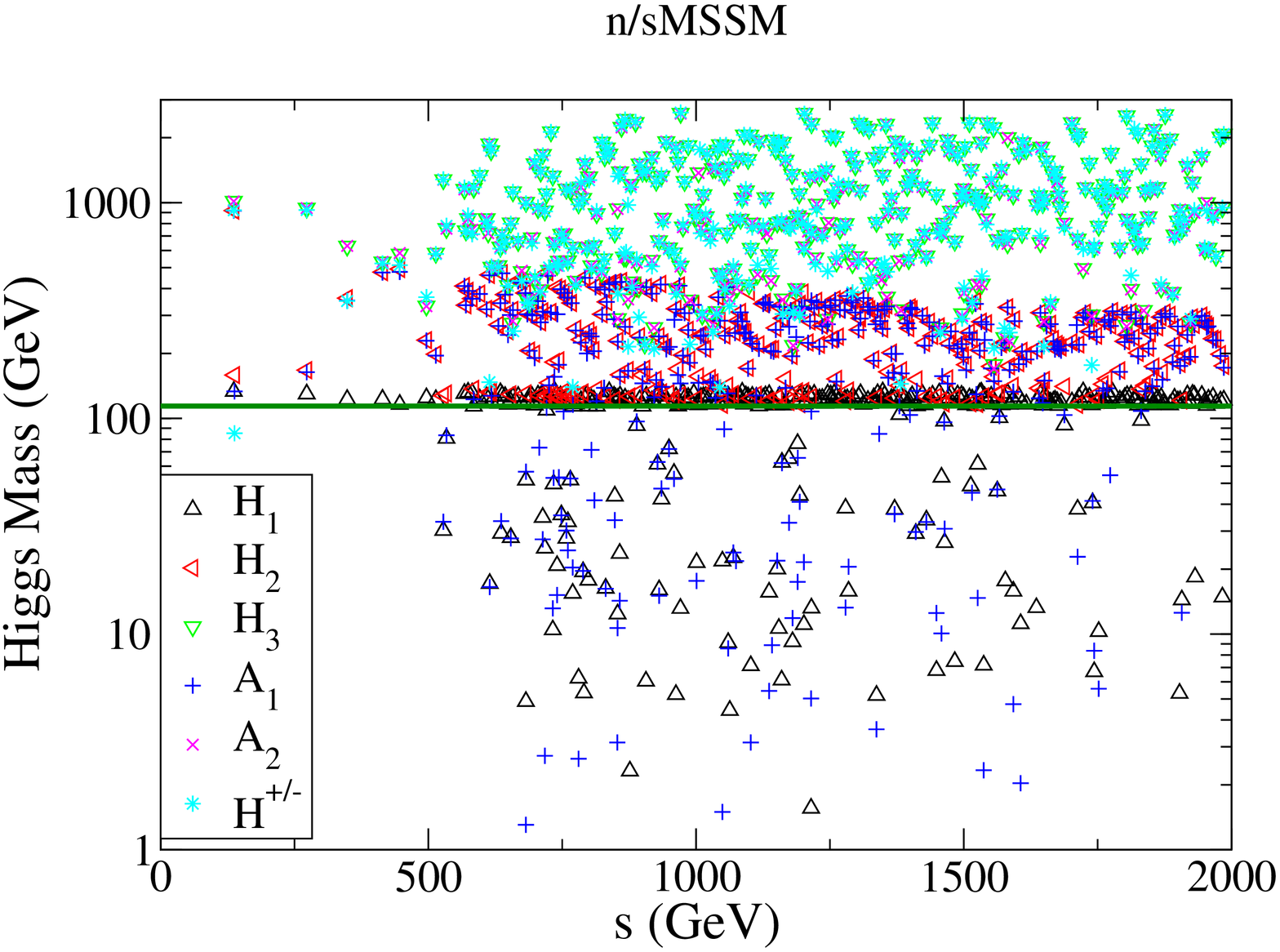}
(a)\hspace{0.48\textwidth}(b)
\includegraphics[angle=-90,width=0.49\textwidth]{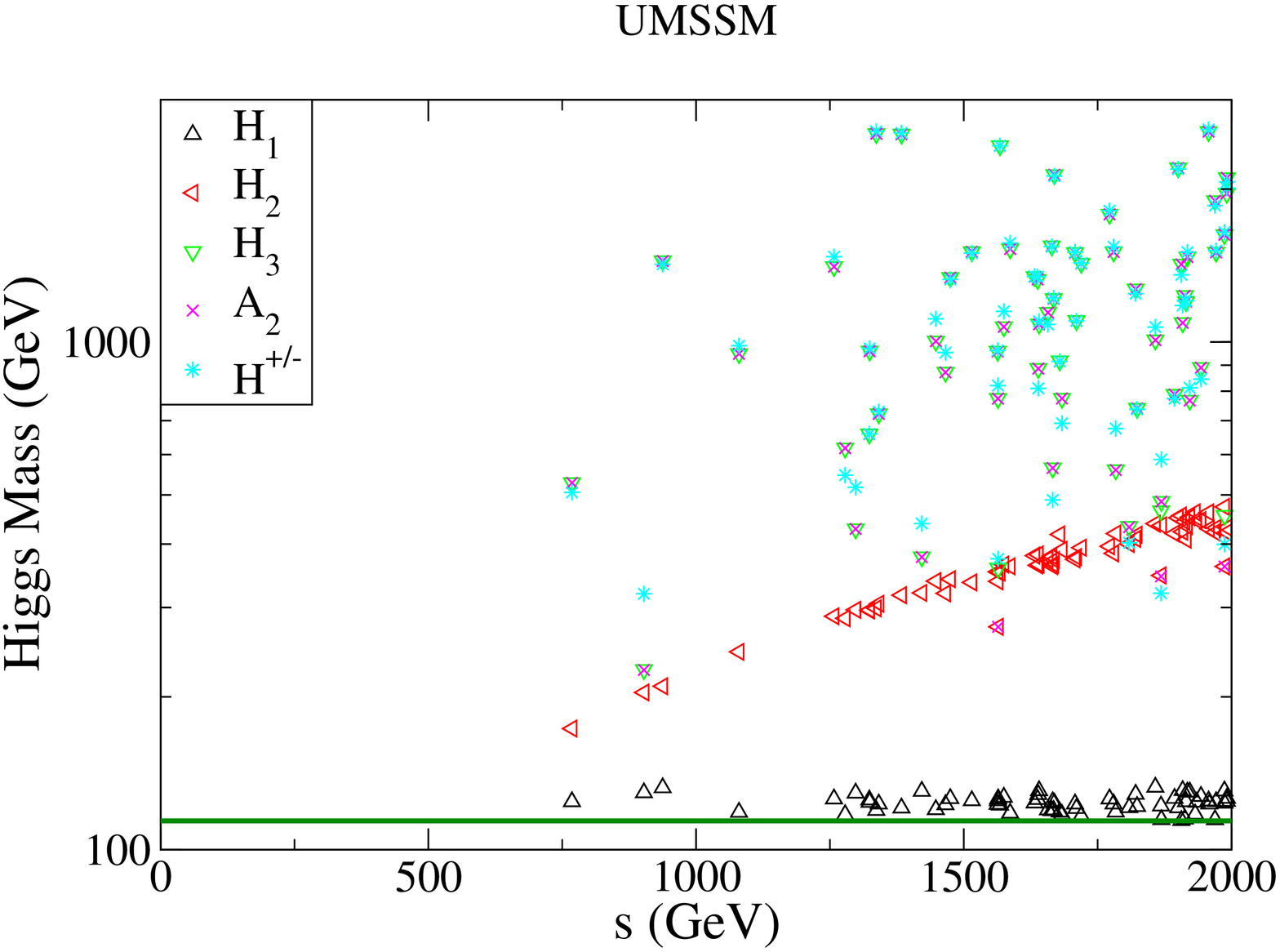}
\includegraphics[angle=-90,width=0.49\textwidth]{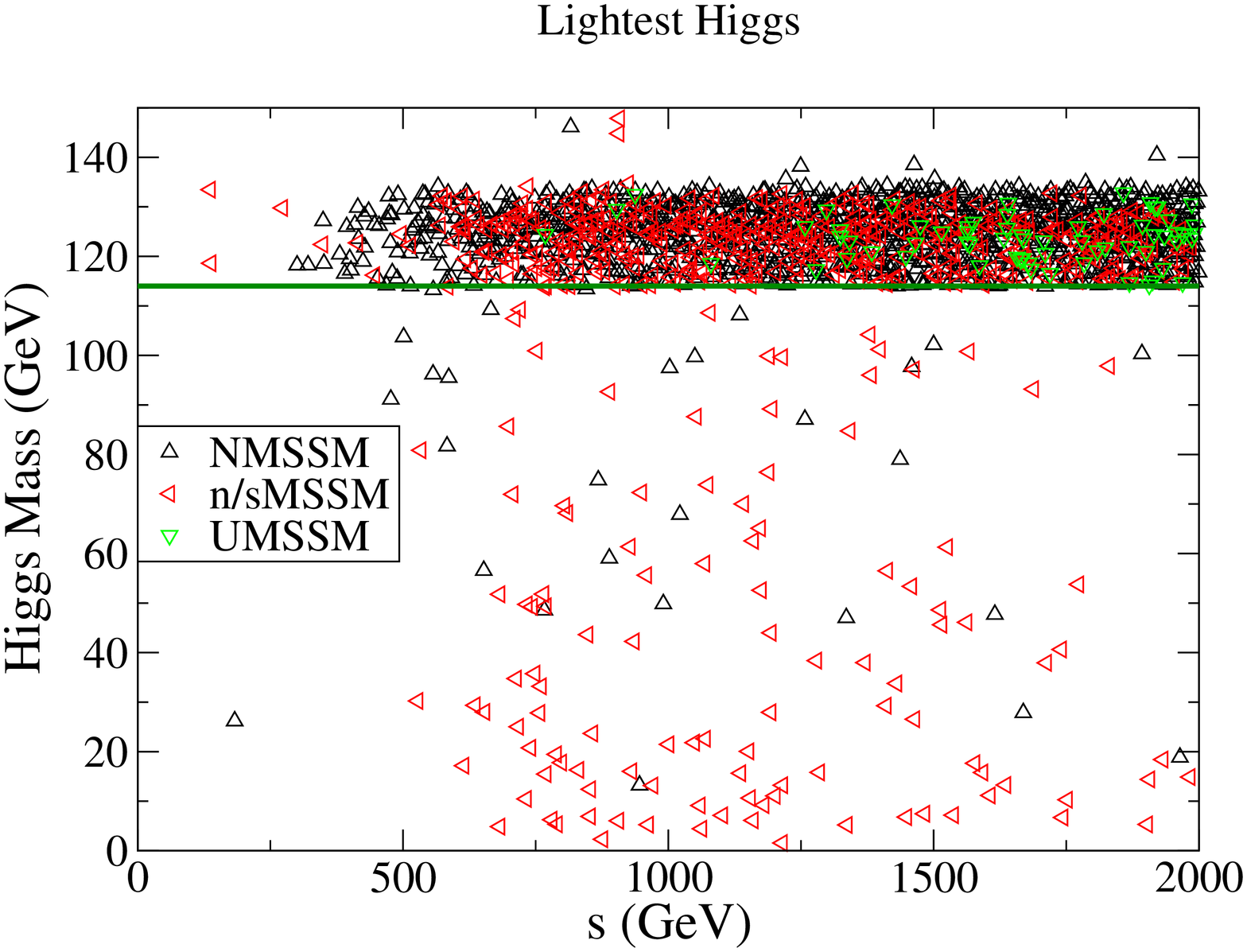}
(c)\hspace{0.48\textwidth}(d)
\caption{Higgs masses vs. $s$ in the (a) NMSSM, (b) n/sMSSM, (c) UMSSM and (d) the lightest CP-even Higgs of all extended models. The vertical line is the LEP lower bound on the mass of the SM Higgs.}
\label{fig:mh-vs-s}
\end{center}
\end{figure}

In the UMSSM, the lightest Higgs mass is concentrated near the LEP limit with $\xi_{\text{MSSM}}$ near one, which is a direct consequence of the high $s$ constraint placed by the strict $\alpha_{ZZ'}$ limit.  This is also seen in Fig \ref{fig:mh-vs-s}, where we plot the Higgs masses versus the singlet VEV.  The lowest allowed point in the UMSSM has $s$ above $\sim 800$ GeV, compared to the other models which allow $s$ to be as low as a few hundred GeV.  By examining Fig. \ref{fig:mh-vs-xi}c and \ref{fig:mh-vs-s}c we see that $M_{H_2}$ varies linearly with $s$ and is characteristically dominantly singlet.  Without the $\alpha_{ZZ'}$ constraint, the $H_1$ and $H_2$ states cross near $s\sim 400$ GeV.  This constraint may be evaded by the fine tuning cases discussed in Section \ref{sect:numeval}.  At this point, the mass eigenstates switch content, below which the lightest Higgs is dominantly singlet, has a mass below the LEP bound, and evades the $ZZH_i$ coupling constraint.  

\begin{figure}[t]
\begin{center}
\includegraphics[angle=-90,width=0.49\textwidth]{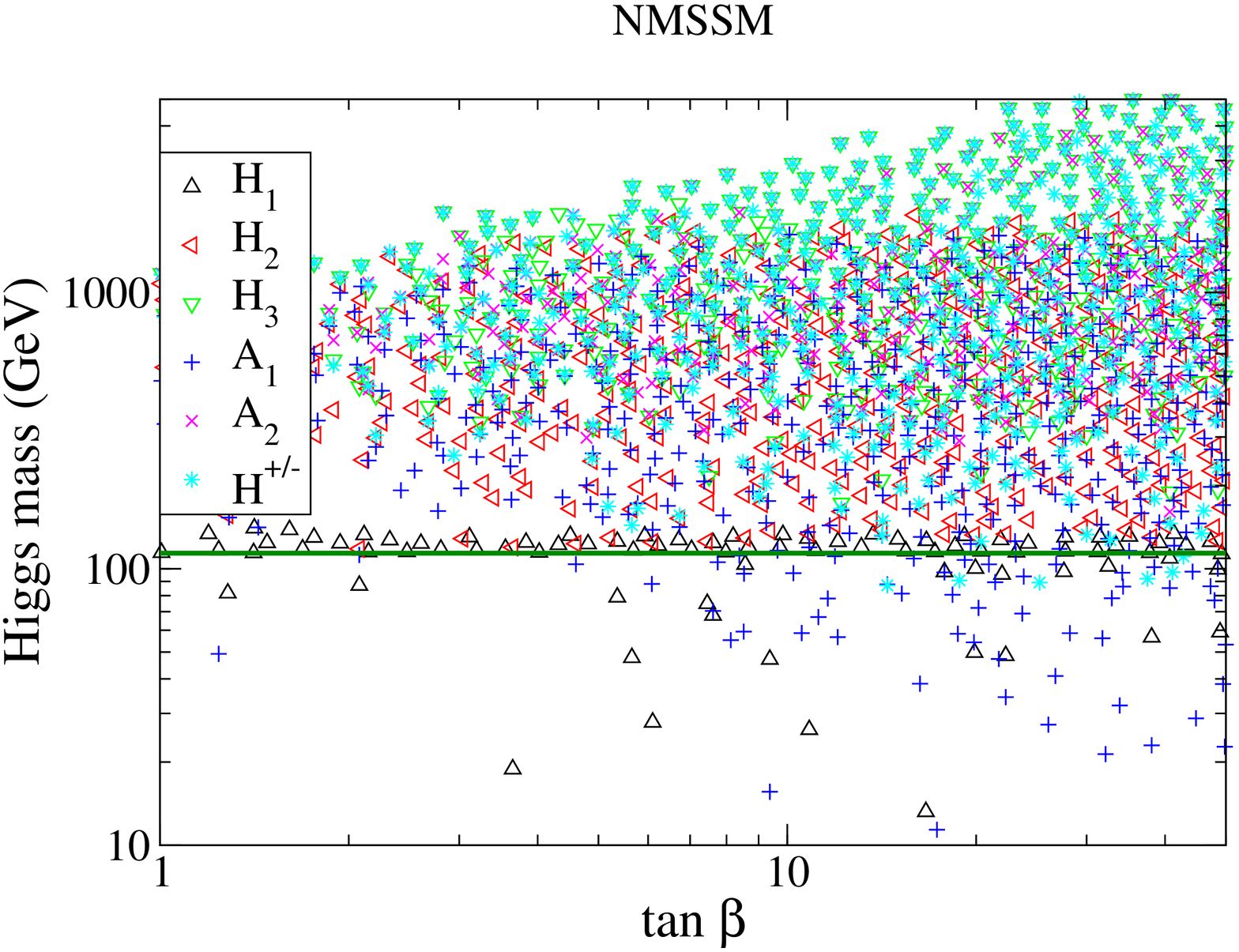}
\includegraphics[angle=-90,width=0.49\textwidth]{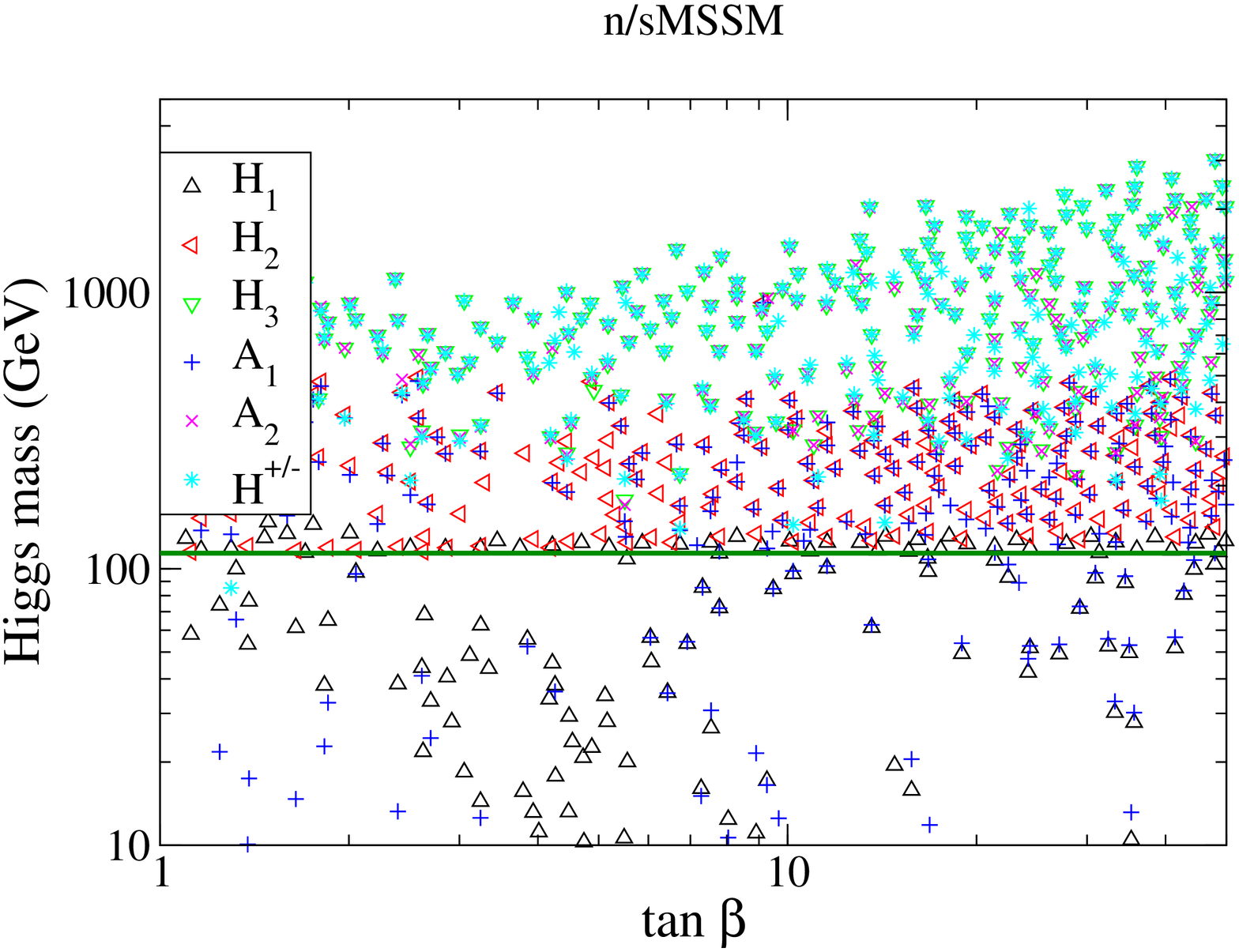}
(a)\hspace{0.48\textwidth}(b)
\includegraphics[angle=-90,width=0.49\textwidth]{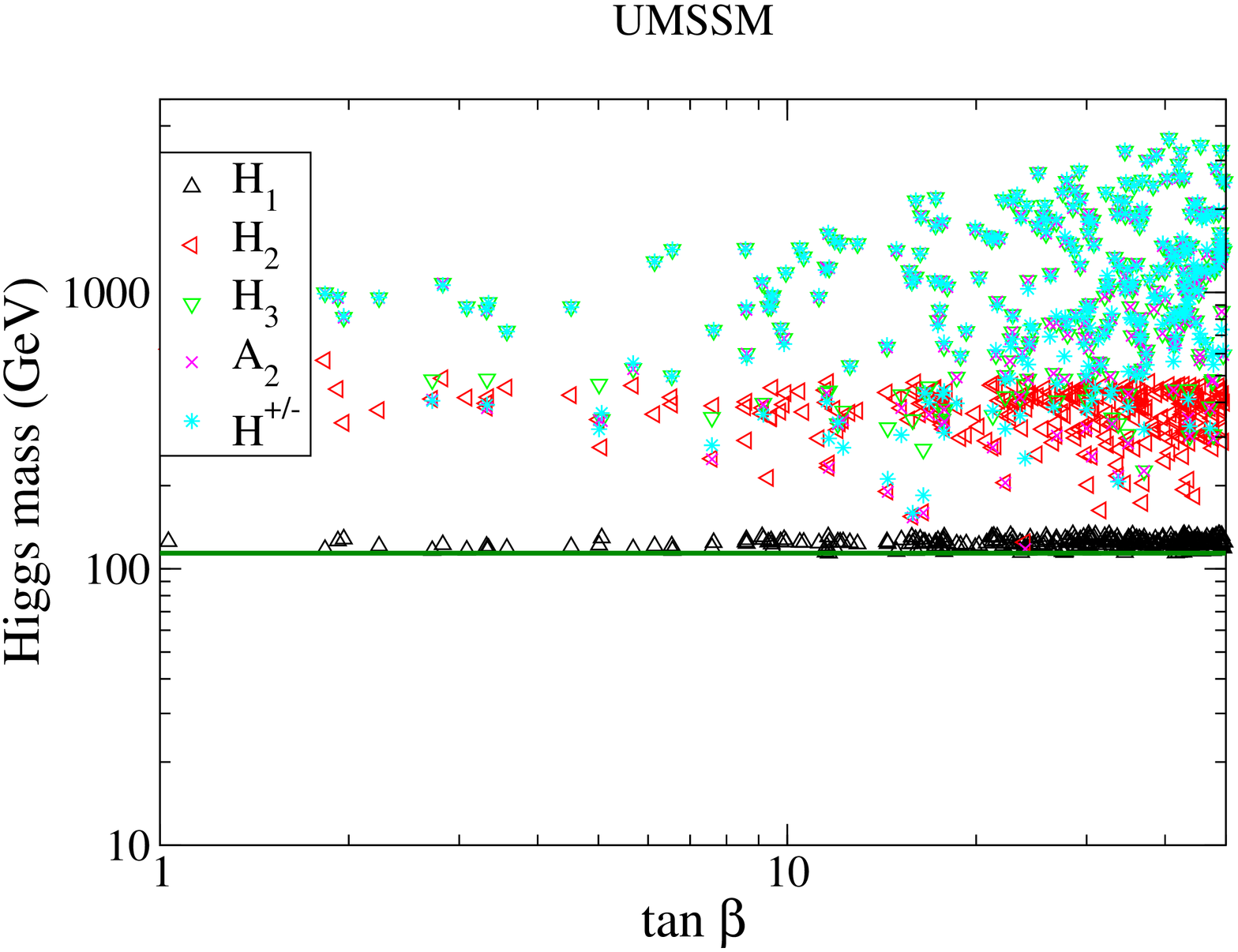}
\includegraphics[angle=-90,width=0.49\textwidth]{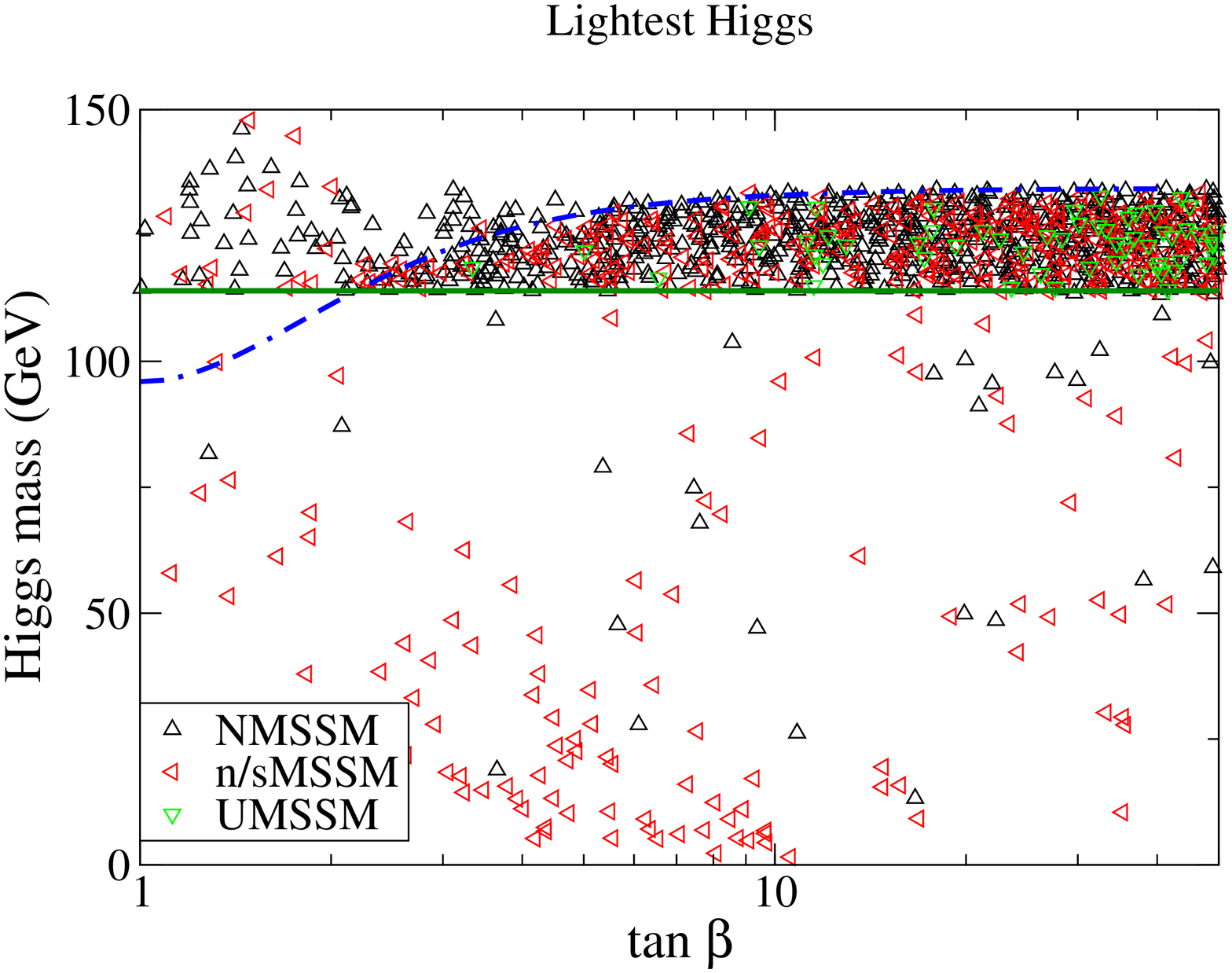}
(c)\hspace{0.48\textwidth}(d)
\caption{Higgs masses vs. $\tan \beta$ in the (a) NMSSM, (b) n/sMSSM, (c) UMSSM and (d) the lightest CP-even Higgs of all extended models where the blue curve shows the theoretical MSSM mass limit with maximal mixing in the stop mass-squared matrix. }
\label{fig:mh-vs-tanb}
\end{center}
\end{figure}

The Higgs mass dependence on $\tan \beta$ has some interesting features, especially that of the lightest Higgs.  We show this dependence in Fig. \ref{fig:mh-vs-tanb} for all the Higgs bosons of each extended MSSM model and separately for the lightest Higgs in all the models considered.  The lightest CP-even Higgs boson mass vs. $\tan \beta$ in each model shown in Fig. \ref{fig:mh-vs-tanb}d has a majority of generated points in the band $114.4\text{ GeV}\lesssim M_{H_1}\lesssim 135\text{ GeV}$ and $\tan \beta \gtrsim 2$.  This is one of the salient features of the MSSM as shown in Fig. \ref{fig:modelscans}a.  The MSSM parameter space has a lower cutoff at $\tan \beta \sim 2$ due to the LEP limit at 114.4 GeV for a SM-like Higgs and is shown in Fig \ref{fig:mh-vs-tanb}d as the intersection of the theoretical MSSM Higgs mass limit shown in blue and the LEP limit in green.  However, the extended-MSSM models may have values of $\tan \beta$ that are below this region.  Since mixing effects can decrease the lightest Higgs mass and thereby satisfy the LEP bounds, a strict bound on $\tan \beta$ cannot be given.  Additionally, an increase in the Higgs mass from the MSSM theoretical limit shown in Section \ref{sect:massbounds} can permit low $\tan \beta$ scenarios which have masses above the LEP limit.

Among these models, the heaviest CP-odd Higgs state follows the same dependence on $\tan \beta$ that was noted above in Section \ref{sect:numeval}.  The heaviest CP-even Higgs and charged Higgs bosons also follow this trend with the charged Higgs boson mass having the same $\tan \beta$ dependence as the CP-odd Higgs mass, see e.g. Eq. (\ref{eq:chghiggs}).  The heaviest CP-even and CP-odd Higgs masses are approximately the same even after radiative corrections.  An explanation is provided by the mass-squared sum rules that each model obeys, namely
\be
\sum_i M_{H_i}^2 - \sum_j M^2_{A_j} = M^2_Z + M_{xMSSM}^2 + \delta M^2,
\label{eq:sumrule1}\ee
The sums are over the massive Higgs bosons, and $M_{xMSSM}$ is a model-dependent mass parameter with values
\bea
M_{NMSSM}^2 &=& 2 \kappa (-h_s v_d v_u + s (\sqrt{2} A_\kappa + s \kappa)),\\
M_{n/sMSSM}^2 &=& 0,\\
M_{UMSSM}^2 &=& M_{Z'}^2.
\label{eq:sumrule2}
\eea
\begin{figure}[t]
\begin{center}
\includegraphics[angle=-90,width=.49\textwidth]{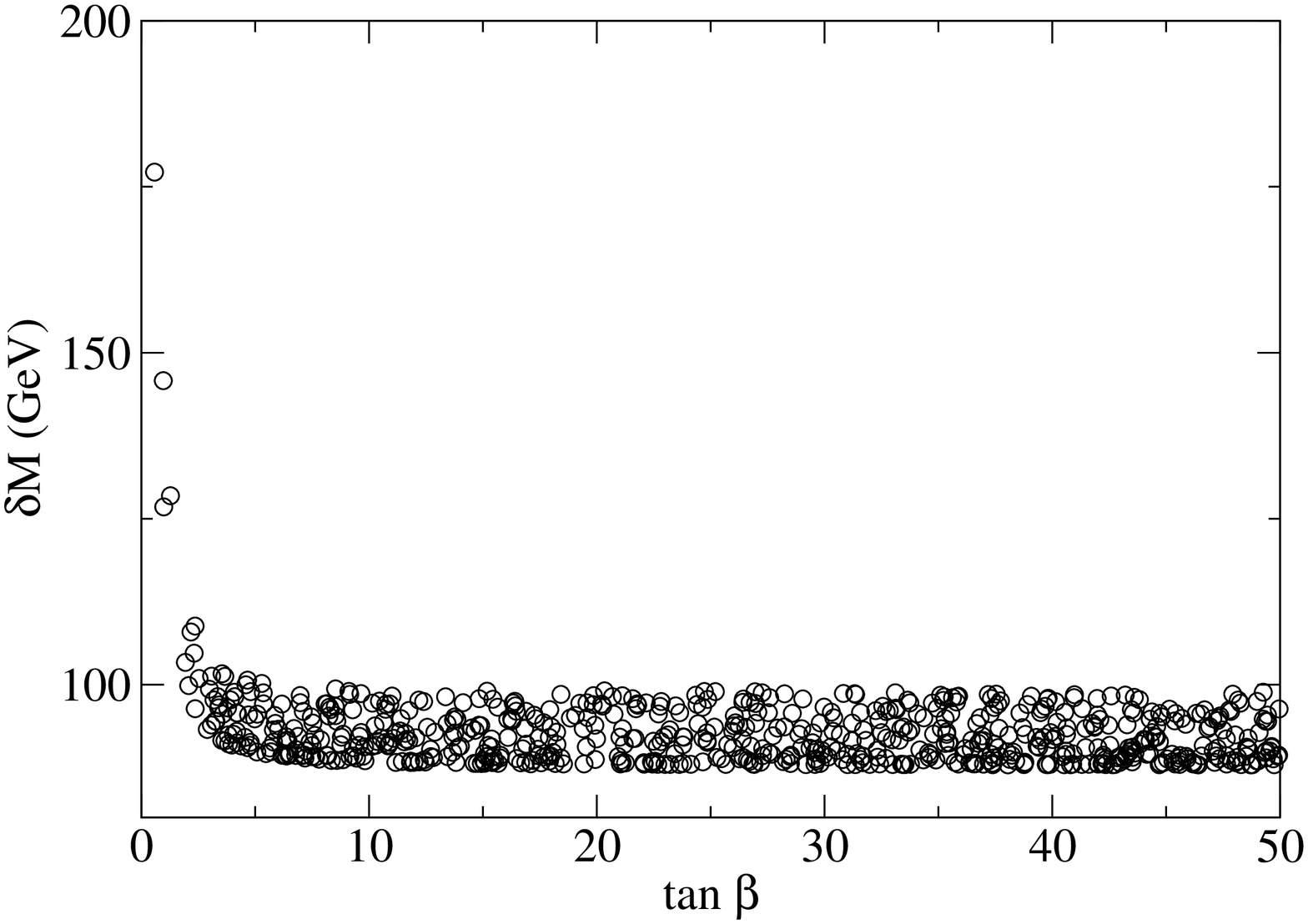}
\includegraphics[angle=-90,width=.49\textwidth]{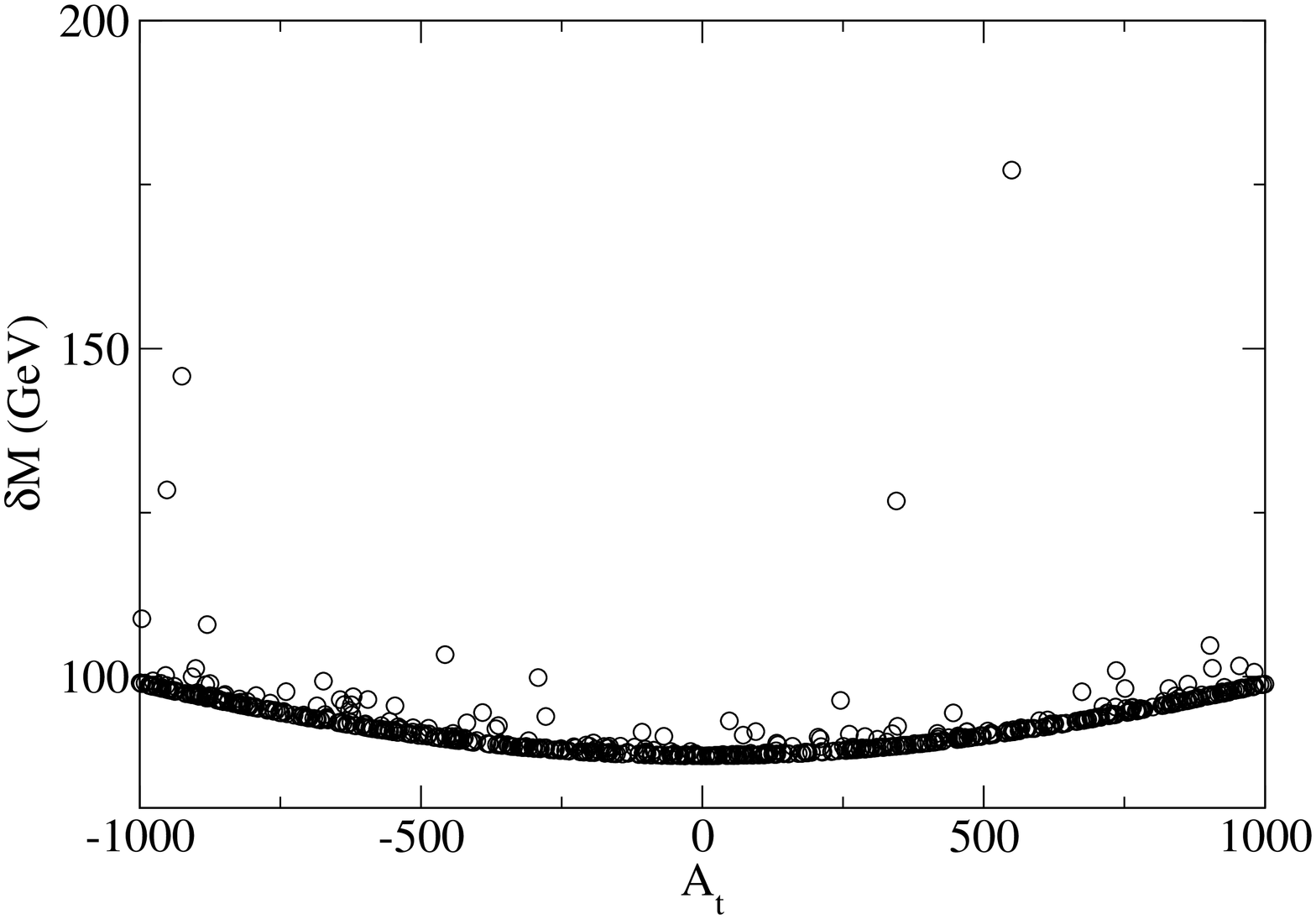}
(a)\hspace{0.48\textwidth}(b)
\caption{Radiative corrections, $\delta M \equiv (\delta M^2)^\half$ to the Higgs mass-squared sum rule vs. (a) $\tan \beta$ and (b) $A_t$.  The radiative corrections introduce a deviation from the sum rule by at most ${\cal O}(100 \text{ GeV})^2$ over most of the range of the scan.  A larger deviation is seen at low $\tan \beta$ due to a larger Yukawa coupling there.}
\label{fig:sumrule}
\end{center}
\end{figure}

The term $\delta M^2$ in Eq. (\ref{eq:sumrule1}) is due to the radiative corrections, and has a value
\be
\delta M^2 = Tr\left[ {\cal M}^1_{+} - {\cal M}^1_{-}\right],
\ee
that gives an estimate of the effect the radiative corrections have on the Higgs masses.  Note that the CP-odd radiative corrections, the ${\cal F}$ terms, are cancelled by equivalent terms in the CP-even mass-squared matrix.  The radiative corrections alter the sum rule by at most ${\cal O}(100 \text{ GeV})^2$ over most of the scanned range, as seen in Fig. \ref{fig:sumrule} where we plot the shift versus both $\tan \beta$ and $A_t$.

The radiative correction contributions to the sum rule are largest for large $A_t$ and small $\tan \beta$.  Since the top quark Yukawa coupling increases when $\tan \beta$ is small, the radiative corrections are enhanced at small $\tan \beta$, causing larger deviations from the sum rule.  Since radiative corrections only affect the sum rule by ${\cal O}(100\text{ GeV})$, any high mass CP-even Higgs boson contribution must be cancelled by a CP-odd Higgs of similar mass.  

\section{Collider Phenomenology}
\label{sect:collpheno}
Higgs boson decays are important to consider as they affect signals at colliders.  Both production and decay modes are relevant in determining whether a given model yields detectable Higgs  physics.  While Higgs searches have been addressed for the NMSSM \cite{ref:studies,ref:Nmssmlhc}, a side-by-side comparison of the NMSSM, n/sMSSM, and UMSSM has not yet been made.  In the above parameter scans, we calculate the partial decay widths relevant to production and branching fraction for decays in these models of various important modes.

\subsection{Higgs Production}

At hadron colliders the dominant production of the lightest Higgs boson in the SM proceeds through $gg$ fusion and/or Weak Boson Fusion (WBF).  The Higgs production cross-sections are directly related to Higgs decay widths when the decay channels are kinematically accessible at the Higgs mass.  Considerable effort has been put into calculating Higgs decays beyond leading order \cite{ref:higgsdecay}.  In the SM, MSSM, and NMSSM, numerical codes have been implemented to calculate these widths precisely \cite{ref:hdecay,ref:nmhdecay}.  We calculate the partial decay widths of $H \to gg$, $WW$, and $ZZ$ in each model via HDECAY \cite{ref:hdecay} with the SM Higgs couplings modified to reflect the model of interest.  Higgs decay to $gg$ occurs via quark and squark loops and is calculated at NLO.  However, to a good approximation, the loops involving heavy squarks are suppressed \cite{ref:gunion86}.  The large squark mass approximation is justified for our assumed values of $M_{\widetilde Q}=M_{\widetilde U}=M_{\widetilde D}=1$ TeV. Therefore, we only consider the SM quark loops in the $H\to gg$ calculation.  Decays to weak boson pairs are calculated at tree-level as the radiative corrections to the width are negligible.  

\begin{figure}[htbp]
\begin{center}
\includegraphics[angle=-90,width=0.49\textwidth]{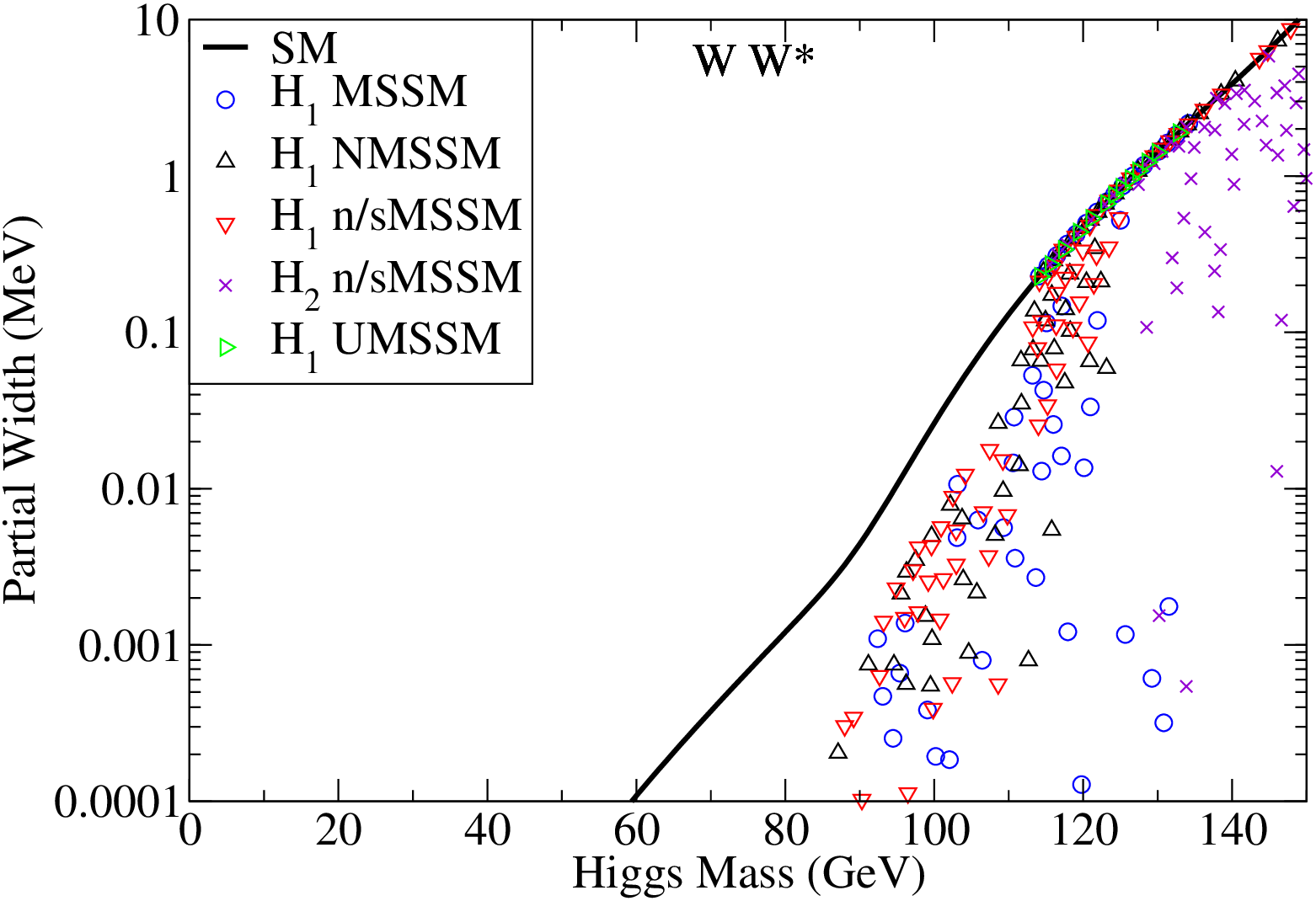}
\includegraphics[angle=-90,width=0.49\textwidth]{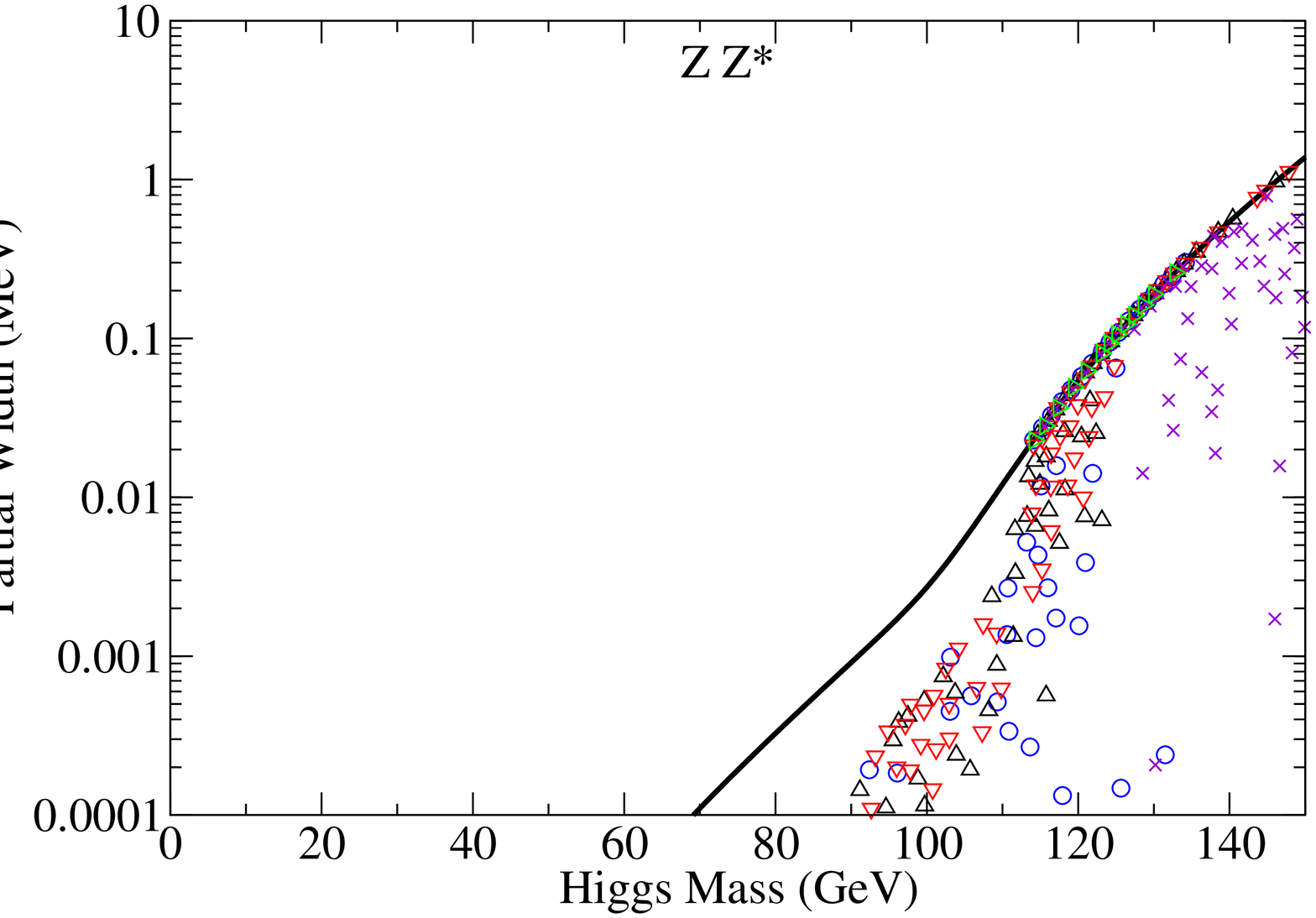}
\includegraphics[angle=-90,width=0.49\textwidth]{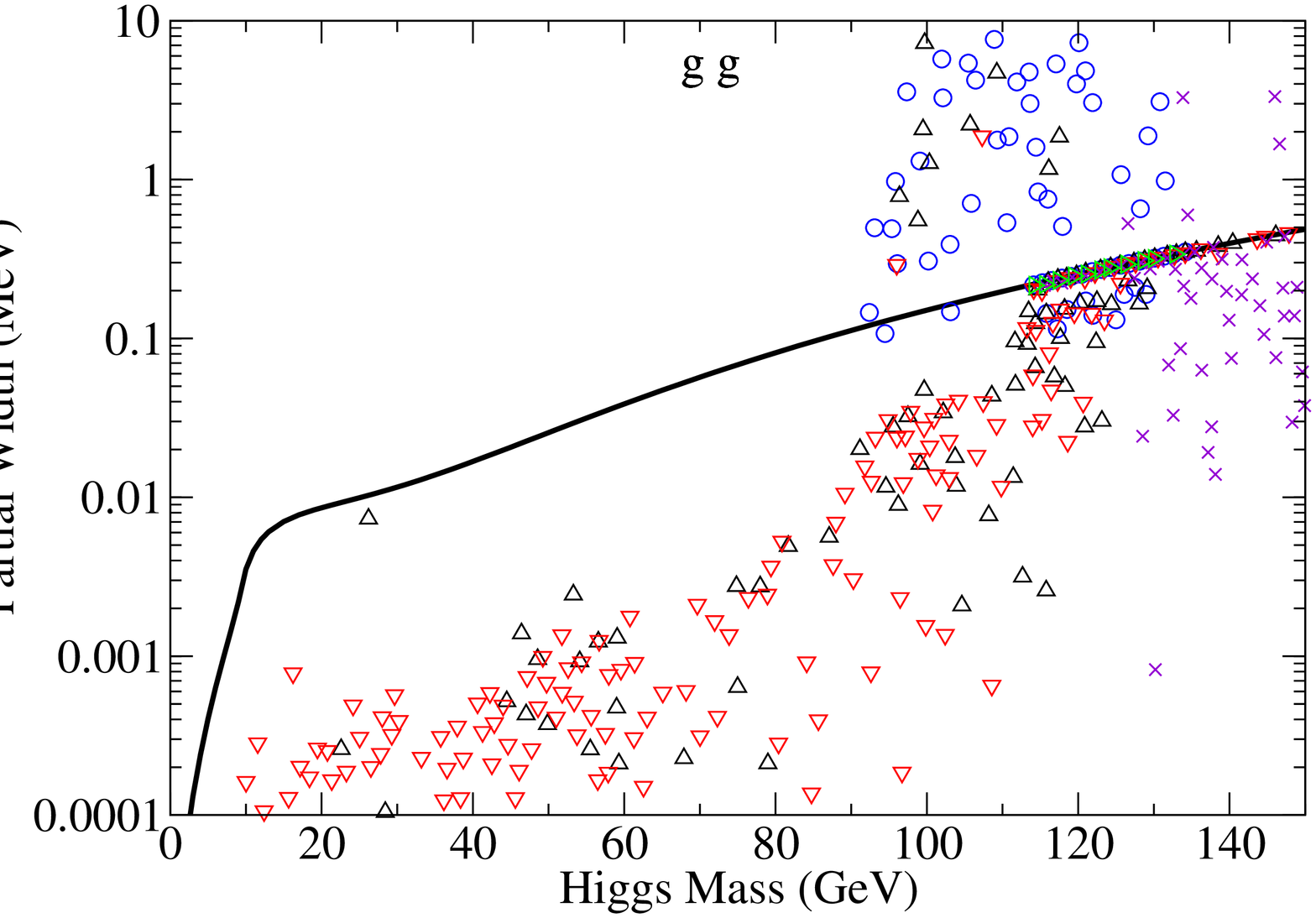}
\caption{Decay widths for $WW^*$, $ZZ^*$, and $gg$ in the MSSM and extended-MSSM models.  Curves denote the corresponding SM width.  For clarity, not all points generated are shown.}
\label{fig:prod}
\end{center}
\end{figure}

In Fig. \ref{fig:prod}, we plot the partial widths of the lightest Higgs boson for the SM, MSSM and extended-MSSM models.  Since the n/sMSSM and, to a lesser extent, the NMSSM contain a very light Higgs with high singlet composition, its decay widths to SM particles are highly suppressed.  However, from Fig. \ref{fig:mh-vs-xi}, the second lightest CP-even Higgs boson in the n/sMSSM has a high MSSM fraction and often has comparable mass to the lightest Higgs bosons in other models.  Hence, we also show the decay width of the second lightest Higgs boson in the n/sMSSM, as it is characteristically similar to the lightest Higgs boson of the MSSM.  The decay widths of the lightest Higgs in the MSSM show a large spread with respect to the SM, associated with low $A_2$ mass: see Fig. \ref{fig:lep}b.  When $M_{A_2} \gg M_Z$, the masses and couplings of the lightest CP-even Higgs approach those of the SM Higgs \cite{ref:barger-phillips-stange}.  In the UMSSM, the $\alpha_{ZZ'}$ limit forces the model to be near the $s$-decoupling limit, resulting in masses, couplings, and decay widths that are close to those of the MSSM.  Consequently, the UMSSM decay widths lie directly on the SM width in Fig. \ref{fig:prod}.

In the models considered, the $H_1$ mass is typically below the $WW$ and $ZZ$ thresholds.  Therefore, the off-shell $WW^*$ and $ZZ^*$ decay widths are evaluated in the MSSM and its extensions\footnote{In this case the decay width cannot be translated directly into production rates since they require transverse and longitudinal polarizations of the $W$-bosons to be treated separately.  However, the gauge coupling is equivalent in either case, and its scaling contains the suppression of the production rate.}.  For the decays of the very light Higgs boson to occur in the n/sMSSM two off-shell gauge bosons are involved, resulting in high kinematic suppression of decay rates.  In all the models considered, the $WW^*$ and $ZZ^*$ partial widths are bounded above by those of the SM.  This is a consequence of the complementarity of the couplings of $H_1$ and $H_2$ to gauge fields in the MSSM.  The gauge couplings in the MSSM follow the relation
\be
(g^{SM}_{VVh})^2=(g^{MSSM}_{VVH_1})^2+(g^{MSSM}_{VVH_2})^2.
\label{eq:coupsumrule}
\ee
More sum rules exist in the MSSM and can be found in Ref. \cite{ref:gunionsum}.  In extended-MSSM models the gauge couplings are related to the SM couplings by
\be
g_{VVH_i}=g^{SM}_{VVh} (R_{+}^{i1} \cos\beta +R_{+}^{i2} \sin\beta),
\label{eq:vvhcoup}
\ee
where $g^{SM}_{ZZh}= {i g_2 M_Z g_{\mu \nu}\over \cos \theta_W}$ and $g^{SM}_{WWh}= i g_2 M_W g_{\mu \nu}$.  The sum rule in Eq. (\ref{eq:coupsumrule}) generalizes to one involving three Higgs couplings.  Therefore, the coupling of the lightest Higgs boson to weak bosons in the MSSM and its extensions is always reduced compared to the SM couplings.  LEP constraints require the $ZZH_i$ coupling to be below the SM coupling when the Higgs mass is below the 114.4 GeV limit.  Associated production $q \bar q \to V^* \to V H_i$ for SM Higgs bosons can be important at the Tevatron and the LHC for low Higgs masses \cite{ref:lhcprodxsect,ref:sugrawgc}.  The corresponding production cross section can be scaled from the SM calculation by the $VVH_i$ coupling in Eq. (\ref{eq:vvhcoup}).

The $H_1\to gg$ partial width governs Higgs boson production via $gg$ fusion at hadron colliders.  We see in Fig. \ref{fig:prod} that the $gg$ partial decay width is typically suppressed in the n/sMSSM for a low mass of the lightest Higgs boson since it is dominantly singlet.  However, there is a trade-off in the production cross-section between smaller $\Gamma(H \to gg)$ and the kinematic enhancement from a lighter $M_{H_1}$ whose interplay is beyond the scope of this paper.  The lightest Higgs in the NMSSM and MSSM and the second lightest Higgs in the n/sMSSM have decay widths to $gg$ that may be either enhanced or suppressed by a few orders of magnitude depending on the Higgs coupling to the internal quarks and their interferences.  However, the lightest Higgs in the UMSSM and the MSSM in the limit of a large CP-odd mass shows no significant deviations from the SM $h \to gg$ decay width.

\subsection{Decay Branching Fractions}
\label{sect:decaybf}
Specific decay modes are important for identifying the Higgs boson at colliders.  We calculate the contributions of $b \bar b$, $c\bar c$, $s \bar s$, $\tau^+ \tau^-$, $\mu^+ \mu^-$, $W W^*$, $Z Z^*$, and  $g g$ to the total decay width of the lightest or MSSM-like Higgs boson in each model using HDECAY after modifying the corresponding Higgs couplings.  In addition, non-SM decays including $\lsp \lsp$ and $A_i A_i$ are calculated since they are often quite light in the extended-MSSM models.  The decays to $\gamma \gamma$, and $\gamma Z$ are also calculated with loops involving quarks, $W^\pm$-bosons, charged Higgs bosons, charginos, and squarks (which decouple for sufficiently large squark mass).  

\begin{figure}[h]
\begin{center}
\includegraphics[angle=-90,width=0.45\textwidth]{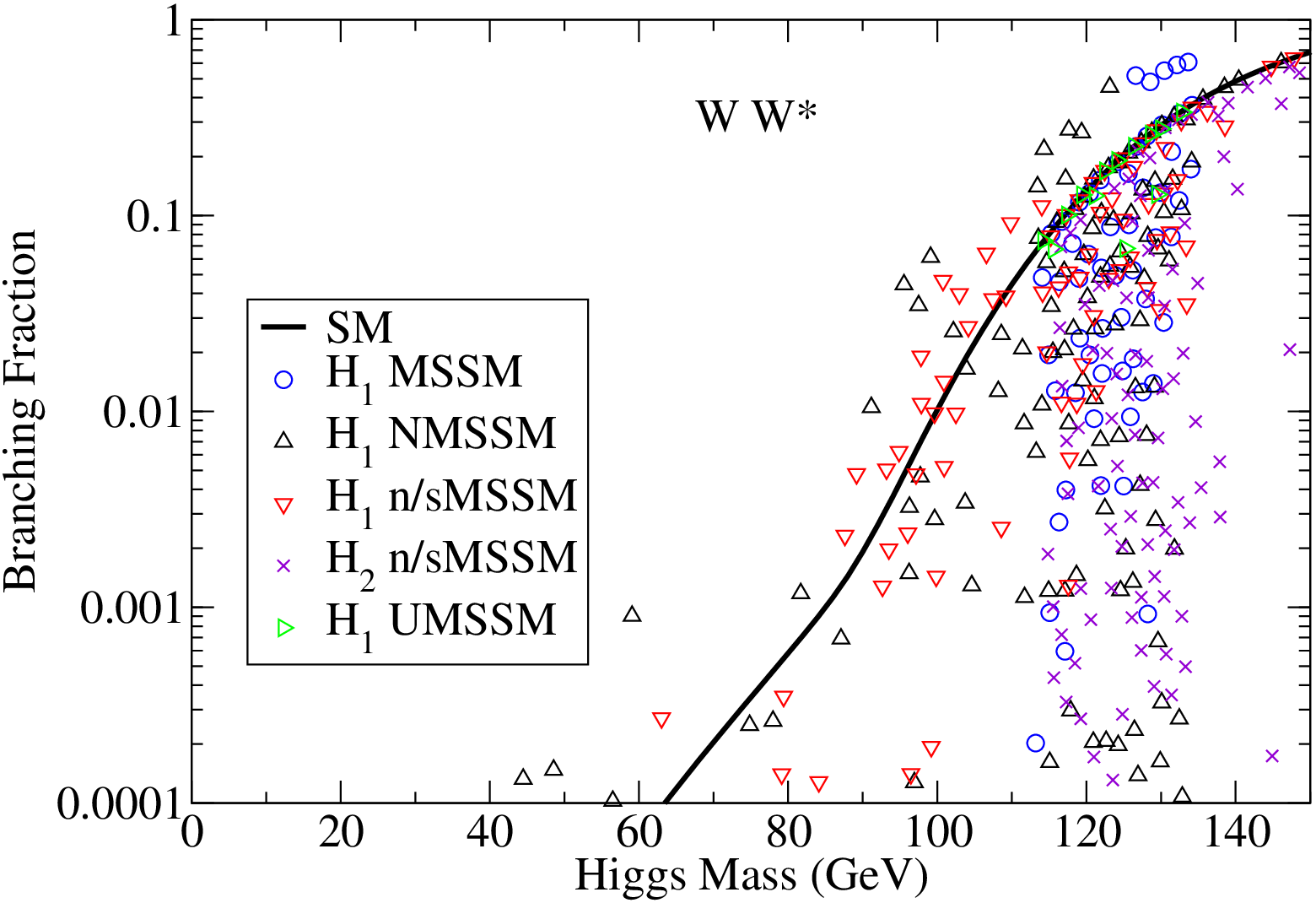}
\includegraphics[angle=-90,width=0.45\textwidth]{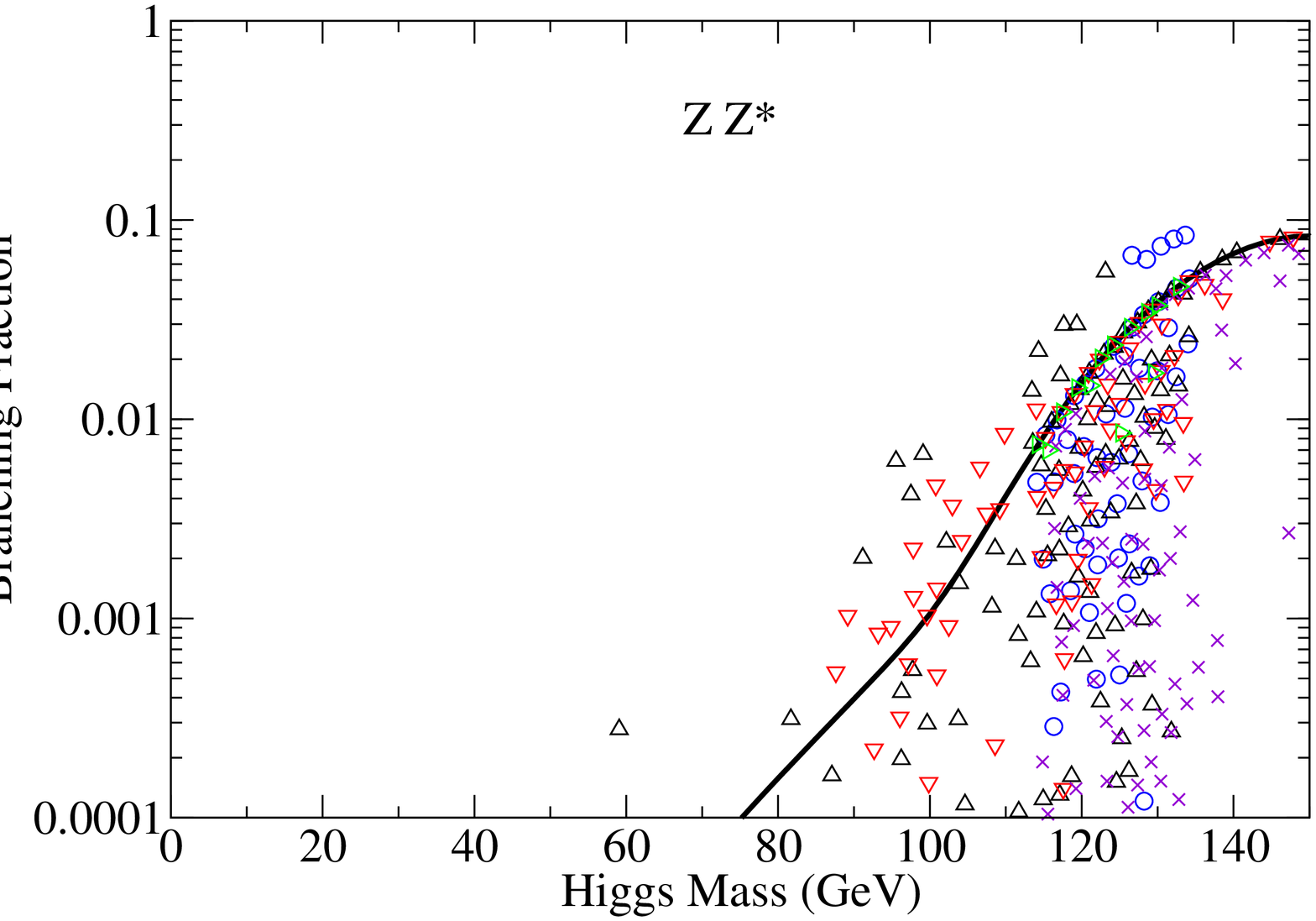}
\includegraphics[angle=-90,width=0.45\textwidth]{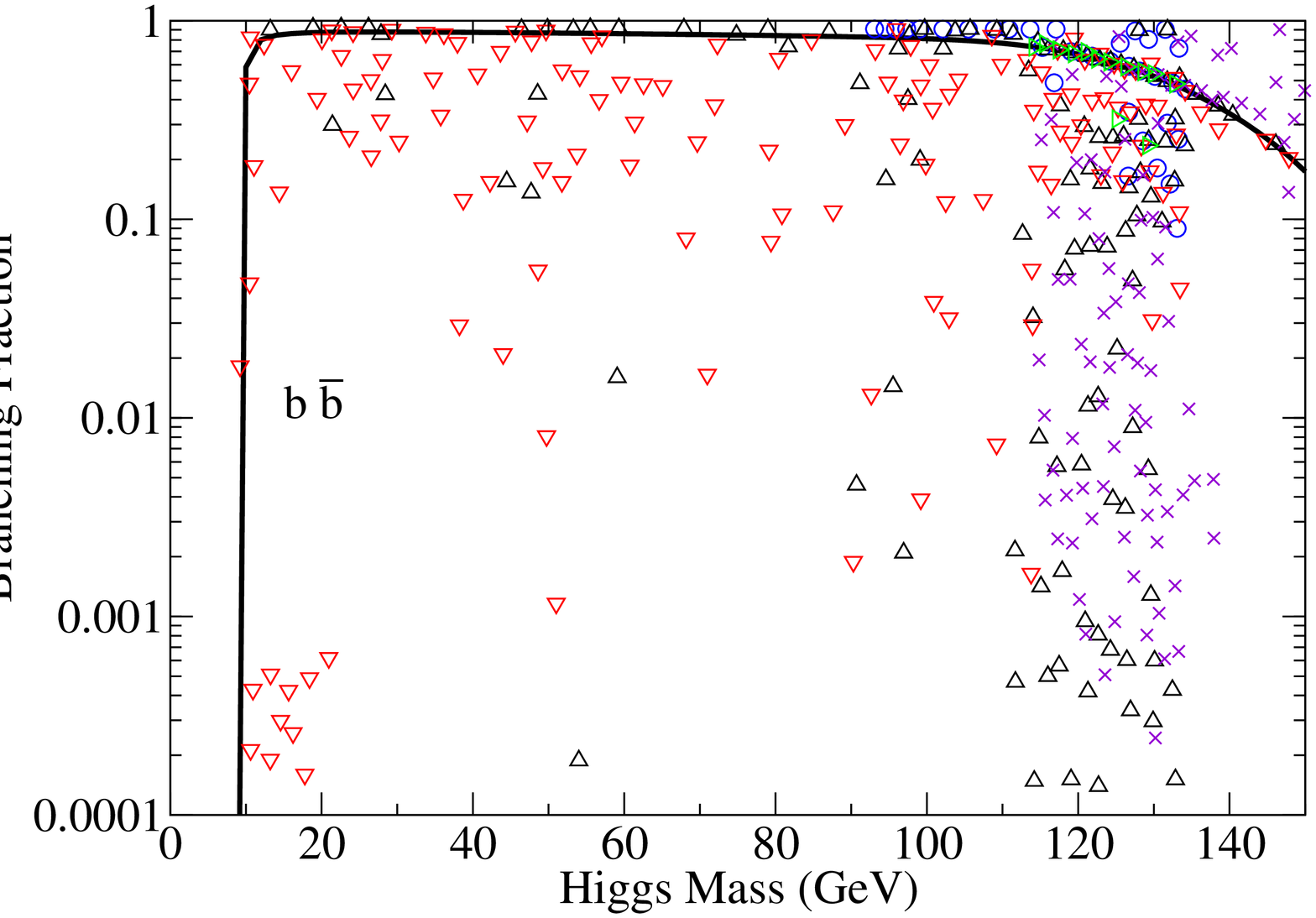}
\includegraphics[angle=-90,width=0.45\textwidth]{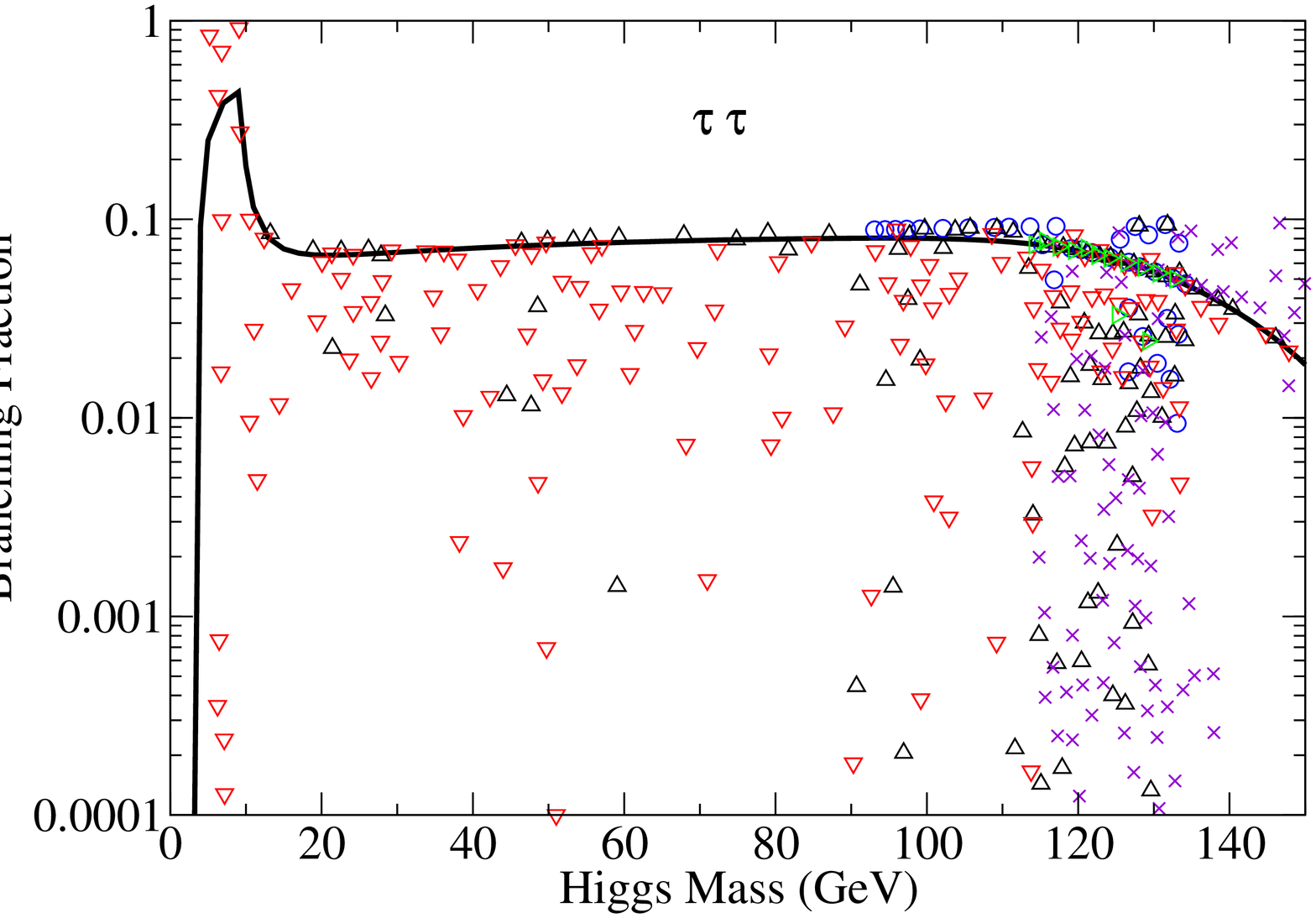}
\includegraphics[angle=-90,width=0.45\textwidth]{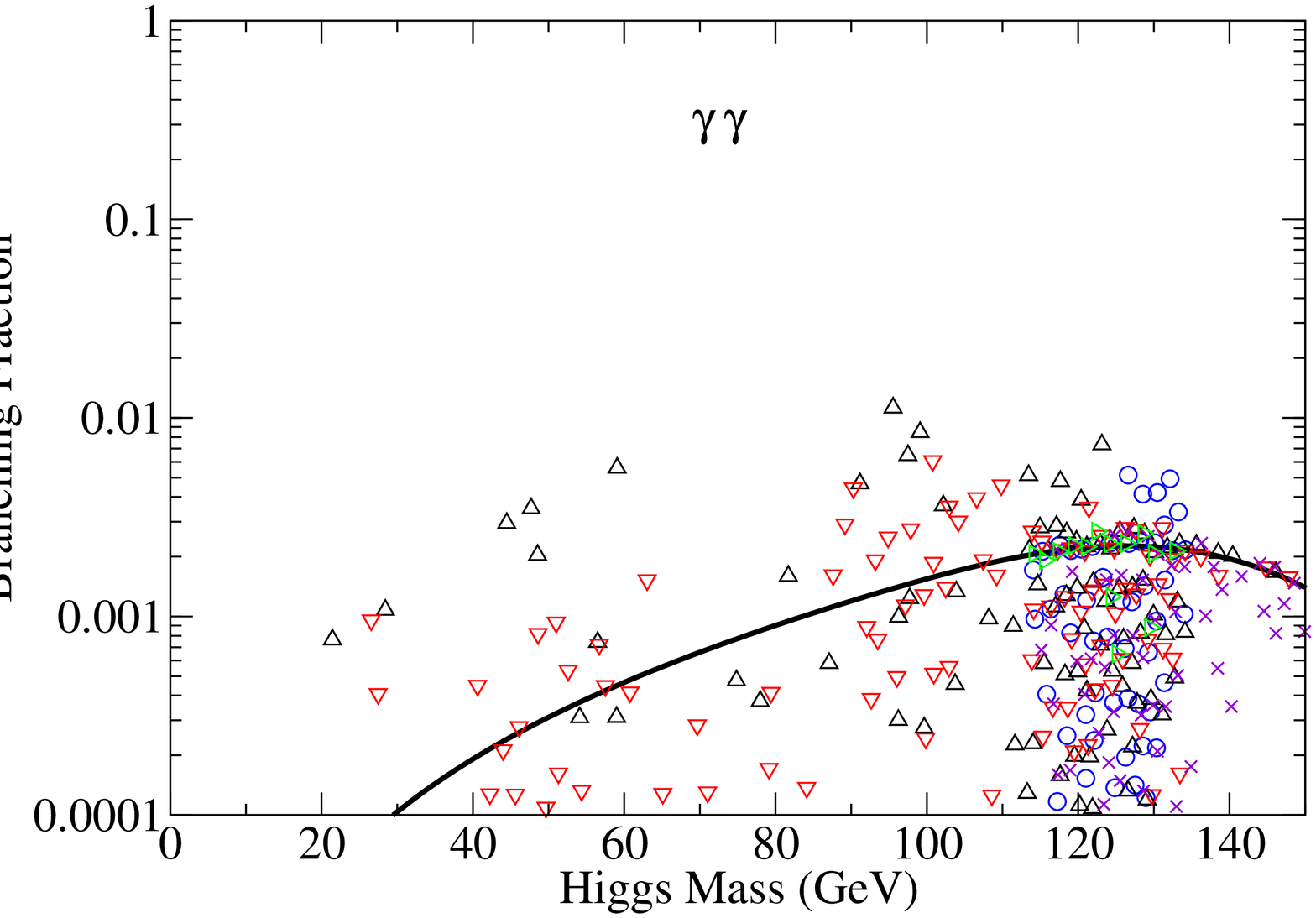}
\includegraphics[angle=-90,width=0.45\textwidth]{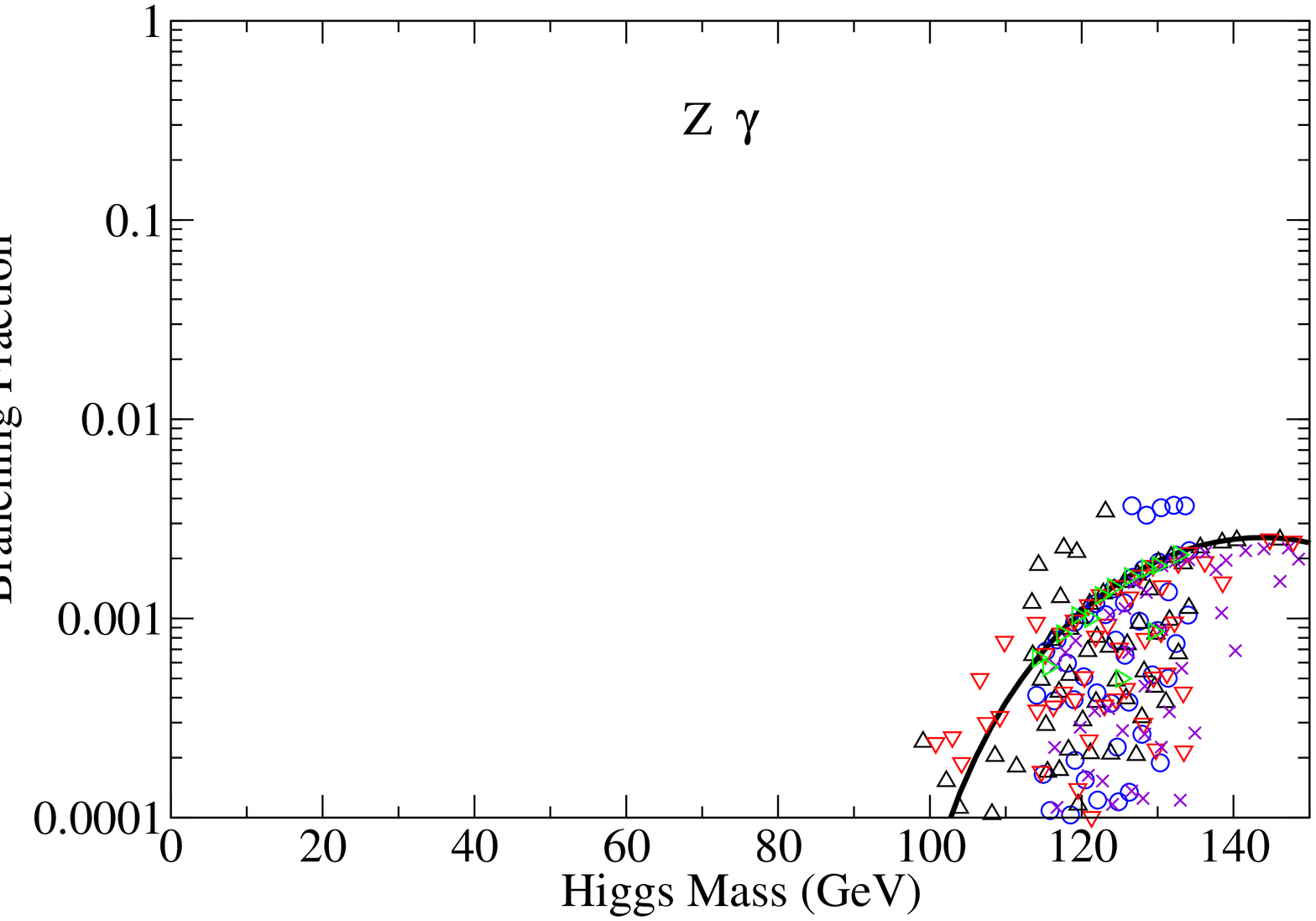}
\caption{Branching fractions for various modes in the MSSM and extended-MSSM models.  Curves denote SM branching fractions.}
\label{fig:decays}
\end{center}
\end{figure}

The branching fractions of representative decays to the SM particles $WW^*$, $ZZ^*$, $b\bar b$, $\tau^+ \tau^-$, $\gamma \gamma$, and $Z \gamma$ are presented in Fig. \ref{fig:decays} for the lightest CP-even Higgs boson in the MSSM, NMSSM and UMSSM, either $H_1$ or $H_2$ in the n/sMSSM, and $h$ in the SM.  

Note that the branching fractions may be larger in the SUSY models than in the SM.  For instance, in the NMSSM the branching fractions to $WW^*$ and $ZZ^*$ can be larger than the corresponding SM branching fractions, as seen in Fig \ref{fig:decays}.  These enhancements are due to the smaller total decay width of the Higgs boson rather than an enhancement of the particular partial width and may aid in the $H \to W^*W^* \to l \bar\nu jj$ and $l \bar\nu \bar l \nu$ discovery modes at the Tevatron \cite{Han:1998ma}.   Since the dominant decay mode is typically to $b\bar b$ in the mass range $2 m_b\lsim M_H\lsim 140 \text{ GeV}$, any decrease in the $b\bar b$ partial width reduces the total Higgs width.  The Higgs boson couplings to fermions are related to the SM values by
\be
g_{ddH_i}=g^{SM}_{ffh} {R_{+}^{i1}\over \cos\beta},\qquad g_{uuH_i}=g^{SM}_{ffh} {R_{+}^{i2}\over \sin\beta},
\ee
where $g^{SM}_{ffh} = -{i g_2 m_f\over 2 M_W}$.  Hence, either suppression or enhancement of the partial decay widths to fermions is possible, but not to an arbitrary degree, since in the MSSM the rotation matrices and $\tan \beta$ obey the tree-level relation
\be
\sin 2\alpha={M_H^2+M_h^2\over M_h^2-M_H^2} \sin 2 \beta,
\ee
where $\sin \alpha = R^{i1}_{+}$ and $\cos \alpha = R^{i2}_{+}$ in the MSSM.  A similar expression holds for the extended models, which restricts the rotation matrix values.  As noted earlier, the couplings converge to SM couplings just as they do for the MSSM in the $s$-decoupling limit \cite{ref:barger-phillips-stange}.

For a SM Higgs boson of mass below 150 GeV, $h\to \gamma\gamma$ is a significant mode for discovery at the LHC.  The branching fraction for this mode can be enhanced significantly due to the modified fermion loops in the n/sMSSM and NMSSM for the same reasons that the $H\to gg$ decay width is enhanced, providing more opportunity for discovery.  The Higgs couplings to $W$-bosons, charginos and $H^\pm$ also affect the $\gamma \gamma$ and $Z\gamma$ branching fractions, shown in Fig. \ref{fig:decays}.  These couplings are reduced from their MSSM values.  However, the reduced couplings may not necessarily lead to a rate suppression as interference effects can enhance the overall partial decay widths.

\subsubsection{Non-SM decays}

\begin{figure}[htbp]
\begin{center}
\includegraphics[angle=-90,width=0.49\textwidth]{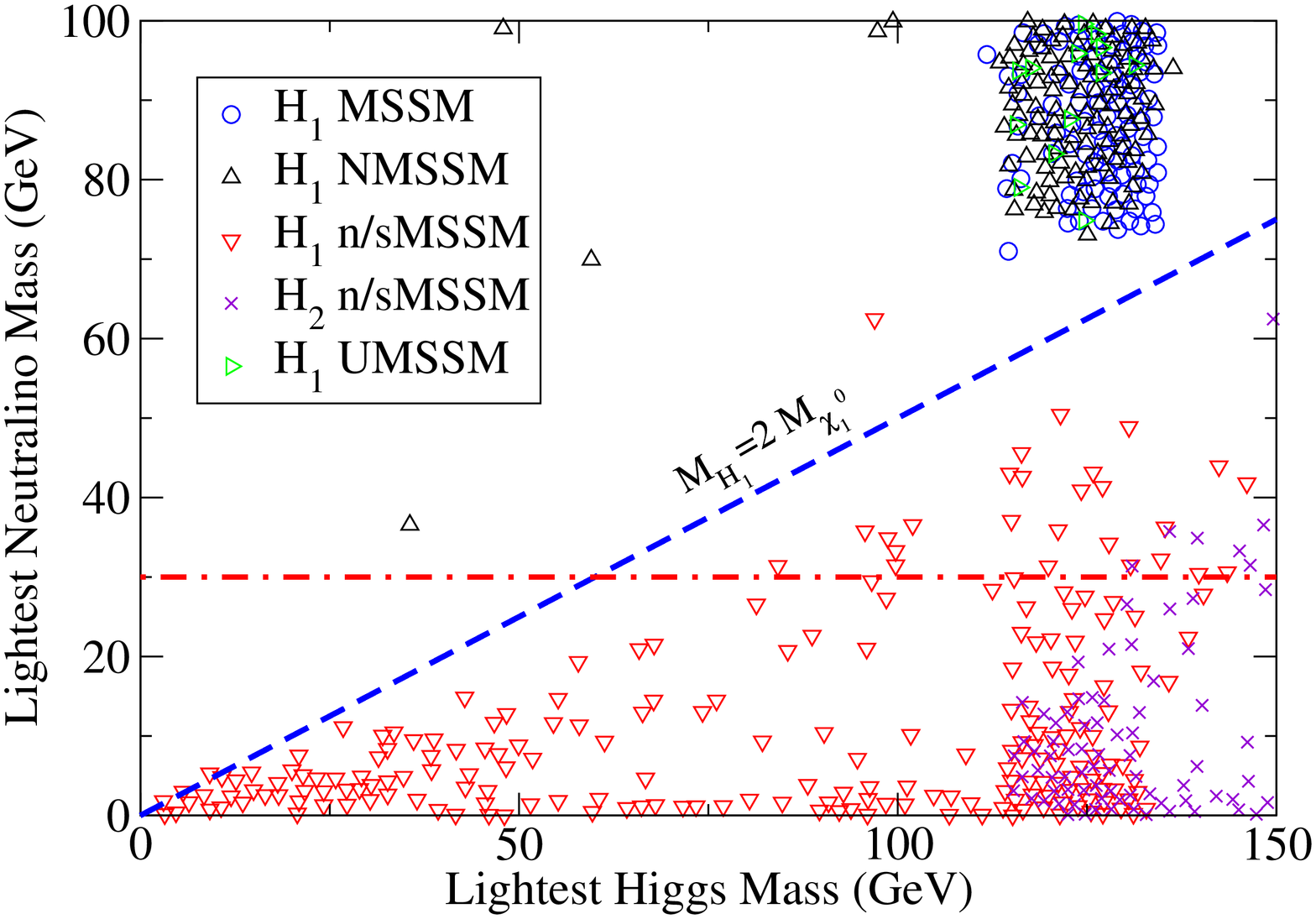}
\includegraphics[angle=-90,width=0.49\textwidth]{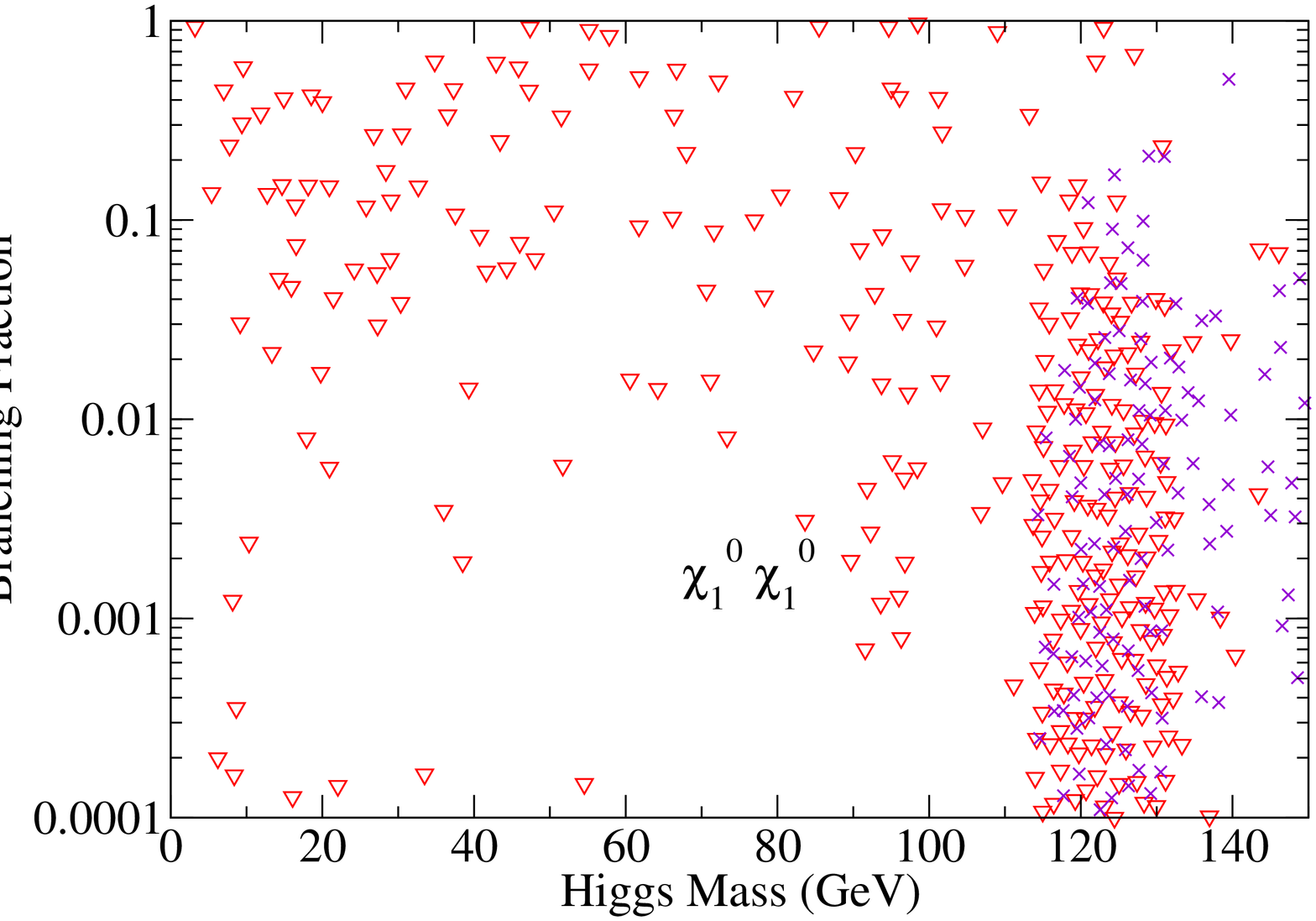}
(a)\hspace{0.48\textwidth}(b)
\caption{(a) $M_{H_i}$ vs. $M_{\lsp}$ in all the models considered.  Points falling below the blue line allow the decay of the lightest CP-even Higgs to two $\lsp$.  (b) Branching fraction of $H_i \to \lsp \lsp$}
\label{fig:lspdecay}
\end{center}
\end{figure}

Decays to non-SM particles can also be important in the extended models.  Since the lightest neutralino is a dark matter candidate, its production at colliders is of great interest to both the particle physics and cosmology communities \cite{dmatcolliders}.  We show in Fig. \ref{fig:lspdecay}a the kinematic region where neutralino production via the decay $H_1 \to \lsp \lsp$ is possible.  The couplings and masses of the lightest neutralino have been investigated for the models considered here \cite{xMSSM_neutralino}, and $M_{\lsp}$ may be quite small in the n/sMSSM \cite{xMSSM_neutralino,Menon:2004wv}.  However, in the n/sMSSM most of the kinematic region is disfavored due to a large $\lsp$ relic density \cite{xMSSM_neutralino}.  This is indicated in Fig. \ref{fig:lspdecay}a below the red horizontal line at $M_{\lsp} = 30$ GeV, which is the lower bound of $M_{\lsp}$ allowed by the dark matter relic density constraint when only the annihilation through the $Z$ pole is considered.  The $Z$ pole is the most relevant channel since the $\lsp$ in this model is very light ($M_{\lsp} \lsim 100$ GeV).  In principle, other annihilation channels such as a very light Higgs may allow the lighter $\lsp$ although the pole will be quite narrow \cite{NMSSM_neutralino}.  Furthermore, in the secluded (sMSSM) version of the model,
it is possible that the $\lsp$ considered here actually decays to a still lighter (almost) decoupled
neutralino, as discussed in Appendix \ref{apx:sumssmdecoup}.

The $\lsp \lsp$ partial decay width is given by
\be
\Gamma_{H_i \to \lsp \lsp} = {1 \over 16 \pi M_{H_i}}  \lambda^{1/2}\left(M^2_{\lsp}/M^2_{H_i},M^2_{\lsp}/M^2_{H_i}\right)\left(M_{H_i}^2-4 M_{\lsp}^2\right) |C_{H_i \lsp \lsp}|^2,
\ee
where the $H_i \lsp \lsp$ coupling is
\bea
C_{H_i \lsp \lsp} &=&\left[(g_2 N_{12}- g_1 N_{11}+ g_{1'} Q_{H_d} N_{16})N_{13}+\sqrt 2 h_s N_{14}N_{15}\right]R_{+}^{i1}\nn\\
&+&\left[(g_1 N_{11}-g_2 N_{12}+ g_{1'} Q_{H_u} N_{16})N_{14}+\sqrt 2 h_s N_{13}N_{15}\right]R_{+}^{i2}\nn\\
&+& \left[g_{1'}Q_S N_{16}N_{15}+\sqrt{2} h_s N_{13}N_{14}-\sqrt{2} \kappa N_{15}N_{15}\right] R_{+}^{i3}.
\eea
where the expression for the NMSSM in Ref. \cite{Choi:2004zx} has been generalized to include the UMSSM while the $H_i \lsp \lsp$ coupling in the n/sMSSM does not contain any model-dependence.  For a particular model, the irrelevant parameters are understood to be set to zero as in Eq. (\ref{eq:potential}).  The lightest Higgs boson in the n/sMSSM can have a high branching fraction to the lightest neutralino as seen in Figs. \ref{fig:lspdecay}b.  In fact, in this model the $\lsp \lsp$ branching fraction can be near 100\%.\footnote{If we do not assume gaugino mass unification and assume that $\mu_{\rm eff}$ is light and $M_{1'}$ is heavy, then $\lsp$ is light and large $\lsp \lsp$ branching fractions are possible in the UMSSM, similar to those found in the n/sMSSM.  For constraints on $M_{\lsp}$ in the MSSM from supernova data see Ref. \cite{Dreiner:2003wh}.}  This $Z$ decay is seen as missing energy and makes Higgs searches difficult at the Tevatron or LHC. It has been explored in the MSSM \cite{Belanger:2001am} and more generally \cite{Davoudiasl:2004aj}.

\begin{figure}[t]
\begin{center}
\includegraphics[angle=-90,width=0.49\textwidth]{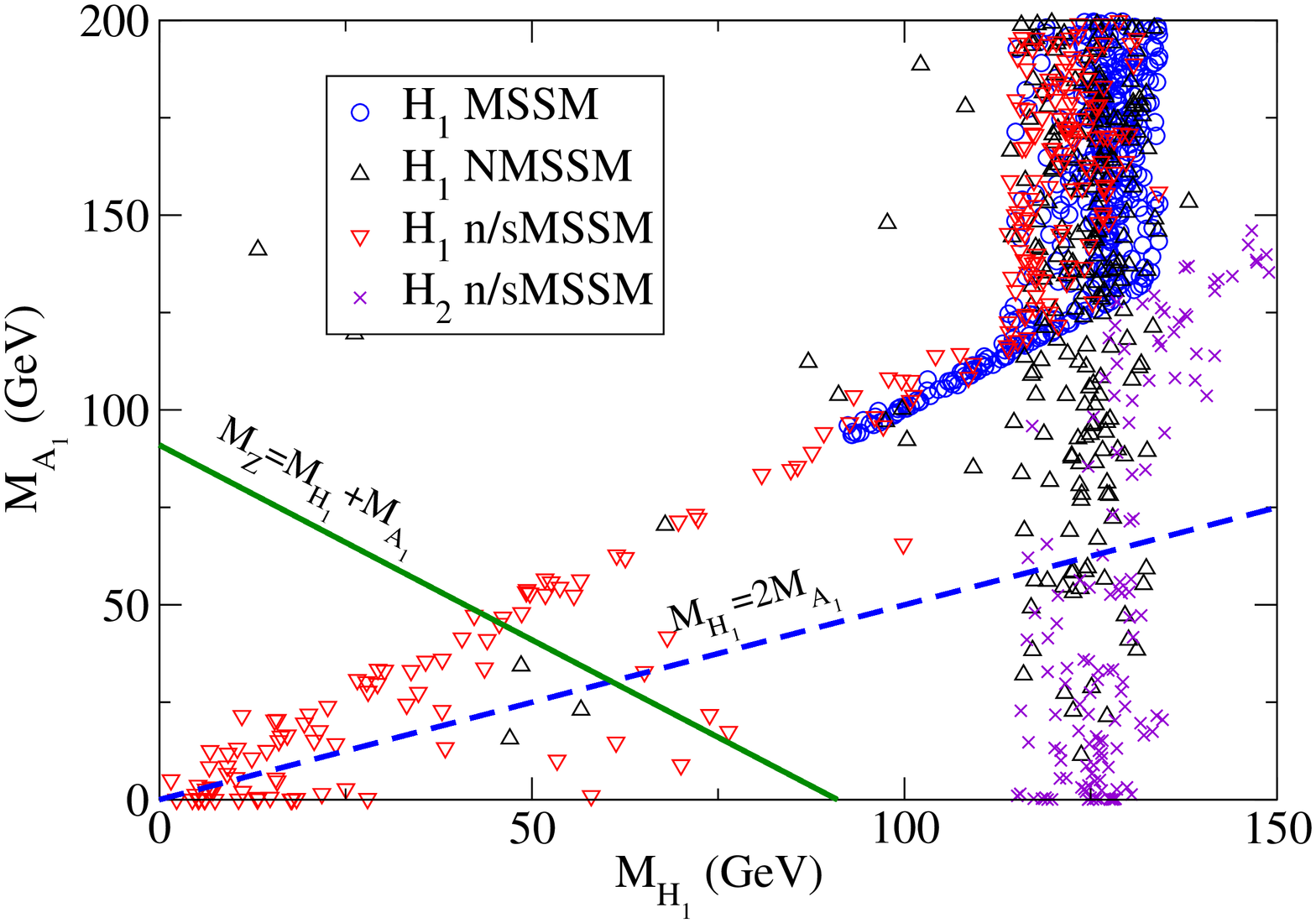}
\includegraphics[angle=-90,width=0.49\textwidth]{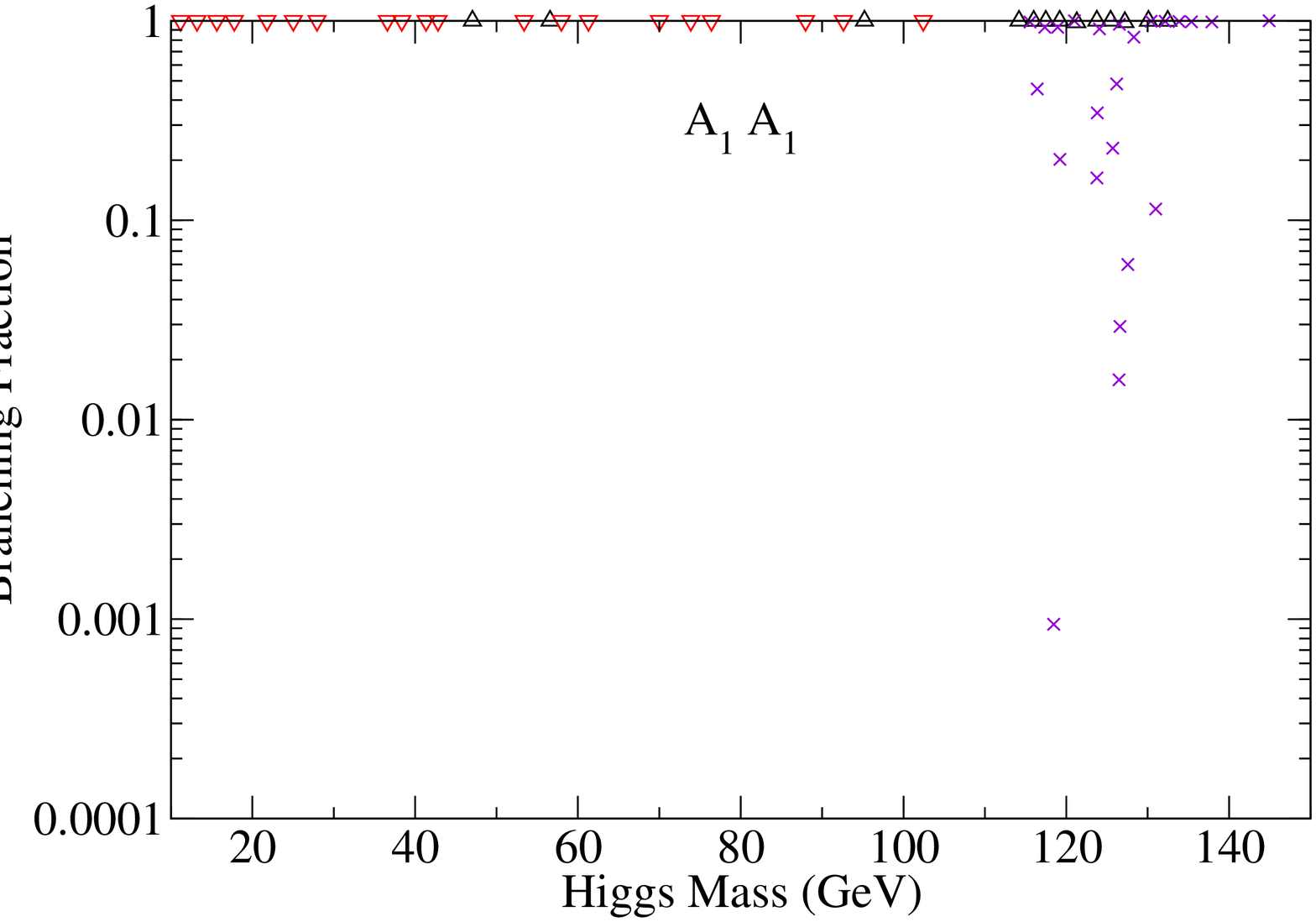}
(a)\hspace{0.48\textwidth}(b)
\caption{ (a)  $M_{H_j}$ vs. $M_{A_i}$ showing the kinematics for decays in extended-MSSM models, where $A_i$ is the lightest nonzero CP-odd state for each model.  $H_i \to A_i A_i$ decays are allowed for regions below the blue-dashed line.  Decays of $Z\to H_j A_i$ are allowed to the left of the green dark line.  (b) $H\to A_i A_i$ branching fraction vs. Higgs mass. The n/sMSSM parameter $\xi_S$ is scanned with a higher density at low $|\xi_S|$ to allow low Higgs masses.}
\label{fig:h1decay}
\end{center}
\end{figure}

In addition to decays to neutralino pairs, decays involving the lightest CP-odd Higgs bosons are relevant in the extended models.  In Fig. \ref{fig:h1decay}a we show the possibilities for decays involving both $A_i$ and $H_j$, where $A_i$ is the lightest nonzero CP-odd state for each model.  The kinematic regions where $Z\to A_i H_1$ and $H_j \to A_i A_i$ are given.  Even though the $Z$ decay is possible in the n/sMSSM and NMSSM, the coupling is suppressed due to the low MSSM fraction of both $A_i$ and $H_1$ seen in Fig. \ref{fig:mh-vs-xi}b.  Also shown is the crossing of states in the n/sMSSM where $H_2$ and $H_1$ switch content and hence their variation with $M_{A_i}$.  The lightest CP-even and CP-odd Higgs masses in both the MSSM and n/sMSSM show a strong correlation below the LEP limit.  
In the MSSM, this is evident from Fig. \ref{fig:lep}b where the reduced $ZZh$ coupling occurs when $\cos^2(\beta-\alpha)$ does not vanish, resulting in a lower CP-even Higgs mass.  The n/sMSSM correlation is more clearly shown in Fig. \ref{fig:mhxis} where the crossing of states at $\xi_S \sim -0.1$ is discussed.

The $H\to A_i A_i$ mode can be significant if allowed kinematically \cite{ref:htoaa} and has been studied in the NMSSM \cite{Dermisek:2005ar} and in the general singlet extended MSSM via an operator analysis \cite{Chang:2005ht}.  Since the lightest Higgs masses are small in the n/sMSSM at low $|\xi_S|$, we scan over this parameter with a higher density in this region to be near the PQ limit, which gives a lightest CP-odd Higgs boson of low mass.  In Fig. \ref{fig:h1decay}a, all the points below the line $M_{H_j} = 2 M_{A_i}$ allow this decay; the corresponding partial width is given by
\be
\Gamma(H_j\to A_i A_i) =  {1 \over 16 \pi M_{H_j}}\lambda^{1/2}\left(M^2_{A_i}/M^2_{H_j},M^2_{A_i}/M^2_{H_j}\right) |C_{H_j A_i A_i}|^2,
\ee
where the $H_j A_i A_i$ coupling,
\be
C_{H_j A_i A_i}=P_{H_j}P_{A_i}P_{A_i} V,
\ee
is determined by the projection operators that parallel the equivalent MSSM operators in Ref. \cite{Gunion:1989we}
\bea
P_{H_j} &=&{1\over \sqrt 2}\left(R_{+}^{j1}{\partial\over\partial{\phi_d}}+R_{+}^{j2}{\partial\over\partial{\phi_u}}+R_{+}^{j3} {\partial\over\partial{\sigma}}\right),\\
P_{A_i} &=&{i\over \sqrt 2}\left(R_{-}^{i1} {\partial\over\partial{\varphi_d}}+R_{-}^{i2}{\partial\over\partial{\varphi_u}}+R_{-}^{i3} {\partial\over\partial{\xi}}\right),
\eea
where we evaluate the potential at the minimum $\phi_{u,d} =\sigma =\varphi_{u,d} =\xi= 0$, where the field values are shifted to the minimum as in Eq. (\ref{eq:fieldexp}).

\begin{figure}[t]
\begin{center}
\includegraphics[angle=-90,,width=0.49\textwidth]{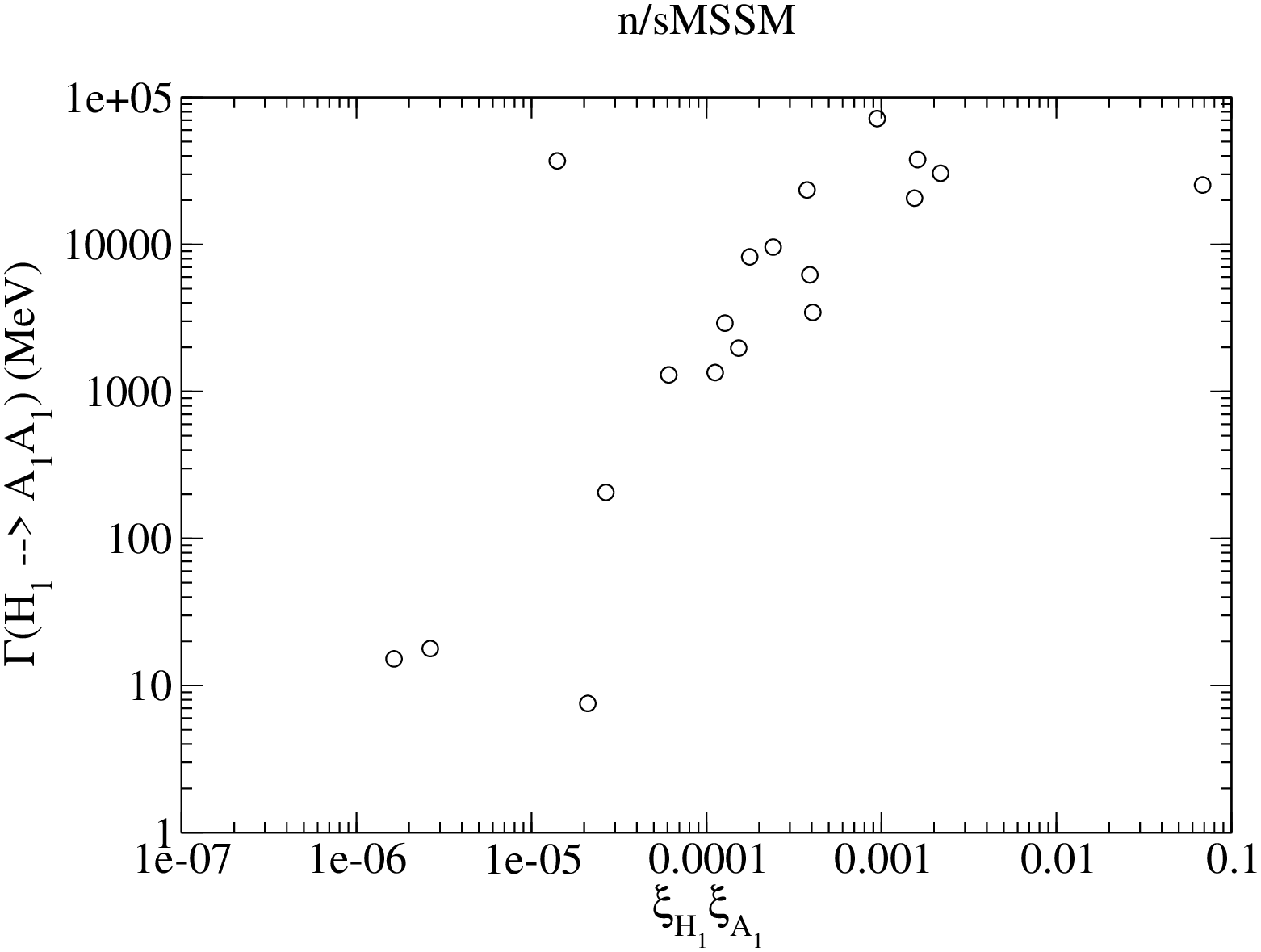}
\includegraphics[angle=-90,width=0.49\textwidth]{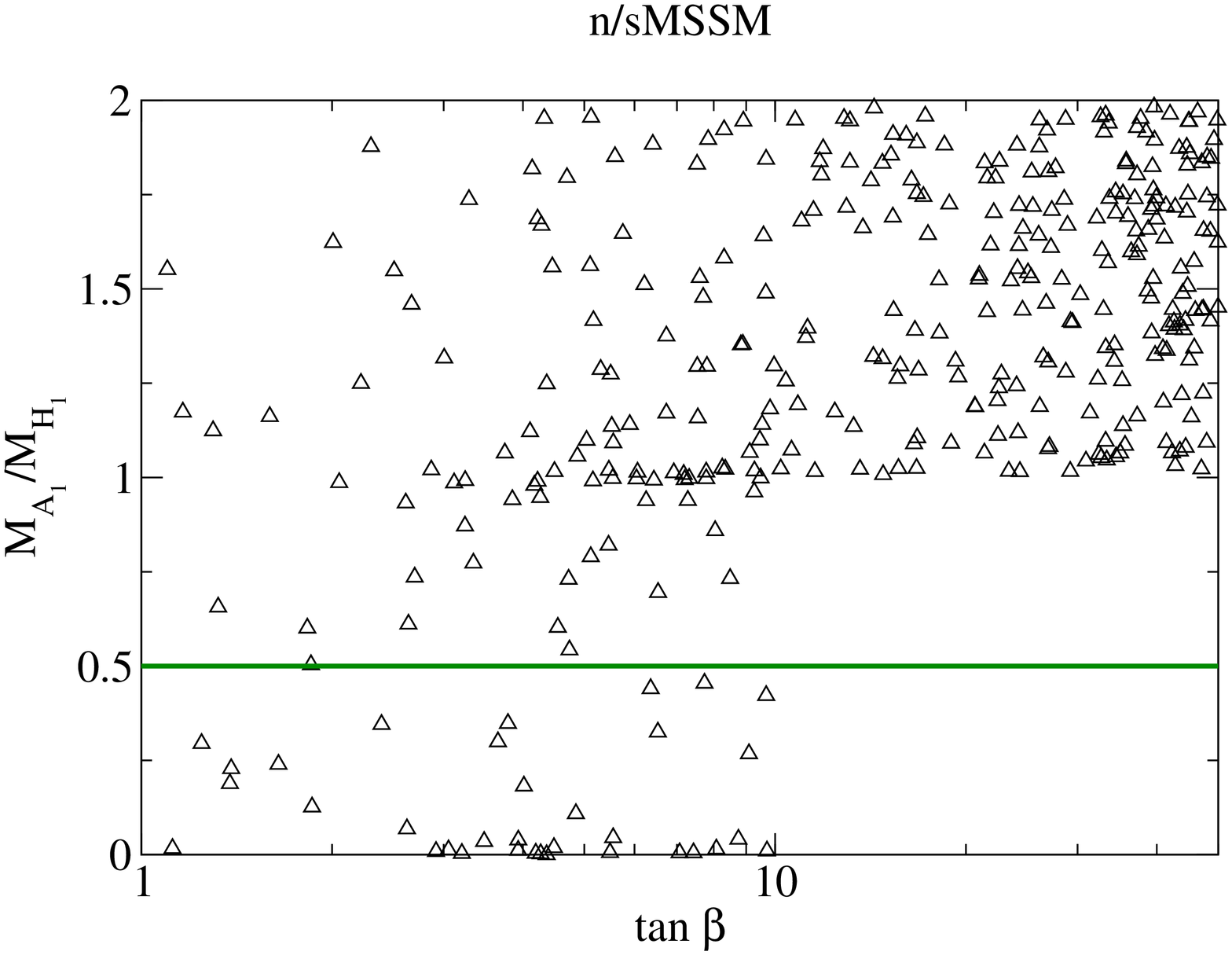}
(a)\hspace{0.48\textwidth}(b)
\caption{(a) $H_1 \to A_1 A_1$ decay width in the n/sMSSM vs. the product of the MSSM fractions of $H_1$ and $A_1$.  (b) $M_{A_1}/M_{H_1}$ vs. $\tan \beta$ in the n/sMSSM. The horizontal line marks the production threshold of $H_1 \to A_1 A_1$.}
\label{fig:AAdecay}
\end{center}
\end{figure}
 
\begin{figure}[t]
\begin{center}
\includegraphics[angle=-90,width=0.49\textwidth]{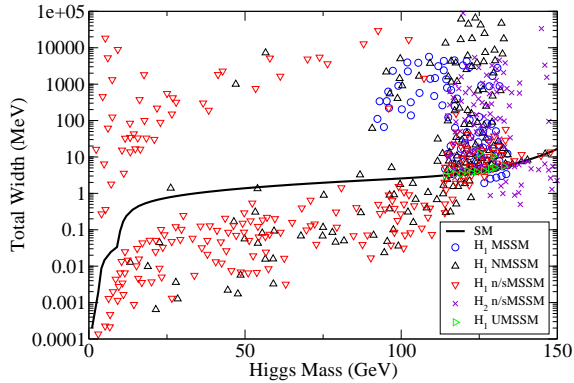}
\caption{Total decay width for each model.  Large enhancements with respect to the SM are largely due to the decays to $A_i A_i$ and $\lsp \lsp$.  }
\label{fig:totwidth}
\end{center}
\end{figure}

The $H_1\to A_2 A_2$ decay is not allowed in the UMSSM.  This is because the $\alpha_{ZZ'}$ constraint often requires a large value of $s$ resulting in a CP-odd mass above the typical CP-even mass, see, e.g., Eq. (\ref{eq:pqlimcpo}).  However, the $A_1 A_1$ decay is kinematically allowed in both the n/sMSSM and NMSSM.  When allowed, this decay mode can be dominant in the NMSSM as seen in Fig. \ref{fig:h1decay}b.  In the n/sMSSM, the $H_1\to A_1 A_1$ decay mode is suppressed since there is no tree-level coupling of three singlet Higgs states in this model and the $H_1$ and $A_1$ states are dominantly singlet.  However, because $H_1$ is not completely a singlet, these decays can still be non-negligible.  In Fig. \ref{fig:AAdecay}a, we show the $H_1\to A_1A_1$ decay width in the n/sMSSM versus the product of the MSSM fractions of $H_1$ and $A_1$.  As both MSSM fractions vanish, $H_1$ and $A_1$ become singlet dominated, giving a vanishing decay width in the n/sMSSM due to the absence of a singlet self-coupling.  Nevertheless, this partial decay width alone characteristically exceeds the total width of the SM Higgs boson as can be seen in Fig. \ref{fig:totwidth}.

 The MSSM-like second lightest Higgs boson in the n/sMSSM also may have a large branching fraction to light $A$ pairs.  Since $H_2$ has a large MSSM fraction, the coupling to the singlet $A_1$ pairs is not suppressed.  In addition, kinematic suppression is absent due to the larger mass of $H_2$.  In the lightest Higgs boson decay to $A_1 A_1$ in the n/sMSSM, low $\tan \beta$ is preferred as shown in Fig. \ref{fig:AAdecay}b, where the horizontal line marks the production threshold.  This decay also requires a low $A_1$ mass and results from the near-Peccei-Quinn limit when $\xi_S \to 0$.  The low $\tan \beta$ preference is a result of the larger $H_1$ mass in this region.  This enhancement is suggested in the one-parameter scans shown in Figs. \ref{fig:modelscans}a and \ref{fig:modelscans1} where the lightest Higgs mass is peaked at low $\tan \beta$ due to the crossing of Higgs states.  In contrast, while the NMSSM's lightest Higgs mass is maximal at low $\tan \beta$, a sharp drop as $\tan \beta$ is increased is not present, yielding little to no correlation of $\tan \beta$ with the existence of this decay mode. 

Finally, we show the total decay width of the light CP-even Higgs bosons to SM modes and $\lsp \lsp$ and $A_i A_i$ in the models considered in Fig. \ref{fig:totwidth}.  The total decay width can be enhanced due to the $\lsp \lsp$ and $A_i A_i$ partial widths.  The total width in the MSSM can be larger than the SM due to the enhanced couplings of the Higgs to $b \bar b$ when away from the MSSM decoupling limit.  In the n/sMSSM, Higgs masses above the LEP bound decaying to $\lsp$ pairs make a contribution to the total width that is no larger than a few MeV.  The $A_i A_i$ decays are responsible for the significantly larger total widths in the NMSSM and n/sMSSM.

\section{Conclusions}\label{sect:concl}

Extensions of the MSSM that include a singlet scalar field provide a natural solution to the undesirable fine-tuning of the $\mu$-parameter needed in the MSSM.  After symmetry breaking, the singlet Higgs obtains a VEV, generating an effective $\mu$-parameter naturally at the EW/TeV scale.  While the extensions to the MSSM that we consider each contain at least one additional singlet field, $S$, the symmetries that distinguish each model and their resulting superpotential terms provide phenomenologically distinct consequences.  We made grid and random scans over the parameter space of each model and imposed the LEP experimental bounds on the lightest CP-even $ZZH_i$ couplings.  The limits on $M_{A_2}$ and $M_{H_1}$ in the MSSM were converted to associated $A_i H_j$ production cross section limits and imposed.  We also imposed constraints from the LEP chargino mass limit and new contributions to the invisible $Z$ decay width.  Within the UMSSM, we enforced an additional constraint on the $Z'$ boson mixing with the SM $Z$.  

We found the following interesting properties of the considered models:

\begin{enumerate}
\item  The lightest Higgs boson can have a considerable singlet fraction in the n/sMSSM and NMSSM.  Since the singlet field does not couple to SM fields, the couplings of the lightest Higgs to MSSM particles are reduced due to the mixing of the singlet field with the doublet Higgs bosons, resulting in the $e^+ e^-$ production cross sections being significantly smaller.  Therefore, in the n/sMSSM and NMSSM, Higgs boson masses that are considerably smaller than the LEP bound on the SM Higgs boson mass are possible.   The upper bound on the lightest CP-even Higgs mass in extended-MSSM models is also relaxed due to the contribution of the singlet scalar through the mixing of the Higgs doublets and the singlet.  The upper limit in parameter scans is increased up to 164 GeV for the NMSSM and 170 GeV for the n/sMSSM.  The lightest CP-even Higgs mass in the UMSSM can be as large as 173 GeV due to additional gauge interactions; however, the lower bound on the lightest CP-even Higgs mass is similar to that of the MSSM.

\item  A common feature of each model is a CP-even Higgs boson with a mass in a range concentrated just above the SM mass limit.  At least two CP-even Higgs bosons must have nonzero MSSM fractions, making the lightest non-singlet dominated Higgs obey limits on the $ZZH_i$ coupling, forbidding masses below 114.4 GeV unless additional doublet mixing occurs.  The lightest Higgs in the NMSSM and n/sMSSM can evade this by singlet-doublet mixing, allowing a Higgs with mass just below the SM limit.

\item  In the $s$-decoupling limit with fixed $\mu_{\rm eff}$ at the EW/TeV scale, two Higgs states correspond to the MSSM Higgs states.  The $s$-decoupling limit in the n/sMSSM and UMSSM yields two CP-even Higgs bosons with similar masses and couplings to those of the MSSM with one extra decoupled Higgs.  The $s$-decoupling limit is often achieved in the UMSSM since the strict $\alpha_{ZZ'}$ mixing constraint must be obeyed, and requires $s$ to be at the TeV scale.   In this case, the mass of the decoupled Higgs scales with $s$.  However, the $s$-decoupling limit is not always required in the UMSSM as either a delicate cancellation of the mixing term in the $Z-Z'$ mass-squared matrix that requires $\tan \beta \sim \sqrt{Q_{H_d} \over Q_{H_u}}$, or a suppression in $Q_{H_u}$ at large $\tan \beta$ can evade the mixing constraint.  These fine-tuning scenarios do allow $s$ to be lower, but does not often result in a dramatically reduced lightest Higgs mass.  In the n/sMSSM, the lightest Higgs boson decouples as it has vanishing mass and is singlet dominated while $H_2$ and $H_3$ correspond to the MSSM Higgs bosons.  However, the NMSSM does not have this behavior.  Although the $s$-decoupling limit provides two MSSM-like Higgs bosons, one becomes massless at tree-level and the other scales as $\sqrt{ \sqrt 2 \kappa s \mu_{\rm eff} \csc{2\beta}}$, while the singlet Higgs boson mass scales with $\kappa$ and $s$.  This departure from the $s$-decoupling behavior of the other models is provided by the cubic self-coupling of the singlet field in the superpotential.

\item  Weak boson couplings  of the Higgs bosons are generally reduced from those of the SM, which translates to lower Higgs production rates.  However, the production rates can be enhanced kinematically since the Higgs mass can be lower than the SM mass limit.  Branching fractions may be larger than in the SM due to the suppression of the total width if the dominant $b\bar b$ decay mode is suppressed by mixing effects.  The $H\to gg$ partial width can be either enhanced or reduced due to both enhancements of couplings to fields running in the loops and their interference effects.  

\item   The branching fraction for $H\to \gamma \gamma$ can be enhanced significantly in the n/sMSSM and NMSSM, providing more opportunity for Higgs discovery.  Interference effects aid the enhancement of the overall decay width, as in $H\to gg$.

\item Non-SM decays can become important if allowed.  The lightest Higgs boson in the n/sMSSM can have a high branching fraction to light neutralino pairs if kinematically allowed.  This decay width can be as large as a few MeV and contribute significantly to the total width of the Higgs.
However, in the n/sMSSM, much of the allowed kinematic region with $M_{\lsp} \le 30$ GeV may be disfavored from the prediction of a high $\lsp$ relic density.  The $H_1 \to A_2 A_2$ decays are not favored in the UMSSM since $s$ must be large to avoid the $Z-Z'$ mixing constraint, which in turn pushes the allowed values of $M_{A_2}$ beyond the allowed region for the decay.  When allowed, this decay mode is dominant in the NMSSM and n/sMSSM.   
 
\item  The total decay width of the lightest Higgs boson can be enhanced by many orders of magnitude due to the large partial widths for the non-SM modes $\lsp \lsp$ and $A_i A_i$.  Decays to $\lsp$ pairs make a contribution to the total width that is no larger than a few MeV, while $A_i A_i$ decays lead to significant total width enhancements.

\end{enumerate}

\appendix

\section{The Secluded $U(1)'$-extended MSSM ($\text{sMSSM}$)}
\label{apx:sumssmdecoup}

In the UMSSM a single standard model singlet field $\hat S$ plays two roles: the expectation 
value of the scalar component generates an effective $\mu$-parameter and also generates
a mass for the $Z'$. There is some tension between maintaining a small enough
$\mu_{\rm eff} = h_s \langle S \rangle$ and generating a sufficiently large $M_{Z'}$ and small $Z-Z'$ mixing. As we have seen, this is most easily resolved by choosing a large value for  $s$ and
a small $h_s$. This typically leads to two CP-even MSSM-like Higgs scalars, one heavy 
CP-even state that is largely singlet, and one  MSSM-like CP-odd Higgs.
Similarly, if gaugino mass unification holds the UMSSM in this limit involves four
MSSM-like neutralinos, and two heavy neutralinos which are mixtures of $\tilde Z'$ gaugino
and $\tilde S$ singlino \cite{xMSSM_neutralino}.

The secluded $U(1)'$-extended model (sMSSM) \cite{SUMSSM} separates
$\mu_{\rm eff} $ from the $Z'$ mass by introducing four standard models 
singlets\footnote{This
is motivated by concrete string constructions, which often have multiple singlets of this kind. See,
e.g., \cite{Cleaver:1998gc}.}, $S, S_1, S_2, S_3$.
All of these are  charged under $U(1)'$ and contribute to $M_{Z'}$,
but only $S$ contributes to $\mu_{\rm eff} $. Moreover, in an appropriate (decoupling) 
limit there is an $F$
and $D$ flat direction along which $S_{1,2,3}$ acquire large vevs, so that
$M_{Z'}$ is naturally much larger than $\mu_{\rm eff} $ near that limit.
The most general form of the model involves a complicated Higgs and neutralino spectrum, which
was studied in Ref. \cite{SUMSSM_Higgs}. Here, we discuss the decoupling limit in which
$\langle S_1\rangle,\langle S_2\rangle,\langle S_3\rangle$ are naturally large (TeV scale),
and show that in that limit the $S_{1,2,3}$ Higgs and neutralino states (and the $\tilde Z'$ gaugino)
approximately decouple, and that the Higgs and neutralino spectrum of the $\hat S$, $\hat H_u$, and
$\hat H_d$ are identical to those of the nMSSM (this was commented on for the
neutralinos in Ref. \cite{xMSSM_neutralino}).

The Higgs part of the superpotential of the general sMSSM is
\be
  W=h_s \hat S \hat H_u \cdot \hat H_d + \lambda_s \hat S_1 \hat S_2 \hat S_3 +W', 
 \label{Wsmssm}
 \ee
where the $U(1)'$ charges satisfy $Q_{H_u}+Q_{H_d}+Q_S=0$ and $-Q_S=Q_{S_1}=Q_{S_2}
=-\frac{1}{2} Q_{S_3}$,
and $W'=\mu_1 \hat S \hat S_1 + \mu_2 \hat S \hat S_2$.
The original form of the sMSSM \cite{SUMSSM} assumed $\mu_{1,2}=0$
because small non-zero values reintroduce a form of the $\mu$-problem\footnote{Initially vanishing
$\mu_i$ are not generated by loop corrections until soft supersymmetry breaking
is turned on because of additional global $U(1)$ symmetries of the superpotential.}.
Setting $\mu_{1,2}=0$ here would lead to the limit of the nMSSM with $\xi_F=0$. 
We have checked that $\xi_F=0$ does not alter the qualitative features of the nMSSM. Nevertheless, we
prefer to allow $\mu_{1,2}\ne 0$ here to show the correspondence with the general nMSSM.
The Higgs potential contributions are
    \begin{eqnarray}
V_F &=& |h_s H_u\cdot  H_d + \mu_1 S_1+\mu_2 S_2|^2 + h_s^2 \left(  |S|^2 |H_u|^2 + |S|^2|H_d|^2\right)
\nonumber\\ &+&
\lambda_s^2 |S_1|^2 |S_2|^2 + |\lambda_s S_2S_3+ \mu_1S|^2 + |\lambda_s S_1S_3+ \mu_2S|^2,
\label{vfsmssm}\end{eqnarray} 
 %
   \begin{eqnarray}
V_D &=& {{G^2}\over 8} \left(|H_u|^2 - |H_d|^2\right)^2 
\nonumber\\&+&
{1\over 2} g_{1'}^2\left(Q_S |S|^2 + Q_{H_u} 
|H_u|^2 + Q_{H_d} |H_d|^2 + \sum_{i=1}^3 Q_{S_i}
|S_i|^2\right)^2, \label{vdsmssm}
\end{eqnarray}  
  \begin{eqnarray}
V_{\rm soft} &=& m_{u}^2 |H_u|^2 + m_{d}^2 |H_d|^2 + m_S^2 |S|^2 +
\sum_{i=1}^3 m_{S_i}^2 |S_i|^2 \nonumber \\
&+&
(A_s h_s S H_u\cdot  H_d - A_{\lambda} \lambda_s S_1 S_2 S_3 +h.c.)
 \nonumber \\
    &+& (m_{S S_1}^2 S S_1 + m_{S S_2}^2 S S_2 
+ m_{S_1 S_2}^2 S_1^{\dagger} S_2 + h.c.).\label{vsoftsmssm}
    \end{eqnarray} 
The soft mass-squares $m^2_{S S_1}$ and $m_{S S_2}^2$
are required to break two unwanted global $U(1)_{PQ}$ symmetries (which are also
broken by $\mu_{1,2}\ne 0$). The $m_{S_1 S_2}^2$ term can lead to CP violation and 
can be ignored (it is useful in considering electroweak baryogenesis \cite{ref:Kang}).
For $\lambda_s=\mu_i=0$ the potential has  $F$ and $D$ flat
directions involving large values for the secluded sector fields $S_{1,2,3}$. If $m^2_{S S_i}=0$
as well the $S_i$ tree-level potential will be unbounded below for appropriate values of the
$S_i$ soft parameters. Turning on small values for these parameters (e.g., $\lambda_s \sim 0.05$)
leads to a natural hierarchy between the ordinary ($H_{u,d}, S$) and secluded sector
fields, with typically $\langle S_i \rangle \sim |m_{S_i}|/\lambda_s$.

It is clear that (\ref{Wsmssm}), (\ref{vfsmssm}), and (\ref{vsoftsmssm})
resemble the nMSSM for the ($H_{u,d}, S$) sector fields provided one identifies
\bea
 \xi_F M_{\rm n}^2 &\equiv& \mu_1 \langle S_1 \rangle + \mu_2 \langle S_2 \rangle,\nonumber \\
 \xi_S M_{\rm n}^3& \equiv&m_{S S_1}^2  \langle S_1\rangle + m_{S S_2}^2  \langle S_2\rangle+ \lambda_s
 \langle S_3 \rangle \left( \mu_1 \langle S_2 \rangle+  \mu_2 \langle S_1
  \rangle\right). \label{muxi} \eea
However, to establish the relation to the nMSSM one must show that the mixing of these
states with the secluded sector fields are small and also that the $U(1)'$ $D$ terms
in (\ref{vdsmssm}) do not significantly affect the masses. 

The decoupling described above does indeed occur in the limit that
$\lambda_s \langle S_i \rangle $, $[m^2_{SS_i} \langle S_i \rangle ]^{1/3}$,
$[\mu_i  \langle S_i \rangle ]^{1/2}$, $h_s A_s$, $|m_{u,d}|$, $|m_S|$,
and the gaugino masses
are all small compared to the $U(1)'$-breaking scale
$M_{Z'}\sim g_{1'}\left[ \sum Q_{S_i}^2 | \langle S_i \rangle |^2\right]^{1/2}$.
This can be seen by explicitly examining the Higgs and neutralino mass matrices, which are given
for $\mu_i=0$ in Ref. \cite{SUMSSM,SUMSSM_Higgs} and easily generalized to include $\mu_i\ne 0$.
The derivation is straightforward but not very enlightening, so we will
just state the results.
\begin{itemize}
\item There are nine neutralinos, consisting of 5 with a mass matrix and composition
similar to the nMSSM, and 4 additional states that have only small mixings with
the nMSSM sector. The latter include two heavy states of masses $\sim \pm M_{Z'}$,
which are admixtures of the $U(1)'$ gaugino $\tilde Z'$ and one linear combination of
the singlinos $\tilde S_i$, and two light states (the orthogonal $\tilde S_i$ states).
The heavy states would be maximal mixtures of $\tilde Z'$ and the singlino state
in the exact decoupling limit, with significant deviations possible away from the
limit. The masses of the light singlinos and the splitting of the heavy pair is comparable
to the electroweak scale, with the typical scale of the former
at $\lambda_s  \langle S_i \rangle$. In some cases, the LSP is actually one of the light neutralinos
from the  (approximately) decoupled sector. The small mixing effects could then lead to
decays of the lightest neutralino from the nMSSM sector, with significant collider and cosmological \cite{Cembranos:2006fp} implications.
\item The CP-even Higgs sector consists of 3 nMSSM-like states, and three additional
states that are mainly in the secluded sector $S_{1,2,3}$. One of the latter has a large mass
of order $M_{Z'}$, generated by the $D$ terms (which yield only one large mass eigenvalue),
while the other two may be much lighter, typically at the electroweak scale
(controlled by $\left[ A_\lambda \lambda_s \langle S_i \rangle\right]^{1/2}$) .
\item The CP-odd Higgs states include two nMSSM-like states, and two that are mainly
from the secluded sector (the third secluded state is the eaten $U(1)'$ Goldstone boson).
The secluded states include one with mass controlled by 
$\left[ A_\lambda \lambda_s \langle S_i \rangle\right]^{1/2}$
and a second that vanishes for $m^2_{SS_i}=\mu_i=0$.

\end{itemize}

\section{Higgs Mass-Squared Matrices}\label{apx:Higgs}
We present the appropriate superpotentials and minimization conditions that are used to determine the tree-level Higgs mass-squared matrices in the MSSM \cite{MSSM_Higgs}, NMSSM \cite{NMSSM_Higgs}, nMSSM  \cite{nMSSM_Higgs}, and the UMSSM \cite{UMSSM_Higgs}.  Additionally, radiative corrections to the mass-squared matrices are obtained via an effective potential whose method is outlined below.
\subsection{Tree-level Higgs potential}\label{apx:tree}
From the tree-level Higgs potential
\be 
V^0 = V_F + V_D + V_{\rm soft},
\ee
the tree-level mass-squared matrices for the CP-even and CP-odd Higgses are obtained by
\be
\left( {\mathcal{M}}_{+}^{0}\right)_{ij}= \left. \frac{\partial ^{2}V^{0}}{\partial \phi _{i}\partial \phi_{j}} \right|_0 ,\qquad \left( {\mathcal{M}}_{-}^{0}\right) _{ij}= \left. \frac{\partial ^{2}V^{0}}{\partial \varphi_{i}\partial  \varphi _{j}} \right|_0,
\ee
where $(\phi_1,\phi_2,\phi_3)$ represent $(\phi_d,\phi_u,\sigma)$, and similarly
for $\varphi_{1,2,3}$.

\subsubsection{MSSM}\label{apx:mssm}
In the MSSM, the superpotential and soft breaking terms are
\be
W = \mu \hat H_u \cdot \hat H_d + h_t \hat U^c \hat Q \cdot \hat H_u,
\ee
\bea
-{\cal L}^{\rm MSSM}_{\rm soft}&=&\left(\sum_{a=1,2,3} \frac{1}{2} M_a\lambda_a\lambda_a + B \mu H_u\cdot H_d + A_t h_t \widetilde U^c \widetilde Q \cdot H_u + h.c.\right) \nn \\
&& +\ M_{\widetilde{Q}}^2|\widetilde{Q}|^2+ M_{\widetilde{U}}^2|\widetilde{U}|^2 + M_{\widetilde{D}}^2|\widetilde{D}|^2+ M_{\widetilde{L}}^2|\widetilde{L}|^2+ M_{\widetilde{E}}^2|\widetilde{E}|^2 \nn \\
&& +\ m_{d}^{2}|H_{d}|^2 + m_{u}^{2}|H_{u}|^2,
\eea
giving a tree-level Higgs potential contributions
\bea
V_F &=& |\mu|^2 \left(|H_d|^2+|H_u|^2 \right) \\
V_D &=& \frac{G^2}{8}\left( |H_d|^2-|H_u|^2 \right)^2 + \frac{g_{2}^2}{2} \left( |H_d|^2|H_u|^2-|H_u \cdot H_d|^2 \right) \\
V_{\rm soft}&=&m_{d}^{2}|H_d|^2 + m_{u}^{2}|H_u|^2 + ( B \mu H_u\cdot H_d + h.c. ).
\eea

We replace the soft mass terms in terms of the VEVs using the potential minimization conditions,
\bea
m_d^2 &=& -\frac{1}{2}\left[\frac{G^2}{4}\right] v_d^2 + \frac{1}{2}\left[\frac{G^2}{4}\right] v_u^2 - \mu^2 + B \mu \frac{v_u}{v_d}, \\
m_u^2 &=&  \frac{1}{2}\left[\frac{G^2}{4}\right] v_d^2 - \frac{1}{2}\left[\frac{G^2}{4}\right] v_u^2 - \mu^2 + B \mu \frac{v_d}{v_u}.
\eea
This results in the tree-level CP-even Higgs mass-squared matrix elements
\bea
({\mathcal{M}_{+}^0})_{11} &=& \left[\frac{G^2}{4}\right] v_d^2 + B \mu \frac{v_u}{v_d}, \\
({\mathcal{M}_{+}^0})_{12} &=& -\frac{G^2}{4} v_d v_u - B \mu, \\
({\mathcal{M}_{+}^0})_{22} &=& \left[\frac{G^2}{4}\right] v_u^2 + B \mu \frac{v_d}{v_u}
\eea
 and tree-level CP-odd Higgs mass-squared matrix elements: 
\be
({\mathcal{M}_{-}^0})_{ij} = B \mu \frac{v_d v_u}{v_i v_j},
\ee
where $v_{1,2}\equiv v_{d,u}$.
One of the CP-odd Higgs bosons is a massless goldstone boson that is absorbed by the $Z$ boson, leaving only one physical CP-odd Higgs with tree-level mass
\be
M_{A_2}^2=B\mu\left({v_d\over v_u}+{v_u\over v_d}\right).
\ee

\subsubsection{NMSSM}\label{apx:NMSSM}
Superpotential and soft breaking terms:
\be
W = h_s \hat S \hat H_u \cdot \hat H_d + \frac{1}{3} \kappa \hat S^3 + h_t \hat U^c \hat Q \cdot \hat H_u,
\ee
\be
-{\cal L}^{\rm NMSSM}_{\rm soft}=-{\cal L}_{\rm soft}+\left( \frac{1}{3} \kappa A_\kappa S^3 + h.c.\right),
\ee
where 
\bea
-{\cal L}_{\rm soft} &=& \left(\sum_{a=1,2,3} \frac{1}{2} M_a\lambda_a\lambda_a  + A_s h_{s} S H_{u}\cdot H_{d}+ A_t h_t \widetilde U^c \widetilde Q \cdot H_u + h.c.\right) \nn \\
&& +\ M_{\widetilde{Q}}^2|\widetilde{Q}|^2+ M_{\widetilde{U}}^2|\widetilde{U}|^2 + M_{\widetilde{D}}^2|\widetilde{D}|^2+ M_{\widetilde{L}}^2|\widetilde{L}|^2+ M_{\widetilde{E}}^2|\widetilde{E}|^2 \nn \\
&& +\ m_{d}^{2}|H_{d}|^2 + m_{u}^{2}|H_{u}|^2 + m_{s}^{2}|S|^2.
\eea

Potential minimization condition:
\bea
m_d^2 &=& -\frac{1}{2}\left[\frac{G^2}{4}\right] v_d^2 + \frac{1}{2}\left[\frac{G^2}{4} - h_s^2\right] v_u^2 - \frac{1}{2}h_s^2 s^2 + \left(\frac{h_s A_s}{\sqrt{2}} + \frac{h_s \kappa s}{2}\right) \frac{v_u s}{v_d}, \\
m_u^2 &=& \frac{1}{2}\left[\frac{G^2}{4} - h_s^2\right] v_d^2 - \frac{1}{2}\left[\frac{G^2}{4}\right] v_u^2 -\frac{1}{2}h_s^2 s^2 + \left(\frac{h_s A_s}{\sqrt{2}} + \frac{h_s \kappa s}{2}\right) \frac{v_d s}{v_u},\\
m_s^2 &=& -\frac{1}{2}h_s^2 v_d^2 - \frac{1}{2}h_s^2 v_u^2 -\kappa^2 s^2 + \left(\frac{h_s A_s}{\sqrt{2}} + h_s \kappa s\right) \frac{v_d v_u}{s} - \frac{\kappa A_\kappa}{\sqrt{2}} s.
\eea

\subsubsection{$\text{n/sMSSM}$}\label{apx:nMSSM}
Superpotential and soft breaking terms:
\be
W = h_s \hat S \hat H_u \cdot \hat H_d + \xi_F M_{\rm n}^2 \hat S + h_t \hat U^c \hat Q \cdot \hat H_u,
\ee
\be
-{\cal L}^{\rm n/sMSSM}_{\rm soft}=-{\cal L}_{\rm soft} + ( \xi_S M_{\rm n}^3 S + h.c.).
\ee
Here $\xi_F $ and $\xi_S$ are model-dependent, dimensionless quantities.

Potential minimization conditions:
\bea
m_d^2 &=& -\frac{1}{2} \left[\frac{G^2}{4}\right] v_d^2 + \frac{1}{2} \left[\frac{G^2}{4}-h_s^2\right] v_u^2 - \frac{1}{2} h_s^2 s^2 + \left( \frac{h_s A_s}{\sqrt{2}} + \frac{h_s \xi_F M_{\rm n}^2}{s} \right) \frac{v_u s}{v_d}, \\
m_u^2 &=&  \frac{1}{2} \left[\frac{G^2}{4}-h_s^2\right] v_d^2 - \frac{1}{2} \left[\frac{G^2}{4}\right] v_u^2 - \frac{1}{2} h_s^2 s^2 + \left( \frac{h_s A_s}{\sqrt{2}} + \frac{h_s \xi_F M_{\rm n}^2}{s} \right) \frac{v_d s}{v_u}, \\
m_s^2 &=& -\frac{1}{2} h_s^2 v_d^2 - \frac{1}{2} h_s^2 v_u^2 + \left( \frac{h_s A_s}{\sqrt{2}} - \frac{\sqrt{2} \xi_S M_{\rm n}^3}{v_d v_u} \right) \frac{v_d v_u}{s}.
\eea

\subsubsection{UMSSM}\label{apx:UMSSM}
Superpotential and soft breaking terms:
\be
W = h_s \hat S \hat H_u \cdot \hat H_d + h_t \hat U^c \hat Q \cdot \hat H_u,
\ee
\bea
-{\cal L}^{\rm UMSSM}_{\rm soft}&=&-{\cal L}_{\rm soft}+ \left( \frac{1}{2} M_{1'}\lambda_{1'}\lambda_{1'} +h.c.\right).
\eea

We do not include the possible terms related to exotic chiral fields to cancel anomalies. The exotic terms depend on details of the model and we assume the masses of the exotics are heavy enough to be decoupled from the EW scale phenomenology that we are interested in.

Potential minimization condition:
\begin{eqnarray}
m_d^{2}&=& - \frac{1}{2}\left[\frac{G^2}{4} + Q_{H_d}^{2} {g_{1'}}^{2}\right] v_d^{2} + \frac{1}{2} \left[\frac{G^2}{4} - h_s^{2} - Q_{H_d} Q_{H_u} {g_{1'}}^{2}\right] v_u^{2} - \frac{1}{2} \left[h_s^{2} + Q_{H_d} Q_{S} {g_{1'}}^{2}\right] s^{2} \nn \\
&& +\ \frac{h_s A_s}{\sqrt{2}} \frac{v_u s}{v_d}, \\
m_u^{2}&=& \frac{1}{2} \left[\frac{G^2}{4} - h_s^2 - Q_{H_d} Q_{H_u} {g_{1'}}^{2}\right] v_d^{2} - \frac{1}{2} \left[\frac{G^2}{4} + Q_{H_u}^{2} {g_{1'}}^{2}\right] v_u^{2} - \frac{1}{2} \left[h_s^{2} + Q_{H_u} Q_{S} {g_{1'}}^{2}\right] s^{2} \nn \\
&& +\ \frac{h_s A_s}{\sqrt{2}} \frac{v_d s}{v_u}, \\
m_s^{2}&=& - \frac{1}{2} \left[h_s^{2} + Q_{H_d} Q_{S} {g_{1'}}^{2}\right] v_d^{2} - \frac{1}{2} \left[h_s^{2} + Q_{H_u} Q_{S} {g_{1'}}^{2}\right] v_u^{2} - \frac{1}{2} Q_S^{2} {g_{1'}}^{2}s^{2} + \frac{h_s A_s}{\sqrt{2}} \frac{v_d v_u}{s}.
\end{eqnarray}

\subsection{One-loop radiative correction to the Higgs potential}\label{apx:loop}
The one-loop correction of the mass-squared matrices for the CP-even and CP-odd Higgses are obtained as
\be
\left( {\mathcal{M}}_{+}^{1}\right)_{ij}= \left. \frac{\partial ^{2}V^{1}}{\partial \phi _{i}\partial \phi_{j}} \right|_0 - \left. \delta_{ij} \frac{1}{v_i} \frac{\partial V^1}{\partial \phi_i} \right|_0,
\qquad \left( {\mathcal{M}}_{-}^{1}\right) _{ij}= \left. \frac{\partial ^{2}V^{1}}{\partial \phi_{i}\partial  \phi _{j}} \right|_0 - \left. \delta_{ij} \frac{1}{v_i} \frac{\partial V^1}{\partial \phi_i} \right|_0.
\ee
The second terms describes the effect of the shift of the Higgs soft mass-squared terms $(m_d^2, m_u^2, m_s^2)$ at the minimum so that we can still use the tree-level values of the minimization condition. The scalar top Yukawa couplings are expected to be dominant over the other model-dependent couplings.  This provides a handle on the radiative corrections that is model-independent.

The  Coleman-Weinberg correction to the scalar potential with the stop 1-loop is given by \cite{Coleman:1973jx}
\begin{eqnarray}
V^{1} & = &  \frac{3}{32 \pi^2} \left[ \sum_{j=1}^{2} m_{\widetilde{t}_{j}}^{4}\left( \ln \frac{m_{\widetilde{t}_{j}}^{2}}{Q^{2}}-\frac{3}{2}\right)
-2\bar{m}_{t}^{4}\left( \ln \frac{\bar{m}_{t}^{2}}{Q^{2}}-\frac{3}{2}\right) \right].
\end{eqnarray}
where \( Q \) is the renormalization scale in the \( \overline{DR} \)
scheme, and \( \bar{m}_{t}^{2} = h_{t}^{2}\left| H_{u}^{0}\right| ^{2}\).

The physical stop masses are given by
\be
m_{\widetilde{t}_{1,2}}^{2} = \frac{1}{2}{\rm tr}{\mathcal{M}}_{\widetilde t}\mp
\frac{1}{2}\sqrt{({\rm tr}{\mathcal{M}}_{\widetilde t})^{2}-4 {\rm det}{\mathcal{M}}_{\widetilde t}}
\ee
The stop mass-squared matrix is
\bea
\mathcal{M}_{\widetilde t} = \left(\begin{array}{cc} M_{LL}^{2} & M_{LR}^{2}\\
{M^{2}_{LR}}^{\dagger} & M_{RR}^{2}\end{array}\right),
\eea
where
\be
M_{LL}^{2}=M_{\widetilde{Q}}^{2}+h_t^{2} |H_u|^2, \qquad M_{RR}^{2}=M_{\widetilde{U}}^{2}+h_t^{2} |H_u|^2,
\ee
and
\bea
M_{LR}^{2}&=&h_t \left(A^*_t H_u^{0*}-\mu H_d^{0}\right)   \quad \mbox{~~~[MSSM]}, \\
          &=&h_t \left(A^*_t H_u^{0*}-h_s S H_d^{0}\right) \quad \mbox{[NMSSM/n/sMSSM/UMSSM]}.
\eea
and is generally valid for complex $A_t,\mu$ and Higgs field values.  However, we assume these parameters are real.  We keep only dominant top Yukawa couplings and neglect gauge couplings and Yukawa couplings of other particles.

\section{Neutralino Mass matrices}\label{apx:neut}

The neutralino sector is extended by the singlino, $\widetilde S$, and in the case of the UMSSM, the $Z'$-ino, $\widetilde Z'$.  The mass matrix for the UMSSM is given by 

\bea
{\cal M}_{\chi^0} = \left( \begin{array} {c c c c c c}
	M_1 	&	0&	{-g_1 v_d/ 2}&	{g_1 v_u / 2}&	0&  0\\
	0 	&M_2&	{g_2 v_d / 2}&	{-g_2 v_u / 2}&	0&  0\\
	{-g_1 v_d / 2} 	&	{g_2 v_d / 2}&	0&	-\mu_{\rm eff}&	-\mu_{\rm eff}v_u/s&  {g_{1'}} Q_{H_d} v_d\\
	{g_1 v_u / 2} 	&	{-g_2 v_u / 2}&	-\mu_{\rm eff}& 0&	-\mu_{\rm eff}v_d/s&  {g_{1'}} Q_{H_u} v_u\\
	0&0&-\mu_{\rm eff} v_u/s&-\mu_{\rm eff} v_d/s&0&{g_{1'}} Q_{S} s\\
	0&0&{g_{1'}} Q_{H_d} v_d&{g_{1'}} Q_{H_u} v_u&{g_{1'}} Q_{S} s& M_{1'}\\
	\end{array} \right).
\eea

For the NMSSM, we have
\bea
{\cal M}_{\chi^0} = \left( \begin{array} {c c c c c}
	M_1 	&	0&	{-g_1 v_d/ 2}&	{g_1 v_u / 2}&	0\\
	0 	&M_2&	{g_2 v_d / 2}&	{-g_2 v_u / 2}&	0\\
	{-g_1 v_d / 2} 	&	{g_2 v_d / 2}&	0&	-\mu_{\rm eff}&	-\mu_{\rm eff}v_u/s\\
	{g_1 v_u / 2} 	&	{-g_2 v_u / 2}&	-\mu_{\rm eff}& 0&	-\mu_{\rm eff}v_d/s\\
	0&0&-\mu_{\rm eff} v_u/s&-\mu_{\rm eff} v_d/s&\sqrt 2 \kappa s\\
	
	\end{array} \right).
\eea
For the n/sMSSM, the neutralino mass matrix is the same as the NMSSM with the limit $\kappa \to 0$ taken.  Note that in the $s$-decoupling limit, neutralino states in addition to those in the MSSM decouple with masses $M_{\chi^0_{5,6}}=\half\left(M_{1'}\pm\sqrt{M_{1'}^2+4 g_{1'}^2 Q_S^2 s^2}\right)$ in the UMSSM, and $M_{\chi^0_{5}}= \sqrt 2 \kappa s$ and $0$ in the NMSSM and n/sMSSM, respectively.

\section{Approximations of the tree-level Higgs mass in various limits}
\label{apx:masses}

We provide approximations to the CP-even Higgs masses and exact CP-odd Higgs masses for the extended models considered.  We replace $h_s$ with $\mu_{\rm eff} / ( \frac{s}{\sqrt{2}})$ and assume $\mu_{\rm eff}$ is at the EW scale in the following derivations.  The approximate mass eigenvalues of the Higgs mass-squared matrix can be found when the mass-squared matrix elements satisfy particular hierarchies.

\subsection{Hierarchy types}
The entries of the matrix $(A,B,C)$ are assumed to have the same order of magnitudes where the powers of $\epsilon \ll 1$ represent the specific hierarchy.  After keeping terms of up to ${\cal O}(\epsilon^2)$, approximate eigenvalues may be analytically given.

Type-1 :
If the dominant terms in the matrix are of the form
\be
M^2 = \left( \ba{cc}
A & \epsilon C^\dagger \\
\epsilon C & \epsilon^2 B
\ea \right),
\ee
where $A$ is a scalar, $B$ is a $2 \times 2$ matrix, and $C$ is a $2 \times 1$ column vector, then the mass-squared matrix can be transformed to an approximate block diagonal form using
\be
V^\dagger = \left( \ba{cc}
1 - \frac{1}{2} \epsilon^2 \Gamma^\dagger \Gamma & - \epsilon \Gamma^\dagger \\
\epsilon \Gamma & 1 - \frac{1}{2} \epsilon^2 \Gamma \Gamma^\dagger
\ea \right),
\ee
where $\Gamma = C/A$ then, 
\bea
V M^2 V^\dagger &=& \left( \ba{cc}
A + \epsilon^2 C^\dagger C / A & {\cal O}(\epsilon^3) \\
{\cal O}(\epsilon^3) & \epsilon^2 (B - C C^\dagger / A)
\ea \right) \nn \\
&=& \left( \ba{ccc}
M_{11}^2 + \frac{(M_{12}^2)^2 + (M_{13}^2)^2}{M_{11}^2} & {\cal O}(\epsilon^3) & {\cal O}(\epsilon^3) \\
{\cal O}(\epsilon^3) & M_{22}^2 - \frac{(M_{12}^2)^2}{M_{11}^2} & M_{23}^2 - \frac{M_{12}^2 M_{13}^2}{M_{11}^2} \\
{\cal O}(\epsilon^3) & M_{23}^2 - \frac{M_{12}^2 M_{13}^2}{M_{11}^2} & M_{33}^2 - \frac{(M_{13}^2)^2}{M_{11}^2}
\ea \right).
\eea
This type of hierarchy is equivalent to that found in the appendix of \cite{Miller:2003ay}.

Type-2 :
We find this method by an analogy of the method for Type-1.
\be
M^2 = \left( \ba{cc}
A & \epsilon C \\
\epsilon C^\dagger & \epsilon^2 B
\ea \right),
\ee
where now $A$ is a $2 \times 2$ matrix, $B$ is a scalar, and $C$ is a $2 \times 1$ column vector, we obtain the unitary transform on $M^2$:
\bea
V M^2 V^\dagger &=& \left( \ba{cc}
A + \frac{1}{2} \epsilon^2 \left( C C^\dagger (A^{-1})^\dagger + 2 A^{-1} C C^\dagger - A^{-1} C C^\dagger (A^{-1})^\dagger A \right) & {\cal O}(\epsilon^3) \\
{\cal O}(\epsilon^3) & \epsilon^2 (B - C^\dagger A^{-1} C)
\ea \right) \nn \\
&=& \left( \ba{cc}
{\cal A} & {\cal O}(\epsilon^3) \\
{\cal O}(\epsilon^3) & {\cal B}
\ea \right),
\eea
with 
\be
V^\dagger = \left( \ba{cc}
1 - \frac{1}{2} \epsilon^2 \Gamma \Gamma^\dagger & - \epsilon \Gamma \\
\epsilon \Gamma^\dagger & 1 - \frac{1}{2} \epsilon^2 \Gamma^\dagger \Gamma
\ea \right),
\ee
and $\Gamma = A^{-1} C$.
The corresponding sub-matrices are given by
\bea
{\cal A}&=&\left( \ba{cc}
M_{11}^2 + \frac{M_{13}^2 (-M_{13}^2 M_{22}^2 + M_{12}^2 M_{23}^2)}{(M_{12}^2)^2 - M_{11}^2 M_{22}^2} & M_{12}^2 - \frac{1}{2} \frac{M_{13}^2 (M_{11}^2 + M_{22}^2) M_{23}^2 - M_{12}^2 ((M_{13}^2)^2 + (M_{23}^2)^2)}{(M_{12}^2)^2 - M_{11}^2 M_{22}^2} \\
M_{12}^2 - \frac{1}{2} \frac{M_{13}^2 (M_{11}^2 + M_{22}^2) M_{23}^2 - M_{12}^2 ((M_{13}^2)^2 + (M_{23}^2)^2)}{(M_{12}^2)^2 - M_{11}^2 M_{22}^2} & M_{22}^2 + \frac{M_{23}^2 (M_{12}^2 M_{13}^2 - M_{11}^2 M_{23}^2)}{(M_{12}^2)^2 - M_{11}^2 M_{22}^2} 
\ea \right), \nn \\
{\cal B}&=&\left( \ba{c}
M_{33}^2 + \frac{(M_{13}^2)^2 M_{22}^2 - 2 M_{12}^2 M_{13}^2 M_{23}^2 + M_{11}^2 (M_{23}^2)^2}{(M_{12}^2)^2 - M_{11}^2 M_{22}^2}
\ea \right) \label{eqn:type2B}.
\eea
The entries of ${\cal A}$ are assumed to be of ${\cal O}(M_{\rm c}^2) + {\cal O}(\epsilon^2 M_{\rm c}^2)$, and ${\cal B}$ is assumed to be of ${\cal O}(\epsilon^2 M_{\rm c}^2)$ where $M_{\rm c}$ is a common mass scale for dominant mass terms.  If this hierarchy is not established, for example by some cancellation which disrupts the hierarchy, e.g. $|(M_{12}^2)^2 - M_{11}^2 M_{22}^2| \ll M_{\rm c}^4$, this approximation may not work.

\subsection{Tree-level mass expressions}

We approximate the CP-even masses in three cases (sometimes with model-specific assumptions):
\begin{enumerate}
\item The $\tan \beta = 1$ limit provides an exact solution.  This case is meaningful since LEP cannot exclude this value for the extended MSSM models where the Higgs doublets can be mixed with a singlet.

\item  Both the large $M_Y^2$ and large $\tan\beta$ limit, where
\be
M_Y^2 \equiv 2 \mu_{\rm eff} A_s \csc 2\beta
\label{eq:my}
\ee
is as the MSSM CP-odd Higgs mass-squared with $\mu_{\rm eff}A_s$ identified as $\mu B$.

\item  Large $s$ limit ($s$-decoupling limit).  This limit provides two MSSM-like Higgs bosons and one singlet dominated Higgs. 
\end{enumerate}
We present only dominant terms in the following CP-even expressions

\subsubsection{MSSM (CP-even)}
The exact eigenvalues can easily be found analytically 
\be
M^2_{H^0_{1,2}} = \frac{1}{2} M_Z^2 + B \mu \csc 2\beta \pm \sqrt{\left( \frac{1}{2} M_Z^2 - B \mu \csc 2\beta \right)^2 + 2 M_Z^2 B \mu \sin 2\beta}.
\ee

\subsubsection{MSSM (CP-odd)}
\be
M^2_{A} = 2 B \mu \csc{2 \beta}
\ee

\subsubsection{Common PQ limit case (CP-even)}
(i) In the $\tan\beta = 1$ limit, the general solutions are given by
\bea
M^2_{H^0} &=& M_Z^2 + 2 \mu_{\rm eff} A_s - \frac{v^2 \mu_{\rm eff}^2}{s^2} \nn \\
M^2_{H^0} &=& \frac{1}{4 s^2} \left( v^2 \mu_{\rm eff} (A_s + 2 \mu_{\rm eff}) \pm v \mu_{\rm eff} (A_s - 2 \mu_{\rm eff}) \sqrt{v^2 + 16 s^2} \right).
\label{eqn:PQ_tanb1}
\eea

(ii) In the large $M_Y^2$ and large $\tan\beta$ limit ($M_Y/M_c \sim 1/\epsilon$ and $\tan\beta \sim 1/\epsilon$), the hierarchy is of Type-1 and results in
\bea
M^2_{H^0}  
&\approx& M_Y^2, \nn \\
M^2_{H^0} 
&\approx& \frac{1}{2} M_Z^2 \pm \frac{1}{2 s} \sqrt{(s M_Z^2)^2 + 16 v^2 (\mu_{\rm eff}^2 - M_Y^2 \cos^2 \beta)^2}. \label{eqn:negative}
\eea
In this limit, the lightest CP-even Higgs mass-squared always has a negative solution which implies this limit is not physical in the PQ limit unless loop corrections can compensate the negative mass-squared.

(iii) In the large $s$ limit ($s /M_c \sim 1 / \epsilon$), the hierarchy is of Type-2.
\bea
M^2_{H^0} &\approx& \frac{1}{s^2} \frac{\mu_{\rm eff} \sec^2 2\beta}{2 G^2 A_s} \left(32 \mu_{\rm eff} ^2 \sin 2 \beta A_s^2-G^2 v^2 \sin^3 2 \beta \left(4 \mu_{\rm eff} ^2+A_s^2\right) \right. \nn \\
&& \left. +2 \mu_{\rm eff}  A_s \left(G^2 v^2-16 \mu_{\rm eff} ^2-2 A_s^2+\cos 4 \beta \left(-G^2 v^2+2 A_s^2\right)\right)\right), \nn \\
M^2_{H^0} &\approx& \frac{1}{2} M_Z^2 + A_s \mu_{\rm eff} \csc 2 \beta \pm \sqrt{ \left( \frac{1}{2} M_Z^2 - A_s \mu_{\rm eff} \csc 2\beta \right)^2 + 2 M_Z^2 A_s \mu_{\rm eff} \sin 2\beta}. \label{eqn:PQ_larges}
\eea
In this limit, the lightest Higgs mass scales as $M_{H_1} \propto 1/s$, which implies that the singlet-dominated Higgs decouples, with mass near zero in the $s$-decoupling limit.  Note that when $\tan\beta \simeq 1$, the $(M_{12}^2)^2 - M_{11}^2 M_{22}^2$ term in the denominator of Eq. (\ref{eqn:type2B}) is of $ {\cal O}(\epsilon^2 M_{\rm c}^4)$ and the above approximation is not valid.  However, one can take the large $s$ limit from the $\tan\beta = 1$ exact solution of Eq. (\ref{eqn:PQ_tanb1}) and arrive at the desired result.  As expected, the other two Higgs bosons have MSSM-like masses.

\subsubsection{Common PQ limit case (CP-odd)}
The PQ symmetry protects one of the CP-odd Higgs masses
\bea
M_{A_1}^2 &=& 0,\nn\\
M_{A_2}^2 &=& 2 \mu_{\rm eff} A_s \csc 2 \beta \left(1 +\frac{v^2}{4 s^2} \sin^2 2\beta \right).
\label{eq:pqlimcpo}
\eea

\subsubsection{NMSSM case (CP-even)}
(i) In the $\tan\beta = 1$ limit:
\bea
M^2_{H^0} &=& M_Z^2 + 2 \mu_{\rm eff} A_s - \frac{v^2 \mu_{\rm eff}^2}{s^2} + \sqrt{2} \kappa \mu_{\rm eff} s, \nn \\
M^2_{H^0} &=& \frac{1}{4 s^2}\left(v^2 \mu_{\rm eff}  \left(A_s +
2 \mu_{\rm eff} \right) + \sqrt{2} s^3 \kappa  A_{\kappa}+ 4 s^4 \kappa ^2 \right. \nn \\
&& \pm \left[ v^2 \mu_{\rm eff}^2 \left(A_s -2 \mu_{\rm eff} \right)^2 \left(v^2 + 16 s^2 \right) +16 s^8 \kappa ^4+8 \sqrt{2} s^7 \kappa ^3 A_{\kappa }+\right. \nn \\
&& \left.\left.2 \sqrt{2} s^3 v^2 \kappa  \mu_{\rm eff}  \left(-2 \mu_{\rm eff} +A_s\right) \left(16 \mu_{\rm eff} +A_{\kappa }\right)+2 s^4 \kappa ^2 \left(4 v^2 \mu_{\rm eff} \left(2 \mu_{\rm eff} +A_s\right)+s^2 A_{\kappa }^2\right) \right]^{1/2} \right). \nn \\
\eea

(ii) In the large $M_Y^2$ and large $\tan\beta$ limit, the hierarchy is of Type-1.
\bea
M^2_{H^0}
&\approx& M_Y^2, \nn \\
M^2_{H^0} 
&\approx& \frac{1}{2} M_Z^2 + \frac{s \kappa}{4} \left(4 s \kappa +\sqrt{2} A_{\kappa }\right) \pm \frac{1}{2 s} \left[ (s M_Z^2)^2+ 16 v^2 \left(\mu_{\rm eff} ^2-M_Y^2 \cos^2 \beta \right)^2 \right. \nn \\
&& \left. + s^3 \kappa  \left(4 s^3 \kappa ^3-4 M_Z^2 s \kappa +\sqrt{2} \left(2 s^2 \kappa ^2- M_Z^2 \right) A_{\kappa }+\frac{1}{2} s \kappa  A_{\kappa }^2\right) \right]^{1/2}.
\eea

(iii) In the large $s$ limit: 
because of the $\kappa$ and $A_\kappa$ terms, the mass-squared matrix is neither Type-1 nor Type-2.
With an additional assumption of $|\kappa|\sim \epsilon$, the leading order terms form a block diagonal matrix.
\bea
M^2_{H^0} &\approx& \frac{1}{2} s \kappa  \left(4 s \kappa +\sqrt{2} A_{\kappa }\right), \nn \\
M^2_{H^0} &\approx& \frac{1}{2} M_Z^2 + \left(A_s + \frac{s \kappa}{\sqrt{2}} \right) \mu_{\rm eff} \csc 2 \beta. \nn \\
&& \pm \sqrt{ \left( \frac{1}{2} M_Z^2 - \left(A_s + \frac{s \kappa}{\sqrt{2}} \right) \mu_{\rm eff} \csc 2\beta \right)^2 + 2 M_Z^2 \left(A_s + \frac{s \kappa}{\sqrt{2}} \right) \mu_{\rm eff} \sin 2\beta}.
\eea
If $ |\kappa|\ll M_{\rm c}/s$ and  $|A_\kappa|\ll s$, we obtain the large $s$ limit of the PQ case of Eq. (\ref{eqn:PQ_larges}), changing the leading order term of the first solution.  If not, the $\kappa$ and $A_\kappa$ terms prevent the Higgs sector from decoupling to two Higgs bosons with MSSM masses and very heavy or light singlet Higgs boson.  Instead, the $s$-decoupling limit yields a massless Higgs, and a divergent Higgs ($M_{H_2} \sim \sqrt{\sqrt 2 \kappa s \mu_{\rm eff}\csc{2 \beta}}$) both with $\xi_{\rm MSSM}=1$ while the singlet dominated Higgs has a mass that scales linearly with $\kappa s$.

\subsubsection{NMSSM case (CP-odd)}
The CP-odd Higgs masses are exactly given at tree-level as
\newpage
\bea
\hspace{-1.3in}M^2_{A_1} &=& \frac{1}{8 s^2 \sin 2\beta} \left( \mu_{\rm eff} A_s (v^2 + 8 s^2) + 2\sqrt{2} s \kappa \mu_{\rm eff} (v^2 + 2 s^2) - v^2 \mu_{\rm eff} \cos 4\beta (A_s + 2\sqrt{2} s \kappa) - 6\sqrt{2} A_\kappa s^3 \kappa \sin 2\beta \right. \nn \\
&& \pm \left[ \left( \mu_{\rm eff} A_s (v^2 + 8 s^2) + 2\sqrt{2} s \kappa \mu_{\rm eff} (v^2 + 2 s^2) - v^2 \mu_{\rm eff} \cos 4\beta (A_s + 2\sqrt{2} s \kappa) - 6\sqrt{2} A_\kappa s^3 \kappa \sin 2\beta \right)^2 \right. \nn \\
&& \left. + 96 s^3 \kappa \mu_{\rm eff} \sin 2\beta \left( 2 A_\kappa s^2 (\sqrt{2} A_s + s \kappa) - 3\sqrt{2} A_s v^2 \mu_{\rm eff} \sin 2\beta \right) \right]^{1/2}.
\eea

\subsubsection{$\text{n/sMSSM}$ case (CP-even)}
\label{apx:lnmhexpr}
(i) In the $\tan\beta = 1$ limit:
\bea
M^2_{H^0} &=& M_Z^2 + 2 \mu_{\rm eff}  A_s - \frac{v^2 \mu_{\rm eff}^2}{s^2} + \frac{2 \sqrt{2} \xi_F M_{\rm n}^2 \mu_{\rm eff}}{s}, \nn \\
M^2_{H^0} &=& \frac{1}{4 s^2}\left( v^2 \mu_{\rm eff}  \left( A_s + 2 \mu_{\rm eff} \right)-2 \sqrt{2} s \xi_S M_{\rm n}^3 \right. \nn \\
&& \left. \pm \left[ v^2 \mu_{\rm eff}^2 \left( A_s - 2 \mu_{\rm eff} \right)^2 \left( v^2 + 16 s^2 \right) -4 \sqrt{2} s v^2 \mu_{\rm eff}  \left( A_s - 2 \mu_{\rm eff} \right) \xi_S M_{\rm n}^3+8 s^2 \xi_S^2 M_{\rm n}^6 \right]^{1/2} \right). \nn \\
\eea

(ii) In the large $M_Y^2$ and large $\tan\beta$ limit, the hierarchy is of Type-1.
\bea
M^2_{H^0} 
&\approx& M_Y^2, \nn \\
M^2_{H^0} 
&\approx& \frac{1}{2} M_Z^2 - \frac{\xi_S M_{\rm n}^3}{\sqrt{2} s}\nn\\
 & \pm& \frac{1}{2 s} \left[ (s M_Z^2 - \sqrt{2} \xi_S M_{\rm n}^3)^2 + 16 v^2 (\mu_{\rm eff}^2 - M_Y^2 \cos^2 \beta)^2 + 4 \sqrt{2} M_Z^2 s \xi_S M_{\rm n}^3 \right]^{1/2}.
\eea

(iii) In the large $s$ limit and small $\xi_S$ limit ($\xi_S M_{\rm n}^3/M_c^3 \sim \epsilon$), the hierarchy is of Type-2.
\bea
M^2_{H^0} &\approx& \frac{1}{s^2} \frac{\mu_{\rm eff} \sec^2 2\beta}{2 G^2 A_s} \left(32 \mu_{\rm eff} ^2 \sin 2 \beta A_s^2-G^2 v^2 \sin^3 2 \beta \left(4 \mu_{\rm eff} ^2+A_s^2\right) \right. \nn \\
&& \left. +2 \mu_{\rm eff}  A_s \left(G^2 v^2-16 \mu_{\rm eff} ^2-2 A_s^2+\cos 4 \beta \left(-G^2 v^2+2 A_s^2\right)\right)\right) - \frac{\sqrt{2} \xi_S M_{\rm n}^3}{s}, \nn \\
M^2_{H^0} &\approx& \frac{1}{2} M_Z^2 + A_s \mu_{\rm eff} \csc 2 \beta \pm \sqrt{ \left( \frac{1}{2} M_Z^2 - A_s \mu_{\rm eff} \csc 2\beta \right)^2 + 2 M_Z^2 A_s \mu_{\rm eff} \sin 2\beta}.
\eea
The additional assumption of a small $\xi_S M_{\rm n}^3$ term provides an additional negative term in $M^2_H$.  This mass is similar to that of the PQ limit in Eq. (\ref{eqn:PQ_larges}) when $\xi_S M_{\rm n}^3/M_{\rm c}^3 \ll M_{\rm c}/s$.  However, as $s$ is increased, the $-\frac{\sqrt{2} \xi_S M_{\rm n}^3}{s}$ term is dominant and the mass diverges from the PQ case.  This limits $\xi_S \lesssim 0$ to avoid negative Higgs mass-squared solutions that are unphysical.  However, effects from radiative corrections may alleviate this condition and allow positive, but small values of $\xi_S$.  
Along with the NMSSM, when $\tan\beta \simeq 1$, the above approximation is not valid, but  the large $s$ limit from the $\tan\beta = 1$ exact solution of Eq. (\ref{eqn:PQ_tanb1}) may be taken to arrive at the desired result.

\subsubsection{$\text{n/sMSSM}$ case (CP-odd)}
The CP-odd Higgs masses is exactly given at tree-level as
\bea
M^2_{A_{1,2}} &=& \frac{1}{8 s^2 \sin 2\beta} \left( \mu_{\rm eff} A_s (v^2 + 8 s^2) - A_s v^2 \mu_{\rm eff} \cos 4\beta + 8\sqrt{2} s \mu_{\rm eff} \xi_F M_{\rm n}^2 - 4\sqrt{2} s \sin 2\beta \xi_S M_{\rm n}^3 \right. \nn \\
&& \pm \left[  \left( \mu_{\rm eff} A_s (v^2 + 8 s^2) - A_s v^2 \mu_{\rm eff} \cos 4\beta + 8\sqrt{2} s \mu_{\rm eff} \xi_F M_{\rm n}^2 - 4\sqrt{2} s \sin 2\beta \xi_S M_{\rm n}^3 \right)^2 \right. \nn \\
&& \left. \left. + 64 s \mu_{\rm eff} \sin 2\beta \left( 2 s (\sqrt{2} A_s s + 2 \xi_F M_{\rm n}^2) \xi_S M_{\rm n}^3 - \sqrt{2} A_s v^2 \mu_{\rm eff} \sin 2\beta \xi_F M_{\rm n}^2 \right) \right]^{1/2} \right). \nn \\
\eea

\subsubsection{UMSSM case (CP-even)}

(i) In the $\tan\beta = 1$ and $Q_{H_d} = Q_{H_u}$ limit:
\bea
M^2_{H^0} &=& M_Z^2 + 2 \mu_{\rm eff} A_s - \frac{v^2 \mu_{\rm eff}^2}{s^2},  \nn \\
M^2_{H^0} &=& \frac{1}{4 s^2} \left(  v^2 \mu_{\rm eff}  \left(A_s + 2 \mu_{\rm eff} \right)+ \frac{1}{2} s^2 \left(4 s^2+v^2\right) g_{Z'}^2 Q_S^2 \pm \left[ v^2 \mu_{\rm eff}^2 \left(A_s - 2 \mu_{\rm eff} \right)^2 \left(v^2 + 16 s^2 \right) \right. \right. \nn \\
&& \left. \left. + s^2 v^2 \left(20 s^2-v^2\right) \mu_{\rm eff}  \left( A_s - 2\mu_{\rm eff} \right) g_{Z'}^2 Q_S^2+ \frac{1}{4} s^4 \left(4 s^2+v^2\right)^2 g_{Z'}^4 Q_S^4 \right]^{1/2} \right).
\eea

(ii) In the large $M_Y^2$ and large $\tan\beta$ limit, the hierarchy is of Type-1.
\bea
M^2_{H^0} 
&\approx& M_Y^2, \nn \\
M^2_{H^0} 
&\approx& \frac{1}{2} M_Z^2 + \frac{1}{2} M_{Z'}^2 \pm \frac{1}{2 s} \left[ s^2 M_Z^4 + s^2 M_{Z'}^4 +16 v^2 \left(\mu_{\rm eff} ^2-M_Y^2 \cos^2 \beta \right)^2+s^2 g_{Z'}^4 \left(v^2 Q_{H_u}^2+s^2 Q_S^2\right)^2  \right. \nn \\
&& \left. + \frac{1}{2} s^2 v^2 g_{Z'}^2 \left(G^2 v^2 Q_{H_u}^2-G^2 s^2 Q_S^2+32 \left(\mu_{\rm eff} ^2-M_Y^2 \cos^2 \beta \right) Q_{H_u} Q_S\right)\right]^{1/2}.
\eea

(iii) In the large $s$ limit:
if we change the basis of the matrix from $\{H^0_d, H^0_u, S\}$ to $\{S, H^0_d, H^0_u\}$, the hierarchy is of Type-1.
\bea
M^2_{H^0} &\approx& M_{Z'}^2, \nn \\
M^2_{H^0} &\approx& \frac{1}{2} M_Z^2 + A_s \mu_{\rm eff} \csc 2 \beta \pm \sqrt{ \left( \frac{1}{2} M_Z^2 - A_s \mu_{\rm eff} \csc 2\beta \right)^2 + 2 M_Z^2 A_s \mu_{\rm eff} \sin 2\beta}.
\eea
In this limit, one of the Higgs masses scales as $M_H \approx M_{Z'}$ which scales as $s$ in the $s$-decoupling limit.  This yields a heavy singlet dominated Higgs which decouples while the other two CP-even states become MSSM-like.

\subsubsection{UMSSM (CP-odd)}
The CP-odd Higgs mass in the UMSSM is equivalent to that in the common PQ limit (e.g. Eq. (\ref{eq:pqlimcpo}))

\section{Additional one parameter plot results}
\label{apx:addparm}

\begin{figure}[ht]
\begin{center}
\vspace*{-.6in}
\includegraphics[angle=-90,width=0.320\textwidth]{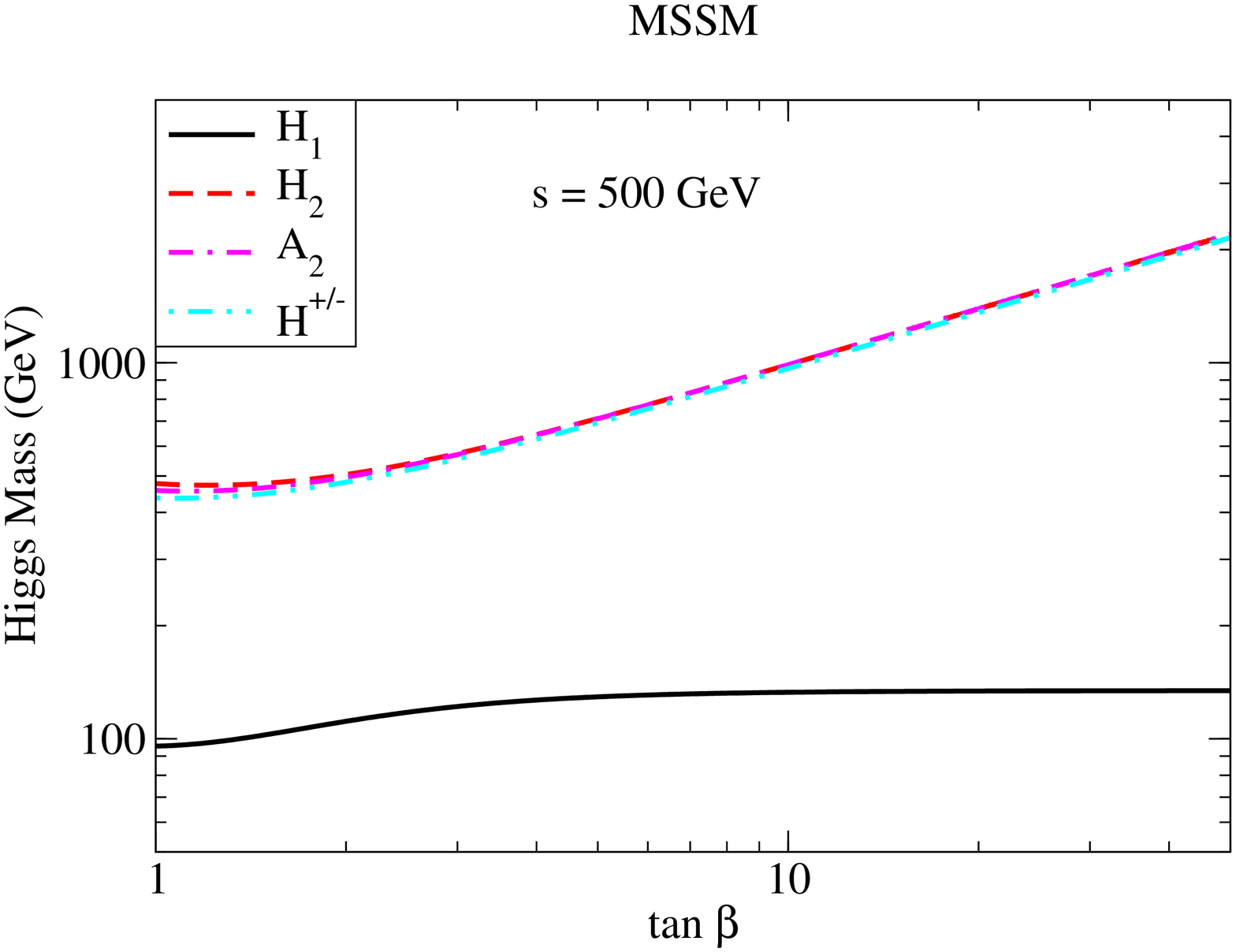}
\includegraphics[angle=-90,width=0.320\textwidth]{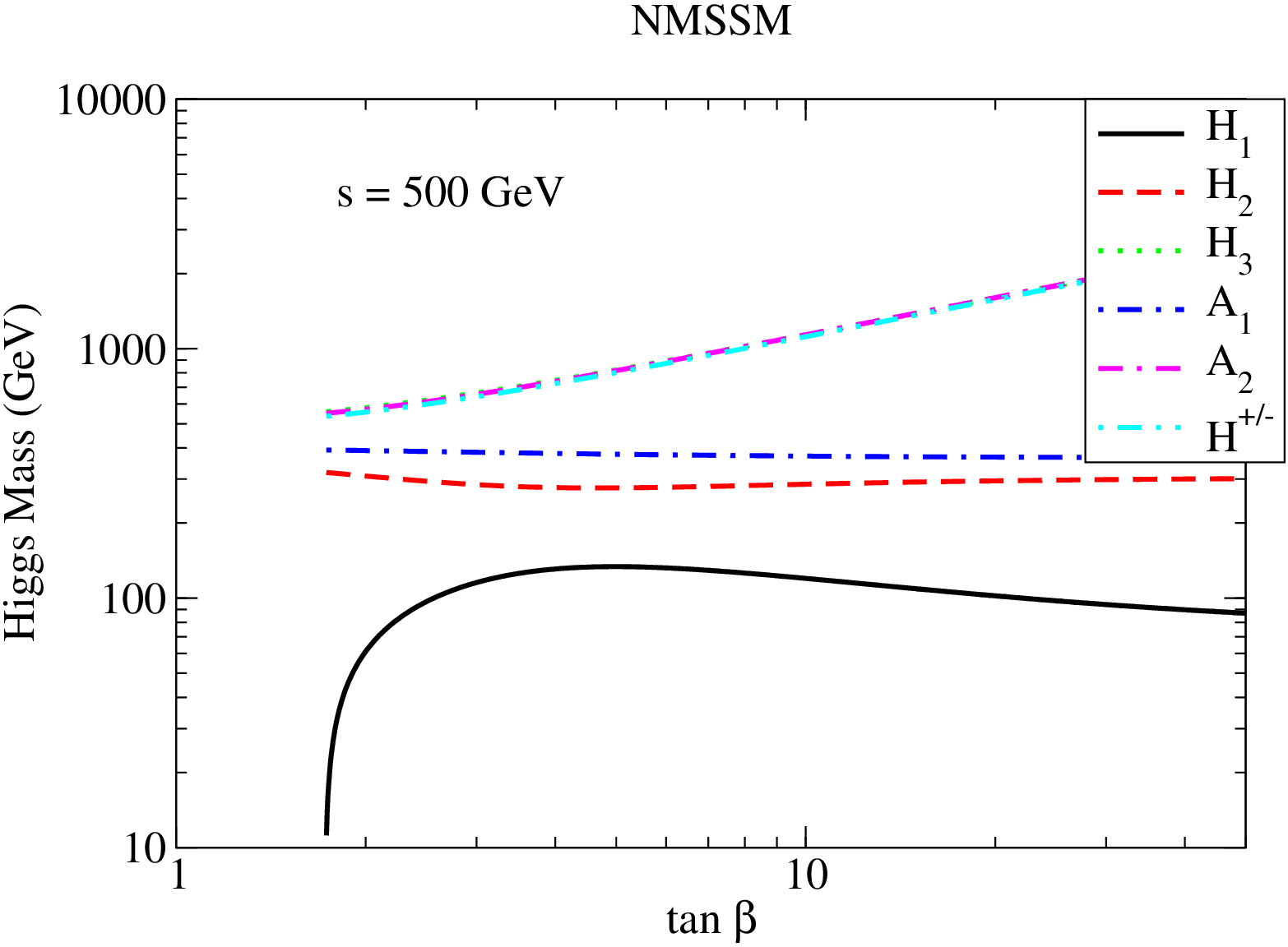}
\includegraphics[angle=-90,width=0.320\textwidth]{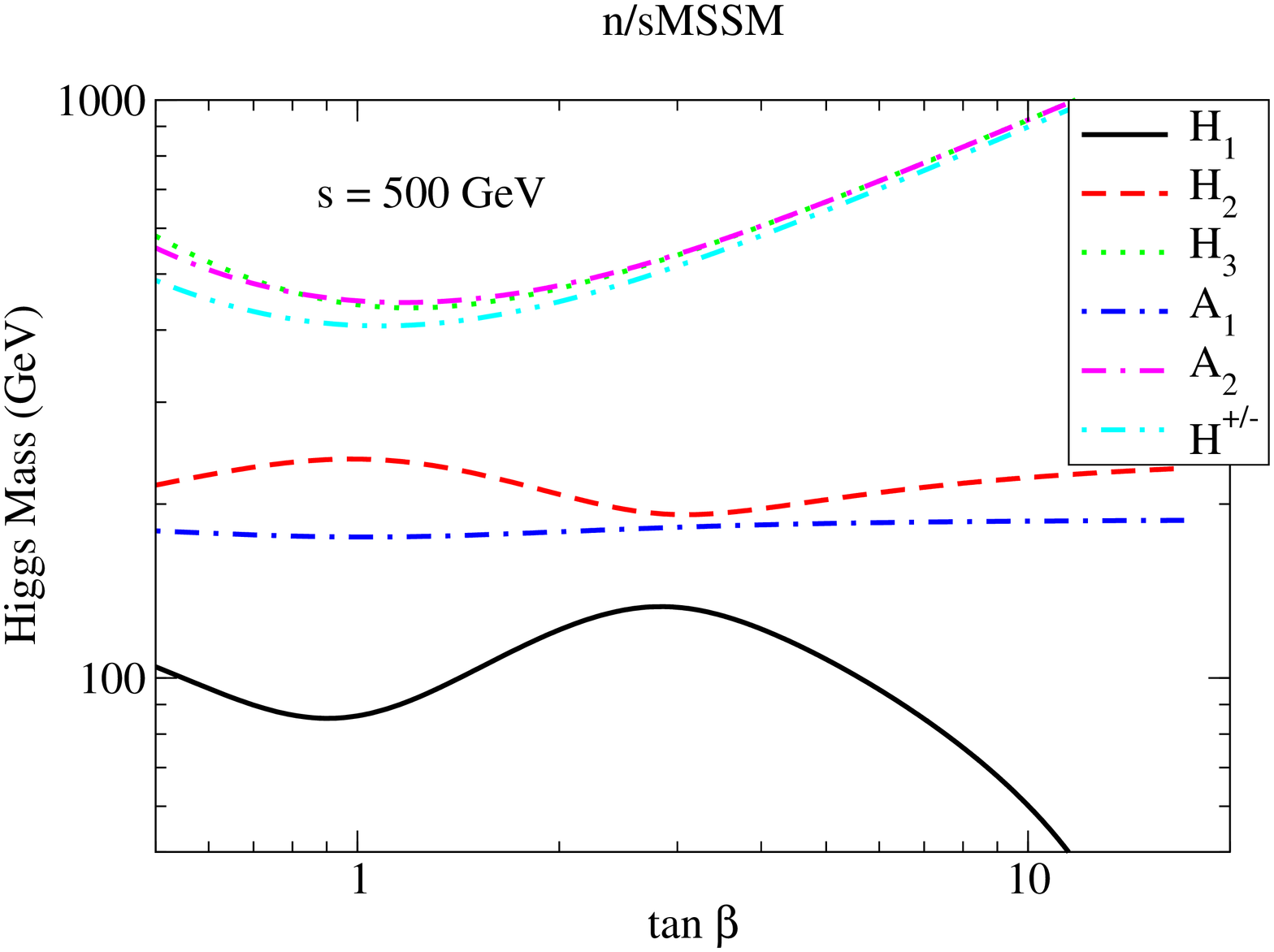}
\includegraphics[angle=-90,width=0.330\textwidth]{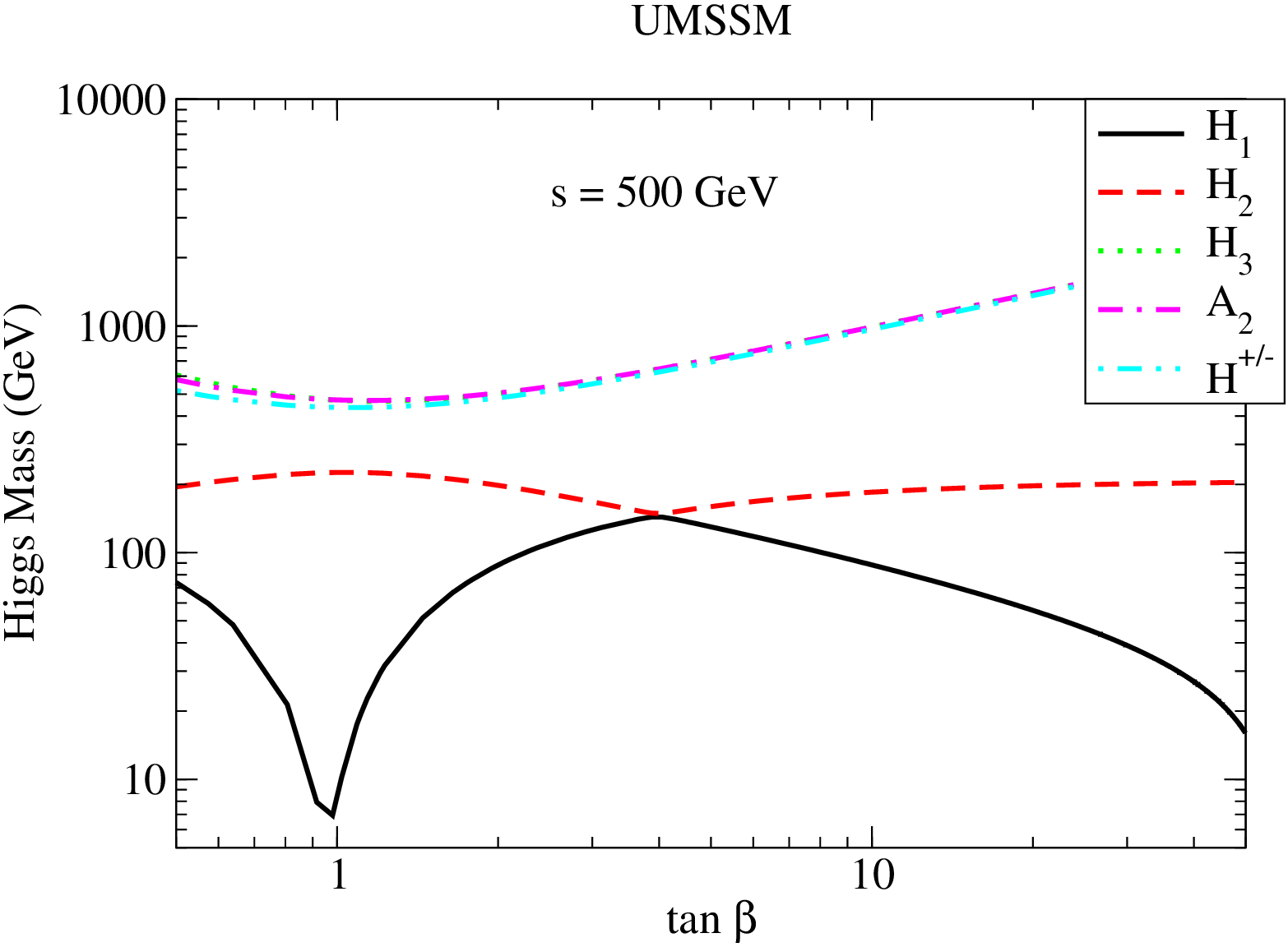}
\includegraphics[angle=-90,width=0.330\textwidth]{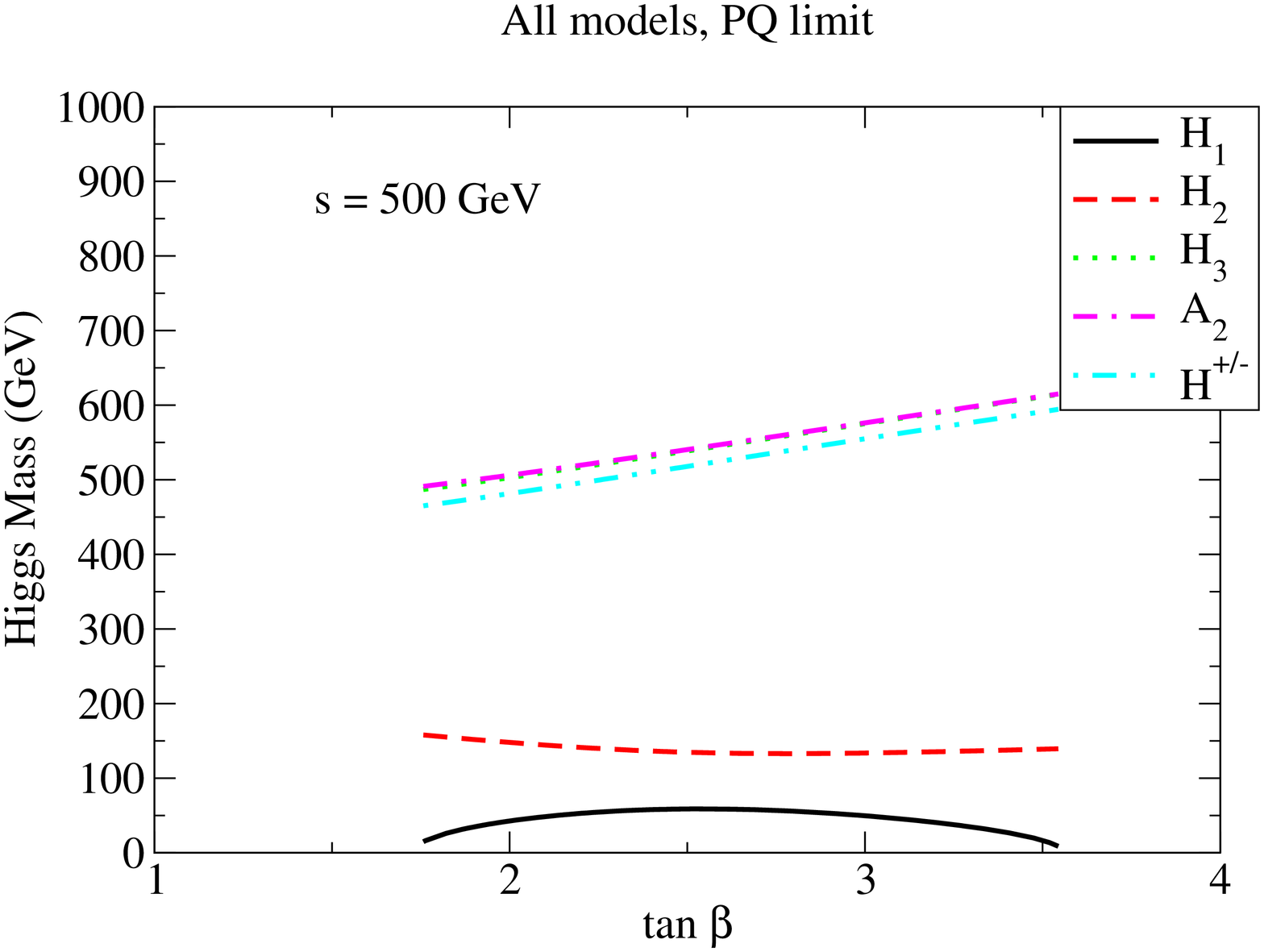}\\
\includegraphics[angle=-90,width=0.320\textwidth]{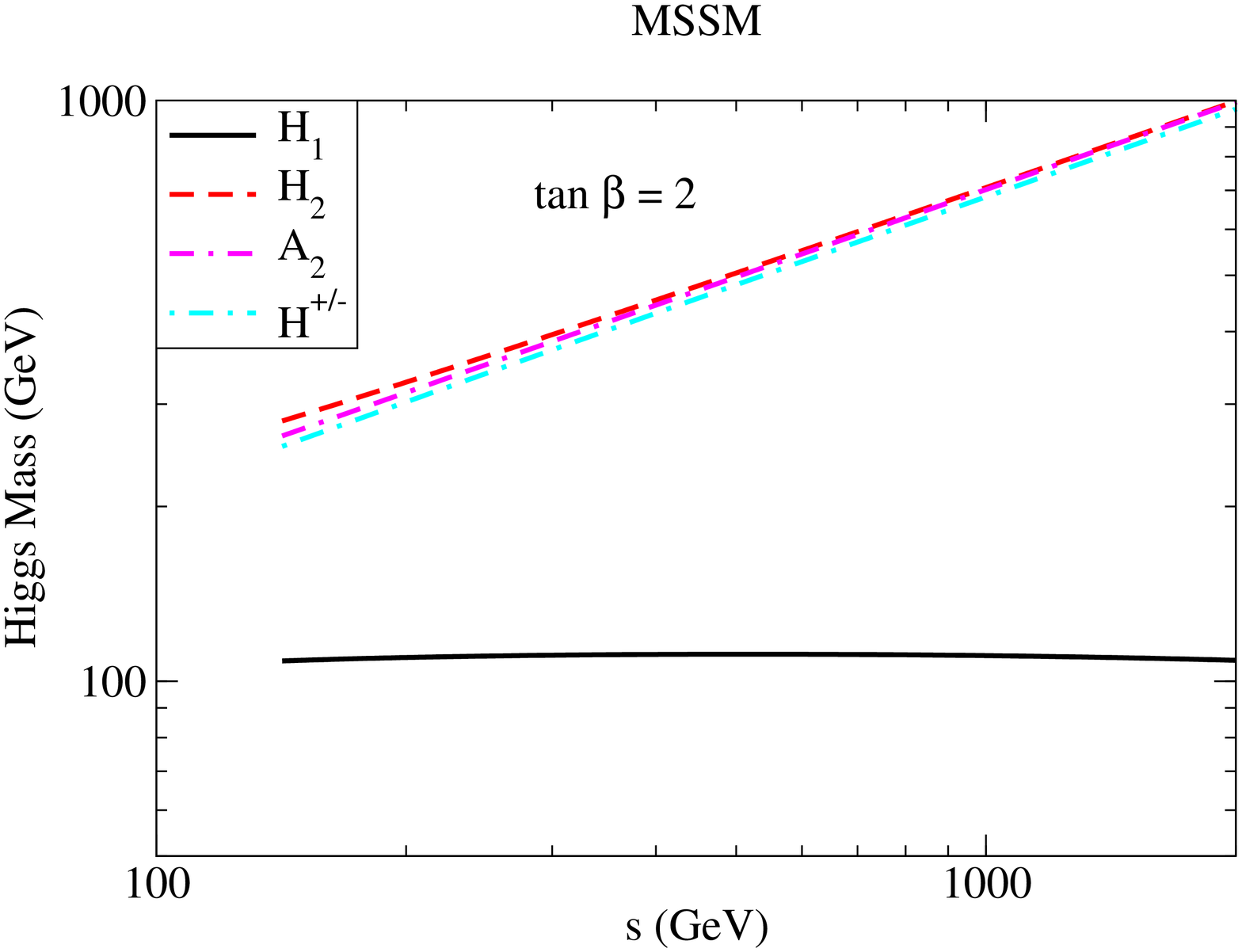}
\includegraphics[angle=-90,width=0.320\textwidth]{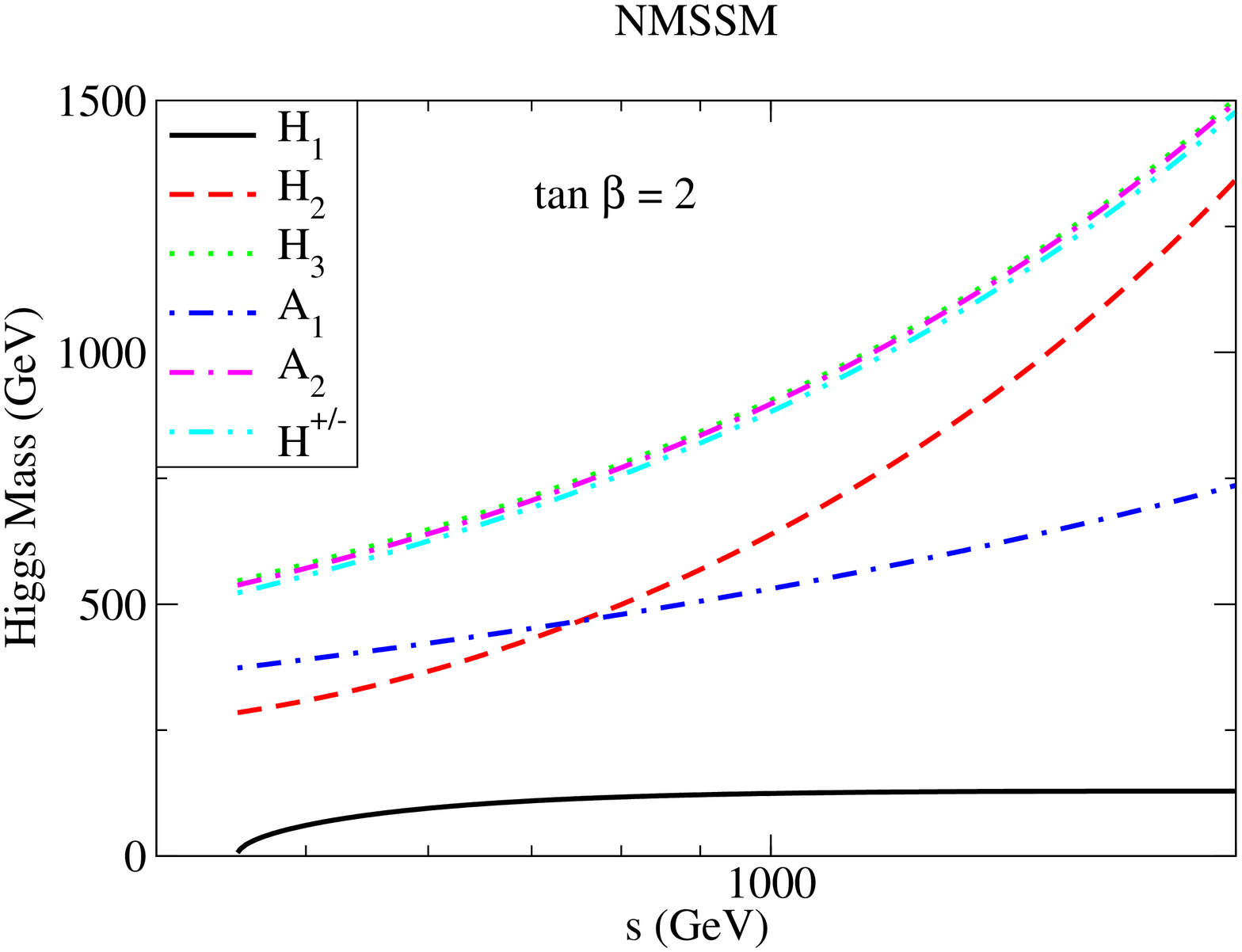}
\includegraphics[angle=-90,width=0.320\textwidth]{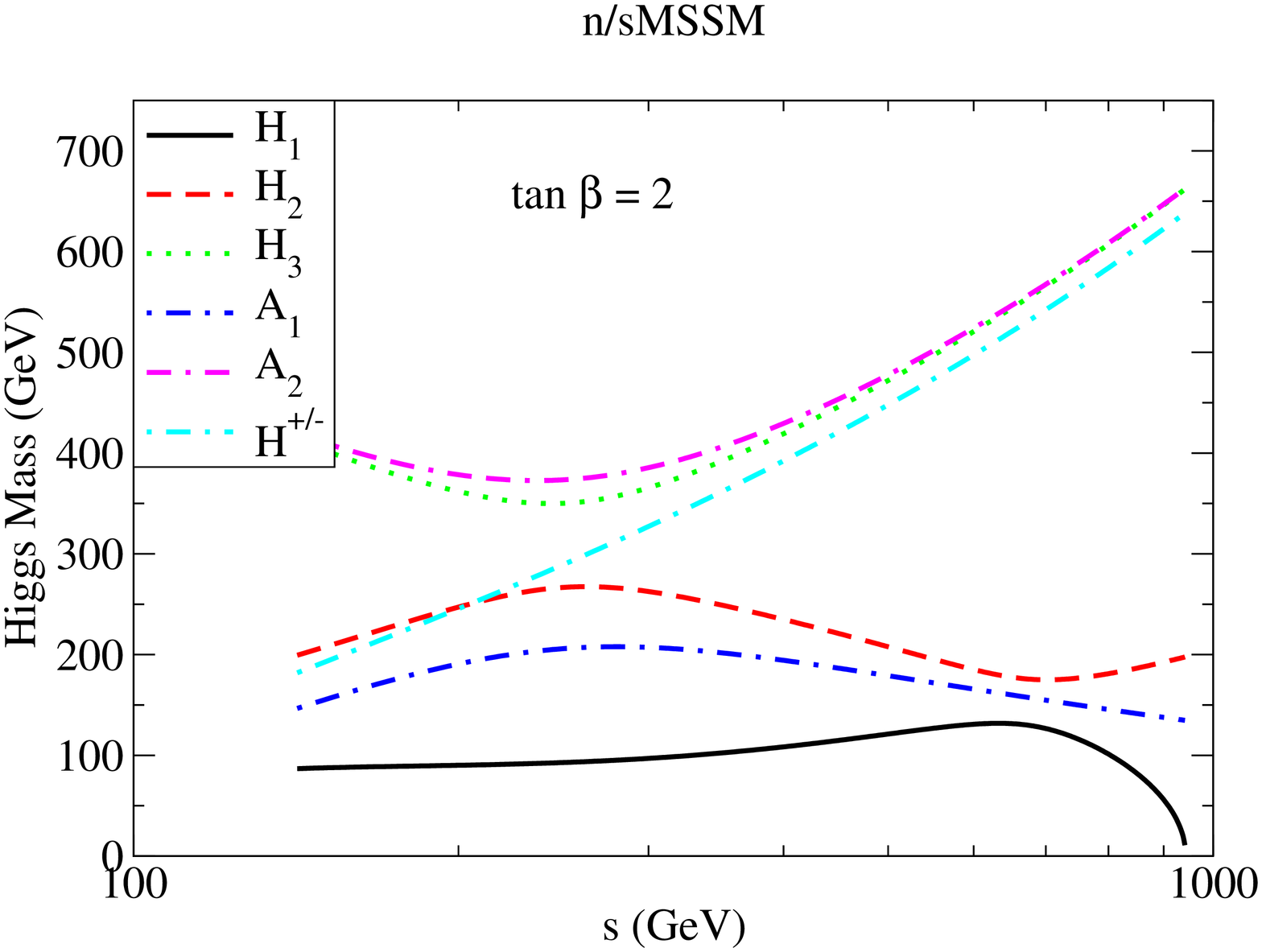}
\includegraphics[angle=-90,width=0.330\textwidth]{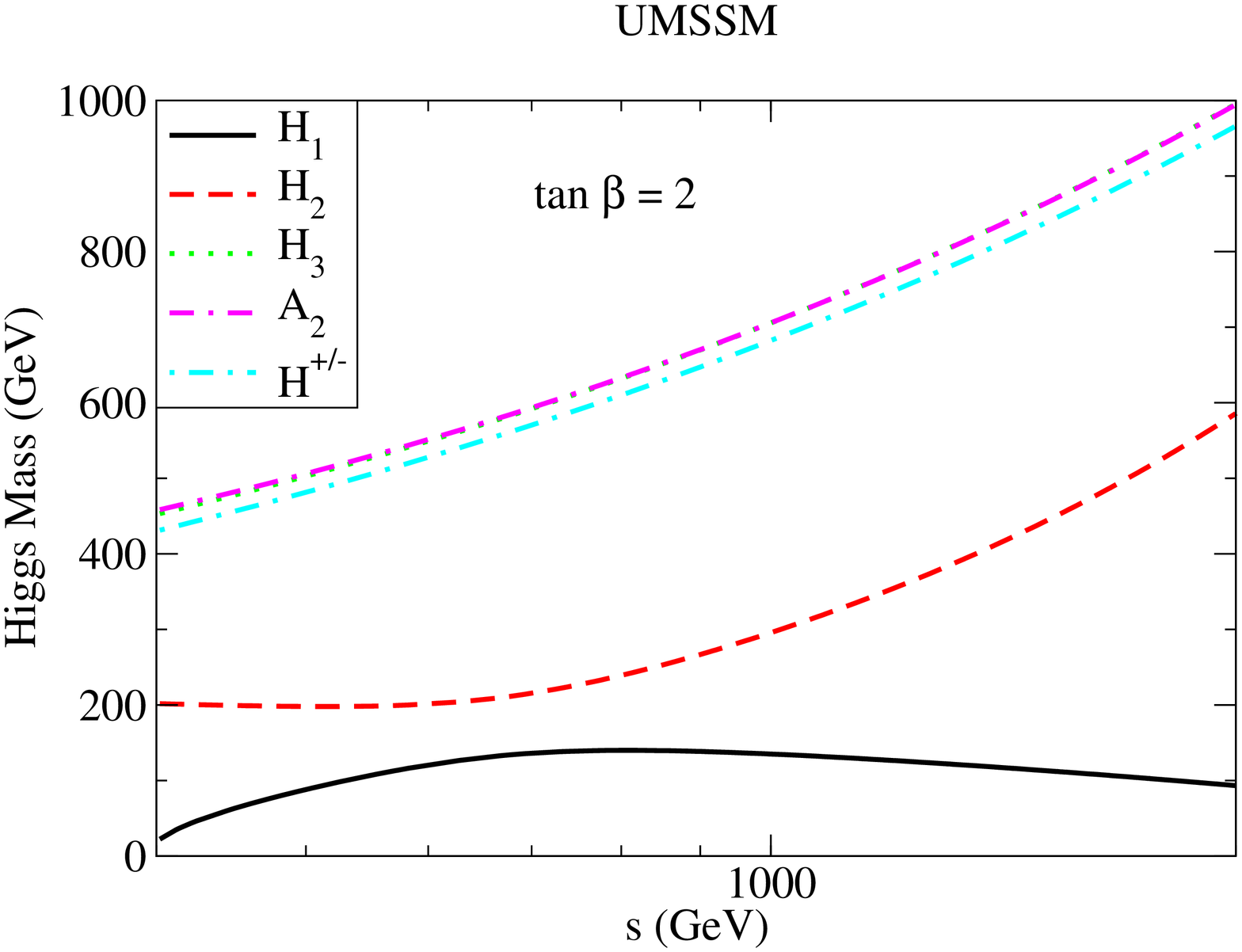}
\includegraphics[angle=-90,width=0.330\textwidth]{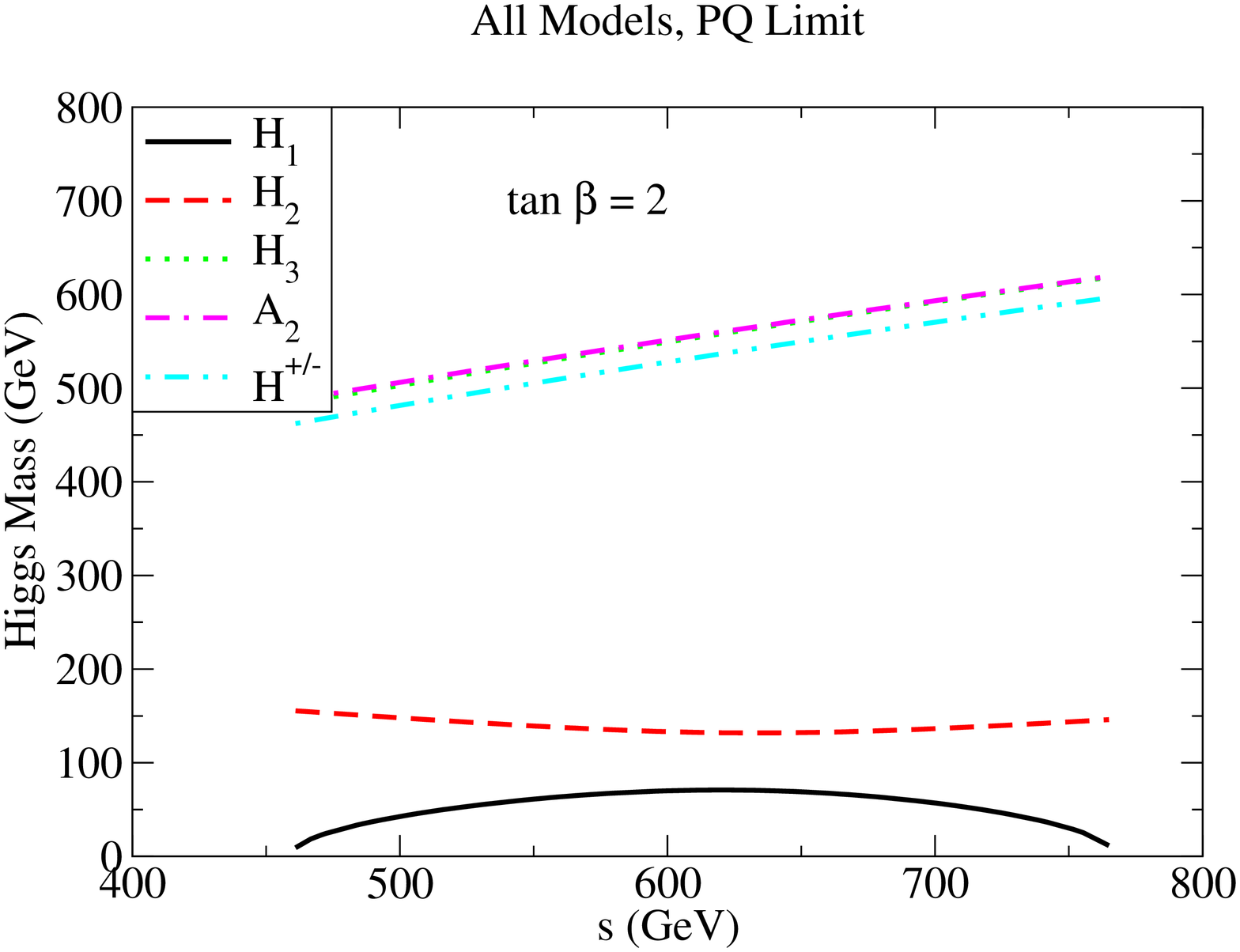}
\caption{Higgs masses vs. $\tan\beta$ and $s$ for the MSSM (with $\mu \equiv h_s s/\sqrt 2$), NMSSM, n/sMSSM, UMSSM, and the PQ limit for the extended models.  Only the theoretical constraints are applied.  Input parameters of  $A_s = 500$ GeV, $A_t = 1$ TeV, $M_{\tilde Q} = M_{\tilde U} = 1$ TeV, $\kappa = 0.5$, $A_\kappa = -250$ GeV, $M_{\rm n}=500$ GeV, $\xi_F = -0.1$, $\xi_S = -0.1$, $h_s = 0.5$, $\theta_{E6} = -\tan^{-1}\sqrt{5\over3}$, and the renormalization scale $Q = 300$ GeV are used.  Note that the $U(1)_{PQ}$ symmetry allows only one CP-odd Higgs boson to be massive.}
\label{fig:modelscans1}
\end{center}
\end{figure}

The one parameter plots performed in section \ref{sect:results} show the lightest CP-even and CP-odd Higgs masses vs. $\tan \beta$ and $s$.  In Fig. \ref{fig:modelscans1}, we show the complete Higgs mass spectrum of each model and the Peccei-Quinn limit versus $\tan \beta$ and $s$.  For comparison, the Higgs mass spectra of the mSUGRA/constrained-MSSM model can be found in Refs \cite{ref:spectcomp}.

In this figure, the level crossings between Higgs states are apparent as $s$ and $\tan \beta$ are varied.  
The general behavior of the heaviest CP-odd mass-squares are similar as they scale as $\tan \beta$ and $s$ with a minimum near $\tan \beta \sim 1$.  The first and second CP-even Higgs states cross at intermediate $\tan \beta$ in all the models for these set of parameters, a feature not present in the MSSM.  This suggests that the lightest CP-even Higgs typically has a maximal mass at some intermediate $\tan \beta$.

The $s$ plots shows a relatively constant lightest CP-even Higgs mass with minor variations when level crossings occur.  Since the mass-squared difference between the charged Higgs and heaviest CP-odd Higgs is, excluding model-dependent variations, on the order of $M_W^2$ both masses have the same dependence on $s$ and $\tan \beta$.  Departures from the similarities in mass signify a model dependence in $M_{H^\pm}$ and $M_{A_2}$.  For large $s$, the behavior of the charged Higgs boson is effectively the same among the models as its mass scales with $s$, see, e.g., Eq. (\ref{eq:chghiggs}).  The heaviest CP-even and CP-odd Higgs bosons masses also scale with increasing $s$ for most of the models in this parameter range.  However, the corresponding n/sMSSM masses decrease for increasing $s$ at low $s$.  This occurs when ${\cal M}_{33}$ becomes the dominant element of the mass-squared matrix.  This occurs for all the models at low enough $s$ (typically beyond the lower limit of $s$ allowed by the theoretical constraints).  However, the $-{\sqrt 2 \xi_S M_{\rm n}^3\over s}$ term specific to the n/sMSSM shifts the point where this cross-over from $M^2_{H_3} \sim s^2$ to $M^2_{H_3} \sim M_{\rm n}^3/s$ occurs to higher values of $s$, resulting in the distinctive departure from the behavior of the other models at low $s$ after theoretical constraints are applied.

\begin{acknowledgments}
This work was supported in part by the U.S.~Department of Energy
under grants Nos. DE-FG02-95ER40896 and DOE-EY-76-02-3071 and in part by the Wisconsin Alumni Research Foundation.  HL was supported by K. Matchev's U.S. DoE Outstanding Junior Investigator award under grant DE-FG02-97ER41209.  VB and PL thank the Aspen Center for hospitality during the initial stages of this work.  VB also thanks the University of Hawaii high energy physics group for hospitality.  We thank Andr$\acute{e}$ Sopczak for kindly providing the LEP constraint data.  HL thanks D. Demir for a useful discussion.
\end{acknowledgments}

\end{document}